\documentclass{aa}  
\usepackage{float}
\usepackage{array}
\usepackage[switch]{lineno}
\usepackage{graphicx,booktabs}
\usepackage{multirow}
\usepackage{color}
\usepackage{txfonts}
\usepackage{numprint}
\usepackage{textcomp,gensymb}
\usepackage{subcaption}
\usepackage{comment}
\usepackage{tabularx}
\usepackage{siunitx}
\usepackage{booktabs}
\usepackage{amsmath}
\usepackage{enumitem}
\usepackage[dvipsnames]{xcolor}
\usepackage{parskip}
\usepackage{rotating}
\usepackage{booktabs}
\usepackage{geometry}
\usepackage{adjustbox}
\usepackage{rotating}
\usepackage{lscape}
\usepackage{stfloats}
\usepackage{placeins}

\usepackage[percent]{overpic}
\usepackage{contour}
\contourlength{0.3pt}
\usepackage{xcolor}

\usepackage{booktabs,longtable}
\usepackage{caption}
\usepackage{txfonts}

\usepackage{hyperref}

\newcommand{\msun}{\ensuremath{M_{\odot}}}

\newcommand{\K}{\mathrm{K}}

\newcommand{\Msun}{\ensuremath{M_\odot}}
\newcommand{\Gauss}{\ensuremath{G}}

\newcommand{\tsmark}[1]{\nobreak\hspace{0.15em}\textsuperscript{\normalfont #1}}

\newcommand{\tsS}[1]{\tsmark{#1}} 
\newcommand{\tsB}[1]{\tsmark{#1}} 
\newcommand{\tsI}[1]{\tsmark{#1}} 
\newcommand{\tsM}[1]{\tsmark{#1}}

\usepackage{afterpage}

\begin{document}

   \title{The universality of the relation between magnetic fields and star formation in galaxies}

   \author{Davide Belfiori
          \inst{\ref{UniBO},\ref{IRA}}   
          \and
          Sergio Martin-Alvarez \inst{\ref{KIPAC}}
          \and
          Enrique Lopez-Rodriguez\inst{\ref{USC}}
          \and
          Rosita Paladino \inst{\ref{IRA},\ref{ESO}}}
        
   \institute{DIFA, Universit\`{a} di Bologna, via Gobetti 93/2, I-             40129 Bologna, Italy \label{UniBO}
              \and
              INAF, Istituto di Radioastronomia, Via Gobetti 101, I-40129 Bologna, Italy \label{IRA}
              \and
              Kavli Institute for Particle Astrophysics \& Cosmology (KIPAC), Stanford University, Stanford, CA 94305, USA\label{KIPAC}
              \and
              Department of Physics \& Astronomy, University of South Carolina, Columbia, SC 29208, USA \label{USC}
              \and 
              ESO, Karl Schwarzchild Srt. 2, 85478 Garching bei München, Germany \label{ESO}
              }

   \date{Received XXXX; accepted XXXX}

\abstract
{The interstellar medium (ISM) is permeated by magnetic fields affecting gas dynamics and star formation. These magnetic fields have been observed to be related to the supernova-driven turbulence. However, whether this scaling is universal across galaxy properties, star formation levels, ISM phases, and energy budgets remains unclear.}
{We quantify the dependence of magnetic fields on the star formation activity, spanning seven orders of magnitude, and encompassing regular and starburst galaxies. In addition, we study the energetic balance of the multi-phase ISM for thermal and non-thermal (turbulent, cosmic-ray) components.}
{We analysed 19 spiral disc galaxies from the cosmological RTnsCRiMHD \textsc{Azahar-a} suite. For each system we built line-of-sight–integrated maps, selected the inner star-forming disc, and measured the median magnetic field strength, star formation rate (SFR), surface density ($\Sigma_{\rm SFR}$), specific SFR (sSFR), as well as the thermal, turbulent, magnetic, and cosmic-ray specific energies.}
{We find an approximately universal magnetic field strength $B$–SFR scaling with \(\alpha \simeq 0.2\text{--}0.3\) across galaxy mass, inclination, and neutral phases (cold and warm). The $B$–\(\Sigma_{\rm SFR}\) slope \(\alpha_{\Sigma} \sim 1/3\) is in agreement with a SN-driven, turbulence-regulated magnetic field. The \(B\)–sSFR relation provides the best separation across galaxy types and multiphase ISM. We find all specific non-thermal energies to increase with SFR, with the magnetic component having the highest slope. Energetically, neutral gas in the simulations is typically turbulence-dominated with a secondary contribution from cosmic rays, and in approximate equipartition with magnetic energies for systems with significant star formation ($\mathrm{sSFR}\gtrsim 0.1\,\mathrm{Gyr^{-1}}$, SFR $\gtrsim 1\, \mathrm{M_{\odot}} \, \mathrm{yr}^{-1}$).
The trends of our simulations are consistent with observations, which show similar slopes ($\alpha\!\approx\!0.25\text{--}0.35$) across a wide range of galaxy types and environments.
}
{Our results show that SN-driven turbulence is the main amplification mechanism yielding the near-universal $B$-SFR. This results in magnetic fields playing a dynamical role in the neutral ISM in galaxies with intense star formation. }
{}

\maketitle

\section{Introduction}
Magnetic fields are ubiquitous in the interstellar medium (ISM)  of external galaxies \citep{Moss2012, Basu2013, Krause2014, Beck2019, Shukurov_Subramanian_2021, Lopez-Rodriguez2022b}. These magnetic fields affect multiple astrophysical processes across different scales, such as inducing gas inflows toward galactic centres \citep{Kim2012} and regulating the collapse of molecular clouds essential for star formation \citep{Hennebelle2019}. Magnetic fields properties such as strength and pitch angles have also been shown to be in a close relationship with the properties of the ISM of the galaxies hosting them \citep{VanEck2015,Chyzy2017,Beck2019, Borlaff2023}.

Polarimetric observations in radio and far-infrared (FIR) are able to constrain magnetic field orientations, depolarisation, and equipartition-level strengths, yet the derived quantities represent line-of-sight–integrated projections of intrinsically three-dimensional field and emission structures \citep{Beck2015}. Cosmological and ISM-scale magnetohydrodynamical (MHD) simulations complement these diagnostics: they follow magnetic amplification and transport in a controlled setting, connect fields to gas phases and feedback channels, and enable energy-budget accounting for magnetic, turbulent, thermal, and cosmic ray (CR) components without assuming equipartition (e.g. \citealp{Rieder2016, Pakmor2017, Pfrommer2017, MartinAlvarez2021, Martin-Alvarez2023, Martin-Alvarez2024}). Increasingly sophisticated and quantitatively calibrated, these simulations now provide an effective laboratory for testing theory against observations.

The leading theory for the origin of galactic magnetic fields resides in the dynamo mechanisms. Initially weak, primordial seed fields are rapidly amplified to strengths of order microgauss within $\lesssim10^8$\,yr by supernova (SN)-driven turbulence and spiral shocks (small-scale dynamo; e.g. \citealp{Rieder2016};  \citealp{Martin-Alvarez2018}; \citealp{Pakmor2024}), producing predominantly tangled fields that saturate near equipartition with ISM turbulence. On gigayear timescales, the large-scale $\alpha$–$\Omega$ dynamo organises the mean field: differential rotation winds poloidal into toroidal components, while helical turbulence regenerates the poloidal field, enabling exponential growth. This process continues until back-reaction (magnetic tension and helicity conservation) limits further amplification, yielding a saturated mean field with energy density comparable to turbulent motions \citep{Subramanian1998, Brandenburg2005, Subramanian2016, Subramanian2019}. Despite this framework, reproducing the full diversity of observed magnetic structures and strengths remains an open, active area of research.

A key observational link to the amplification mechanisms resides in the correlation between magnetic field strength ($B$) and star formation activity. 
This connection ultimately arises from the empirical Kennicutt–Schmidt relation that relates the star formation rate surface density ($\Sigma_{\rm SFR}$) to the amount of available gas: \(\Sigma_{\rm SFR}\propto \Sigma_{\rm gas}^{N}\), where \(\Sigma_{\rm gas}\) represents the total (atomic + molecular) gas surface density, and \(N \simeq 1\text{--}1.5\) in nearby galaxies \citep{Schmidt1959, Kennicutt1998}.
Stellar feedback associated with this star formation injects mechanical energy into the ISM through SN explosions and stellar winds, driving turbulence that both regulates the gas dynamics and amplifies magnetic fields through the small-scale dynamo. 
In a steady state, the magnetic energy constitutes a roughly constant fraction of the turbulent energy, 
so that \(B \propto \sqrt{\rho}\,v_{\rm turb}\), with \(\rho\) the gas density and \(v_{\rm turb}\) the turbulent gas velocity. 
If turbulence is approximately feedback-regulated, such that \(v_{\rm turb}\) varies only weakly with \(\Sigma_{\rm SFR}\), 
then \(B\propto \sqrt{\rho}\propto \sqrt{\Sigma_{\rm gas}}\), 
which combined with the Kennicutt–Schmidt relation gives 
\(B \propto \Sigma_{\rm SFR}^{\,1/(2N)}\). 
For a classical index of \(N=3/2\), this yields \(B \propto \Sigma_{\rm SFR}^{1/3}\) \citep{Schleicher2013}.

Radio polarimetric observations find a similar relation for spiral galaxies ($B \propto \Sigma_{\mathrm{SFR}}^{0.34}$, \citealt{Niklas1997}) and dwarf irregular galaxies ($B \propto \Sigma_{\mathrm{SFR}}^{0.25}$, \citealt{Chyzy2011}). 
Using spatially resolved low-frequency radio continuum from second release of the Low Frequency Array Two-metre Sky Survey (LOFAR LoTSS-DR2), \citet{Heesen2022} found a sub-linear $B$–$\Sigma_{\rm SFR}$ relation across 39 nearby galaxies: the local slope $0.18$ steepens to $\simeq0.22$ for galaxy-averaged values and to $\simeq0.28$ after correcting for cosmic-ray transport. 
However, a comprehensive analysis incorporating physically motivated models is still needed to fully understand the origin and universality of this relation.

In this work, we use the \textsc{Azahar} cosmological MHD framework to quantify how magnetic fields are related to star formation activity and their energy balance within the multi-phase ISM. We focus on theory–observable links such as the $B$–SFR and $B$–$\Sigma_{\rm SFR}$ scalings, and on energy-budget diagnostics (e.g. the balance among magnetic, turbulent, thermal, and CR components). The analysis is phase-resolved, with results reported separately for distinct ISM phases.

The paper is organised as follows. In Section \ref{sec:methods} we describe the properties of the simulation set-up used for this work, as well as the algorithm used to select the simulated galaxy sample and the disc regions. Our main results are explored in Section \ref{sec:results}, while we summarise and conclude our work in Section \ref{sec:conclusions}. The maps of the simulated galaxies used for this work are shown in Appendix \ref{sec:gal_maps}. The tables summarising the estimates for our simulated galaxies and the observed samples used for comparison are reported in Appendix \ref{sec:tables}. We review the robustness of our results with respect to varying galaxy inclination in Appendix \ref{sec:faceon_test}.

\begin{figure*}[htbp]

\centering
\begin{minipage}[t]{0.66\textwidth}
  \vspace{0pt} % ensure top alignment
  \includegraphics[width=1.01\textwidth, height=0.97\textwidth]{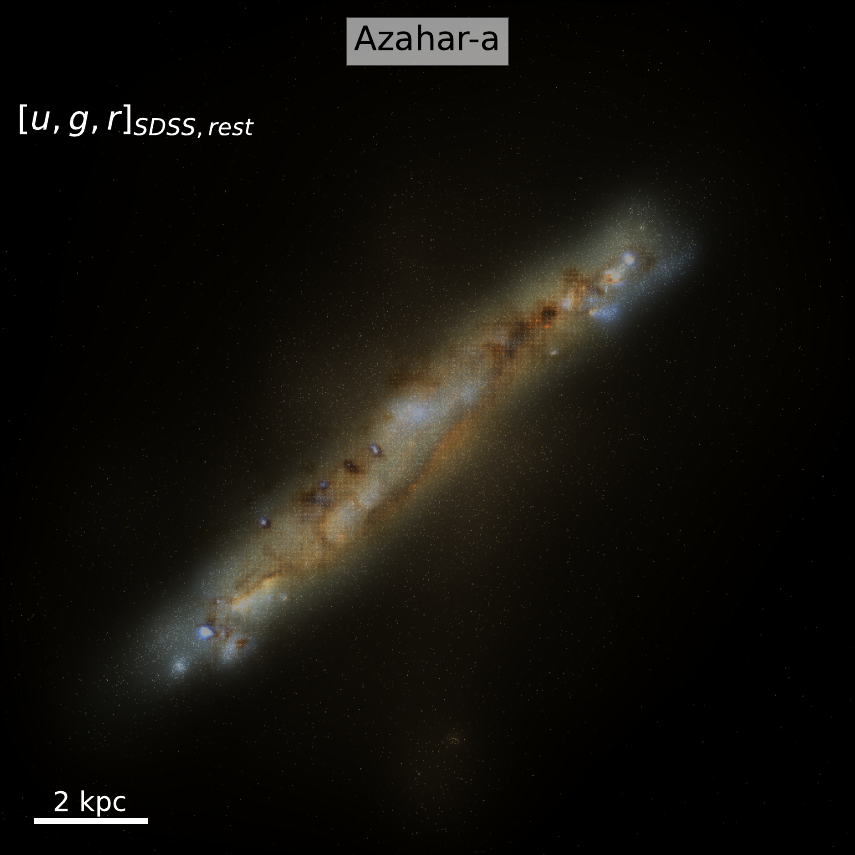}
\end{minipage}%
\begin{minipage}[t]{0.33\textwidth}
  \vspace{0pt} % ensure top alignment
  \includegraphics[width=\textwidth]{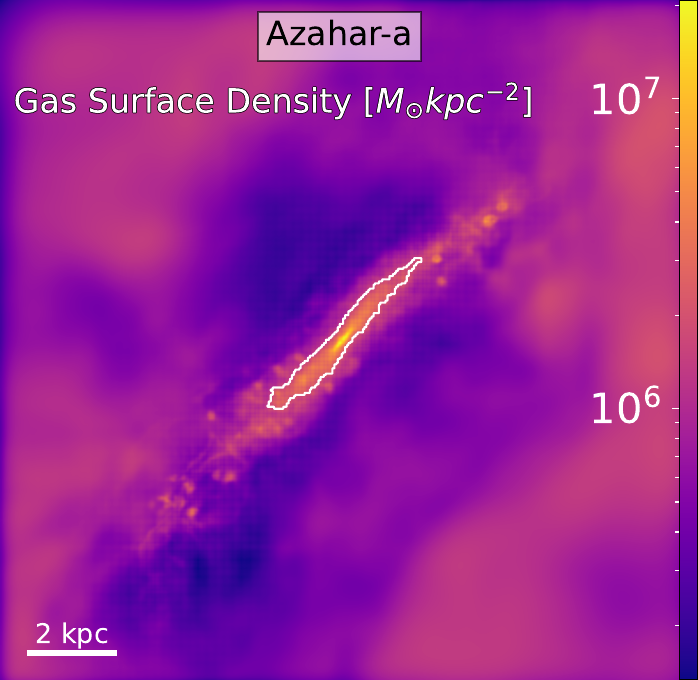}\\[-0.2em]
  \includegraphics[width=\textwidth]{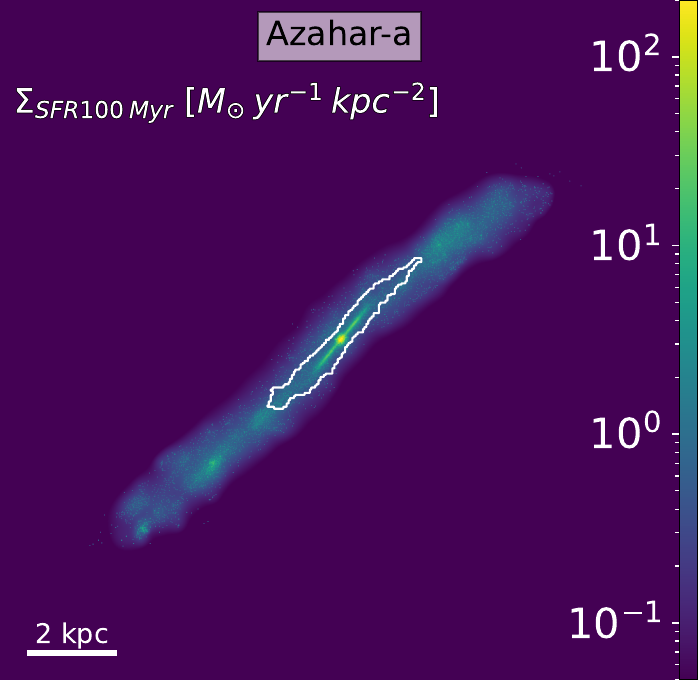}
\end{minipage}

\vspace{-0.2em}

\begin{minipage}[t]{0.33\textwidth}
  \includegraphics[width=\textwidth]{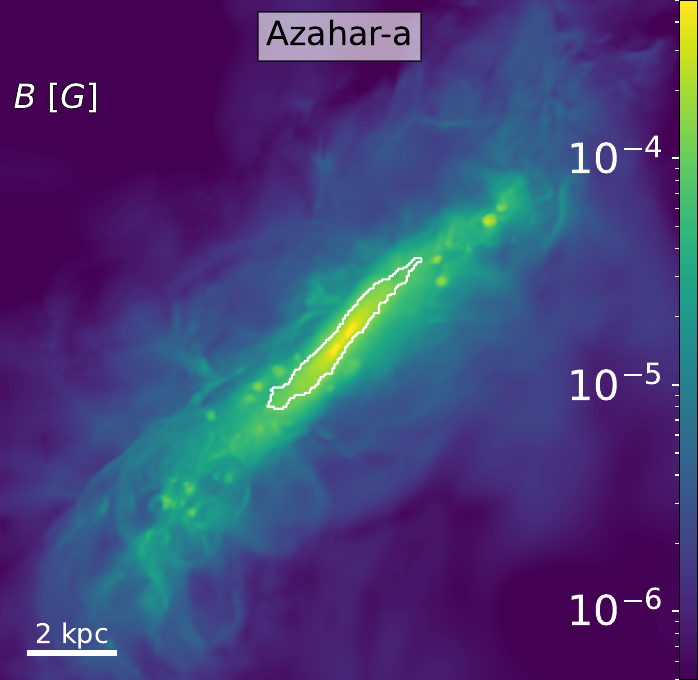}
\end{minipage}%\hspace{-1mm}
\begin{minipage}[t]{0.33\textwidth}
  \includegraphics[width=\textwidth]{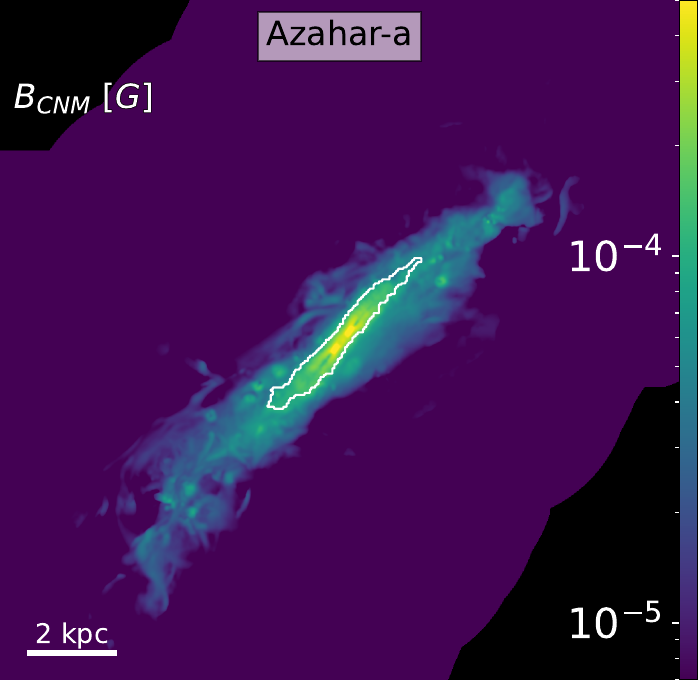}
\end{minipage}%\hspace{-1mm}
\begin{minipage}[t]{0.33\textwidth}
  \includegraphics[width=\textwidth]{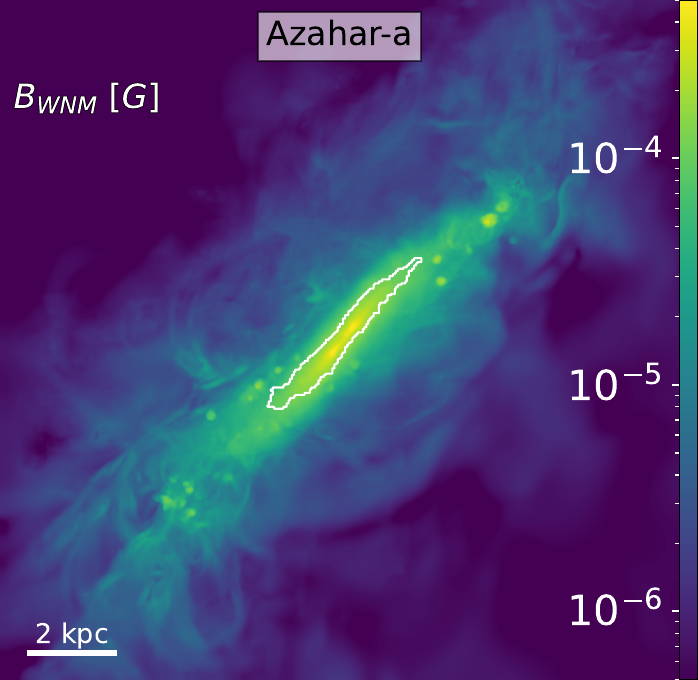}
\end{minipage}

\caption{Multi-panel view of the simulated galaxy \textsc{Azahar-a}. 
The large left panel shows the rest-frame mock optical (SDSS $ugr$) image of the system. 
The two upper right panels display the gas surface density and the SFR surface density, averaged over the past $100$~Myr. 
The three bottom panels show the total magnetic field strength for the full gas, CNM, and WNM components.
White contours in all panels mark the disc region adopted for the analysis. }
\label{fig:Azahar-a_images}
\end{figure*}

\section{Numerical methods}
\label{sec:methods}

\subsection{The \textsc{Azahar} model}
\label{ss:Azahar_methods}

We used the \textsc{Azahar} cosmological simulation \citep{Martin-Alvarez2025AAS} to study the relation between magnetic field and star formation in galaxies. \textsc{Azahar} is a `full-physics' model using high-resolution galaxy formation suite incorporating the non-thermal physical processes of MHD, radiative transfer (RT), and CRs.
The \textsc{Azahar} simulations are generated with a customised version of the publicly available \textsc{RAMSES} code \citep{Teyssier2002}.
This code solves for the coupled evolution of baryons and dark matter using an Eulerian grid-based approach for the gas, and a collisionless particle-based scheme for stars and dark matter. These components are coupled through a gravitational potential solved on the mesh. The \textsc{RAMSES} code employs an octree structure Eulerian grid to resolve the magneto-hydrodynamical evolution of the gas component, incorporating an adaptive mesh refinement (AMR) scheme to increase the resolution in regions of interest. The simulations feature an extensive convex hull zoom region of approximately 10~comoving Mpc, which we used for our study. This region allows the AMR scheme to refine down to a minimum cell-size spatial resolution of $20$~pc, and has particle mass resolutions of $m_\mathrm{DM} \sim 4\,\times\,10^{5}$\,\msun~and $m_\mathrm{*} \sim 4\,\times\,10^{4}$\,\msun~for dark matter and stellar components, respectively.

We solved the evolution of magnetic fields through a constrained transport (CT) method \citep{Teyssier2006, Fromang2006}, which ensures the solenoidal condition ($\nabla \cdot \mathbf{B} = 0$) to machine precision. We assumed ideal MHD, with the magnetic field initialised to a uniform Biermann battery-like kinematic seed strength of $B \sim 10^{-20}\,\mathrm{G}$ \citep{Martin-Alvarez2018}. In addition to this primordial seed, SN feedback further amplifies the field by injecting small-scale magnetic loops \citep{Martin-Alvarez2021}, providing a sustained source of magnetisation.
Radiative transfer uses the implementation by \cite{Rosdahl2013}, with a configuration resembling that of the \textsc{SPHINX} simulations \citep{Rosdahl2018,Rosdahl2022}, modelling three frequency bins corresponding to the ionisation of H~\textsc{i} ($13.6\,{\rm eV} \lesssim \epsilon_{\rm photon} \lesssim 24.59\,{\rm eV}$), He~\textsc{i} ($24.59\,{\rm eV} \lesssim \epsilon_{\rm photon} \lesssim 54.42\,{\rm eV}$), and He~\textsc{ii} ($54.42\,{\rm eV} \lesssim \epsilon_{\rm photon}$). The zoom sub-volume reionisation is primarily driven by self-consistent star formation, and incorporates a background model by \cite{Haardt1996}.
We employed the CR solver implementation by \citet{Dubois2016} with the improvements by \citet{Dubois2019}, but note that the RTnsCRiMHD model has the anisotropic CR streaming deactivated (nsCR). The CR solver accounts for hadronic and Coulomb CR cooling, and anisotropic diffusion with a constant diffusion coefficient of $\kappa_{\parallel} = 3\times 10^{28} \mathrm{cm}^2 \mathrm{s}^{-1}$ \citep{Ackermann2012, Cummings2016, Pfrommer2017}.

The model incorporates gas cooling above and below $10^4$~K through interpolated \textsc{CLOUDY} tables \citep{Ferland1998} and following \citet{Rosen1995}, respectively. 
Furthermore, star formation is modelled via a magneto-thermo-turbulent scheme \citep{Kimm2017, Martin-Alvarez2020}, enabled at the highest level of refinement when the gravitational pull overcomes local turbulent, thermal, and magnetic support. The star formation rate (SFR) follows a Schmidt law \citep{Schmidt1959, Krumholz2012}: $\dot{\rho}_\star = f_{\mathrm{ff}} \frac{\rho_\mathrm{gas}}{t_{\mathrm{ff}}}$, where $\rho_\mathrm{gas}$ is the gas density and $f_{\mathrm{ff}}$ is the local efficiency derived from turbulent models \citep{Federrath2012}. 
Supernova feedback is implemented through mechanical injection \citep{Kimm2014}, whereby star particles deposit mass, momentum, and energy. Each SN assumes a Kroupa initial mass function (IMF) \citep{Kroupa2001}, releasing a total energy of $\sim 10^{51}$~erg per $10~M_\odot$, returning $\eta_{\mathrm{SN}} = 0.213$ of the mass to the ISM, with $\eta_{\mathrm{metals}} = 0.075$ as metals. The SNe inject magnetic energy through local small-scale loops \citep{Martin-Alvarez2021}, with an injection energy of $0.01\, E_{\mathrm{SN}}$ per event, leading to magnetisations comparable to SN remnants \citep{Parizot2006}, and consistent with values used by other authors \citep{Beck2013, Butsky2017, Vazza2017}. The SNe are the exclusive source of CR energy, injecting 10\% of the SN energy \citep{Morlino2012, Helder2013}. The RTnsCRiMHD model of the \textsc{Azahar} simulations has already been described and studied by \citet{Witten2024}, \citet{Dome2025} and \citet{Yuan2025}, and will be introduced in further detail in Martin-Alvarez et al. (in prep.).

In addition to \textsc{Azahar}, we complemented our analysis with a suite of zoom-in simulations of a Milky-Way-like galaxy, referred to as \textsc{nut}. This suite is based on the initial conditions (ICs) introduced by \citealp{Powell2011}, and used here to probe seed-field and SN-injection dependencies. The full \textsc{nut} set-up is summarised in Section~\ref{sec:NUTsetup}.

\subsection{The simulated galaxy sample} \label{The simulated galaxy sample}
For our analysis, we selected a subset of galaxies from the \textsc{Azahar} catalogues (v4.7) corresponding to the population present in the simulation at $z = 3$.
At this redshift, the simulation provides a population of galaxies spanning high SFRs and SFR surface densities, comparable to those of local starburst systems, which would be rare at lower redshift in a cosmological volume of this size.
These galaxies will be compared to an observational compilation of nearby systems spanning a wide range of SFRs and morphologies (described in Section~\ref{subsec:B_vs_SFR}).

To ensure a consistent comparison with this observational sample, which spans SFRs in the range $2.8\times10^{-3} < \mathrm{SFR} < 155\,\mathrm{M_\odot\,yr^{-1}}$, we adopted a minimum SFR threshold of $10^{-3}\,\mathrm{M_\odot\,yr^{-1}}$ and included all simulated systems with higher SFRs.
Additionally, we want to identify similarities with the starburst galaxies in our polarisation observations, which are predominantly viewed edge-on. We therefore selected only edge-on projections of the simulated galaxies. This choice also facilitates a clean separation between the star-forming disc and the galactic outflow regions, which is essential for measuring the magnetic field structure in the disc. In this context, we excluded interacting systems and galaxies lacking a coherent rotating disc, since their disturbed morphologies prevent a reliable distinction between disc and outflow components. 
The edge-on orientation of the galaxy, as provided by the galaxy catalogues, corresponds to a line of sight perpendicular to a fixed, arbitrary vector and the angular momentum of the galaxy as computed from the total baryons, with their centring calculated using a shrinking sphere method \citep{Martin-Alvarez2023}.
The final sample is therefore restricted to systems undergoing predominantly secular evolution. This selection yields a population of relatively stable, disc-dominated galaxies whose internal ISM properties are more directly comparable to those of star-forming galaxies in the local Universe.

The resulting sample consists of 19 galaxies with stellar masses of $2.41\times10^{8} \le M_\star/\mathrm{M_\odot} \le 2.07\times10^{10}$ 
(median $1.05\times10^{9}\,\mathrm{M_\odot}$, $\sigma = 4.5\times10^{9}$),
SFRs averaged over the past $100$~Myr of
$1.22\times10^{-3} \le \mathrm{SFR}_{100}/(\mathrm{M_\odot\,yr^{-1}}) \le 6.02$ 
(median $6.6\times10^{-2}$, $\sigma = 1.4$),
specific SFRs of
$4.63\times10^{-3} \le \mathrm{sSFR}_{100}/(\mathrm{Gyr^{-1}}) \le 8.51\times10^{-1}$ 
(median $7.6\times10^{-2}$, $\sigma = 1.7\times10^{-1}$),
and SFR surface densities of
$1.5\times10^{-2} \le \Sigma_{\mathrm{SFR}}/(\mathrm{M_\odot\,yr^{-1}\,kpc^{-2}}) \le 1.59$
(median $3.45\times10^{-1}$, $\sigma = 4.6\times10^{-1}$).
In Appendix \ref{sec:tables}, Table ~\ref{tab:gal_properties} shows the unique galaxy ID, their stellar mass, SFRs, specific SFRs, and SFR surface density.

We computed projected maps for the gas surface density, SFR surface density, density-weighted magnetic field strength, and total energies associated with magnetic fields, CRs, thermal motions, and turbulence\footnote{Turbulent energies were computed following the small-scale motion separations described in \citep{Martin-Alvarez2022}.}.
For each quantity, we also generated ISM phase-separate maps by isolating the contributions from the cold and warm neutral media (CNM and WNM, respectively). Cells were separated into phases according to their temperature, $T$, and self-consistent hydrogen ionisation state, $x_\mathrm{HII}$, following
\begin{itemize}[label=-]
    \item Cold neutral medium (CNM): \(T \le 200\,\mathrm{K}\);
    \item Warm neutral medium (WNM): \(200\,\mathrm{K} < T \le 10^6\,\mathrm{K}\), with the WNM gas density defined as the neutral fraction of the total warm gas density, \(\rho_\text{gas,WNM} = \rho_\text{gas,warm}\,(1 - x_\text{HII})\);
\end{itemize}
\noindent
as employed in \citet{Martin-Alvarez2024}.

To separate the galaxy disc from galaxy outflows, we used the gas surface density maps to define a disc mask for each galaxy. 
In order to define the mask, we followed the approach employed in section 2.3 of \citealt{Barnouin2023}: we estimated a characteristic background level of diffuse gas from the surface density distribution, defined as the most populated bin in the intensity histogram. The disc region was then identified as the contiguous area where the gas surface density exceeds ten times this background level. This assures a robust noise estimation for galaxies with very different properties, such as size and gas surface density distribution.

As an illustrative example, Figure~\ref{fig:Azahar-a_images} presents a multi-panel view of the simulated galaxy \textsc{Azahar-a}, the most massive system in the \textsc{Azahar} suite.  
The large left panel shows the mock optical ($ugr$) image of the galaxy for reference. These were generated with our modified version of the \textsc{sunset} code\footnote{\textsc{sunset} is part of the publicly available library within the \textsc{ramses} distribution.}, which produces synthetic broadband images by assigning single stellar population (SSP) emission models to each stellar particle in the simulation, and mapping their emission through telescope-filter transmissivity curves. During this process, we also applied line-of-sight dust attenuation, computed from a 3D dust distribution. The two upper right panels display the gas surface density and the SFR surface density averaged over the past $100$~Myr.  
The three bottom panels show the magnetic field strength for the full gas, CNM, and WNM components.  
The white contours in the maps delineate the inner, star-forming galactic disc region.
Figures \ref{fig:Bfieldsmaps_all_edgeon} and \ref{fig:Bfieldsmaps_all_faceon} in Appendix \ref{sec:gal_maps} present the B field maps of the whole simulated galaxy sample in both the face-on and edge-on projection. The B field morphology is also shown through white streamlines calculated using the density-weighted
magnetic field for the entire column displayed along the line of sight.

All physical quantities extracted from the simulations were analysed as follows, unless otherwise specified.
The SFRs extracted were measured on 100~Myr timescales, by selecting stellar particles with ages $\leq$~100~Myr. The magnetic field strength and specific energy measurements were computed on the disc regions of the simulated galaxies (see Section \ref{The simulated galaxy sample} for regions definitions). 
We fitted power--law relations $y = a\,x^{b}$ in log--log space using ordinary least squares (\texttt{numpy.polyfit}) within a paired bootstrap. Specifically, we drew $N_{\mathrm{boot}}=3000$ resamples of the $(x,y)$ pairs with replacement; for each resample we refitted $Y=A+bX$, where $Y \equiv \log_{10} y$, $X \equiv \log_{10} x$, and $A \equiv \log_{10} a$, and stored the slope and intercept. 
The quoted slope and intercept values correspond to the median of their respective bootstrap distributions. The associated uncertainties were derived from the dispersion of these distributions.
We also computed the central 95\% prediction intervals, shown as a shaded band in the plots.
Finally, per-point interquartile ranges (IQRs), defined as the interval between the 25th and 75th percentiles, are displayed as error bars in the plots for magnetic field strength estimates.

We outline the physical properties of our galaxy sample in Table ~\ref{tab:gal_properties}, which shows the unique galaxy ID, their stellar mass, SFRs, specific SFRs, and SFR surface density.
We show the median value of the magnetic field for the different gas phases for each galaxy in Table ~\ref{tab:Btot_disk}.

\subsection{Magnetisation model variations}
\label{sec:NUTsetup}

To assess the robustness of our results with respect to the magnetisation scenario, in addition to \textsc{Azahar}, we used a Milky Way-like cosmological simulation suite following the formation of a spiral galaxy under different magnetisation models \citep{Martin-Alvarez2024}. The ICs for these simulations are the \textsc{nut} ICs \citep{Powell2011}, studying a halo with dark matter mass of $M_\text{vir} (z = 0) \simeq 5 \times 10^{11}~\Msun$. The simulation box is 12.5 cMpc per side, with a spherical zoom region of approximate diameter $4.5$~cMpc. Its mass and stellar resolutions are $m_\text{DM} \simeq 5 \times 10^4~\Msun$ and $m_{*} \simeq 5 \times 10^3~\Msun$, respectively.

A full description of the galaxy formation physics in the models is provided in \citet{Martin-Alvarez2020} and \citet{Martin-Alvarez2021}. In short, they feature the physics described above for the RTnsCRiMHD model of \textsc{Azahar}, with the exception of no RT and no CRs. The suite of models employed here spans four simulations, each with a different magnetisation. Models labelled MB20 ($B_0 = 10^{-20}\,G$), MB11 ($B_0 = 10^{-11}\,G$), and MB10 ($B_0 = 10^{-10}\,\Gauss$) have magnetic fields exclusively sourced through their ICs, with a uniform field of strength $B_0$. In addition to $B_0 = 10^{-20}\,G$, the MBinj model also features magnetic fields injection through SNe, following the SN modelling described in Section~\ref{ss:Azahar_methods}.

\section{Results}
\label{sec:results}

We measured the magnetic field strength relations with star formation activity across the \textsc{Azahar} sample, which spans SFRs from $\mathrm{SFR} = 1.22 \times 10^{-3} \, \mathrm{M_\odot\,yr^{-1}}$ to $6.22~\mathrm{M_\odot\,yr^{-1}}$ and stellar masses from $M_\star = 2.41\times10^{8}$ to $2.07\times10^{10}~\mathrm{M_\odot}$. The SFR and stellar masses of the sample are shown in Table \ref{tab:gal_properties}.
We compared our simulations with heterogeneous observational samples (Section~\ref{subsec:B_vs_SFR}), and discuss our inferred scalings in the context of the $\sim\!1/3$ prediction by saturated dynamo models. We investigate the energetic relevance of magnetic fields in galactic discs, comparing the specific energies ($\mu_X\!\equiv\!U_X/M_{\rm gas}$) for magnetic, thermal, turbulent, and CR components (Section~\ref{subsec:energies_vs_sfr}). The magnetic field strength and specific energies scaling with star formation were studied as a function of the ISM phases (Section~\ref{subsec:phase_B_vs_SFR}).
Finally, we measured the magnetic field strength relation with star formation using the \textsc{nut} models featuring seed field magnetisation (Section \ref{sec:bsfr_model_dependence}).

\begin{figure}[htbp]
\centering
\includegraphics[width=\linewidth]{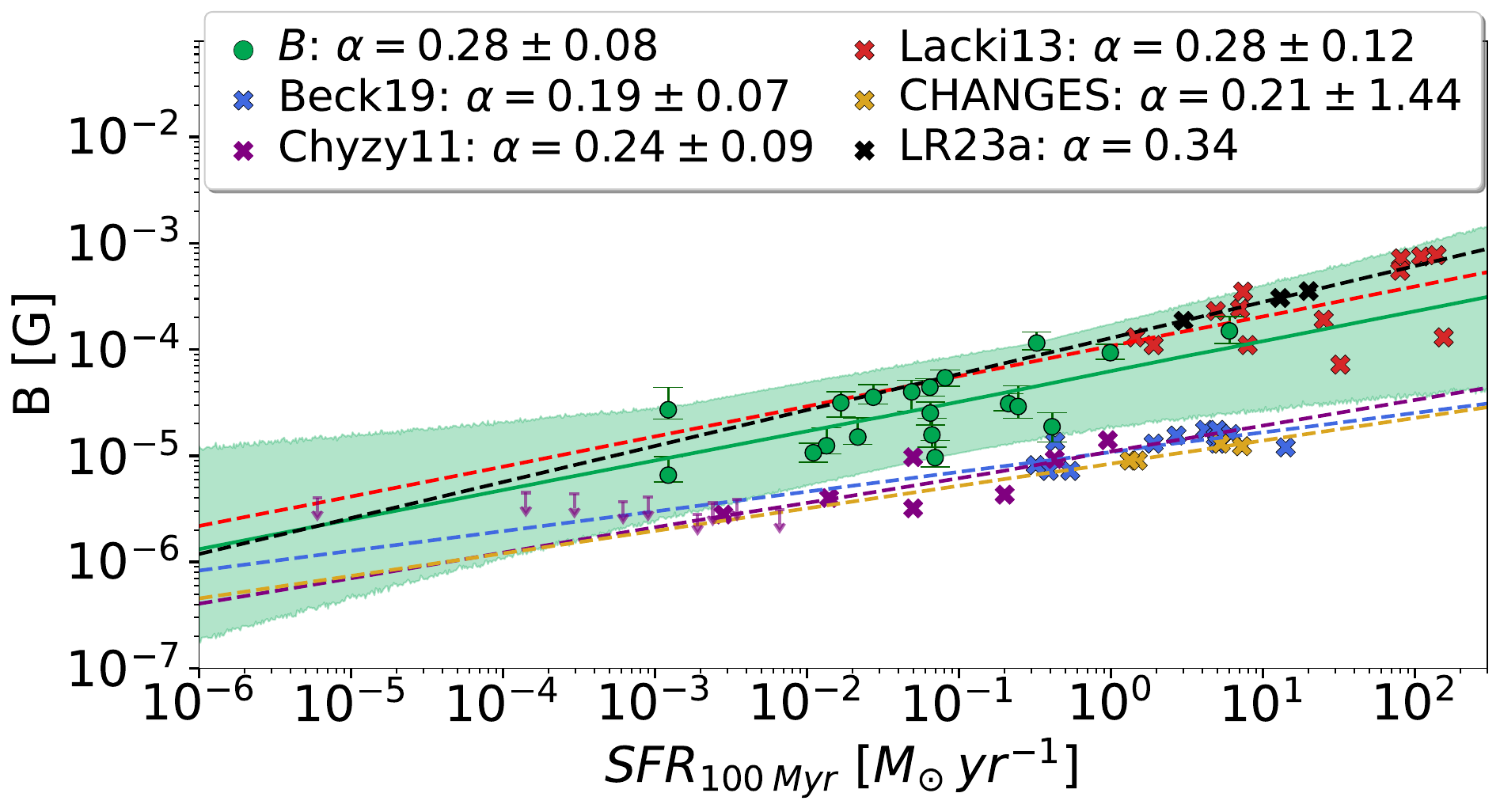}
\includegraphics[width=\linewidth]{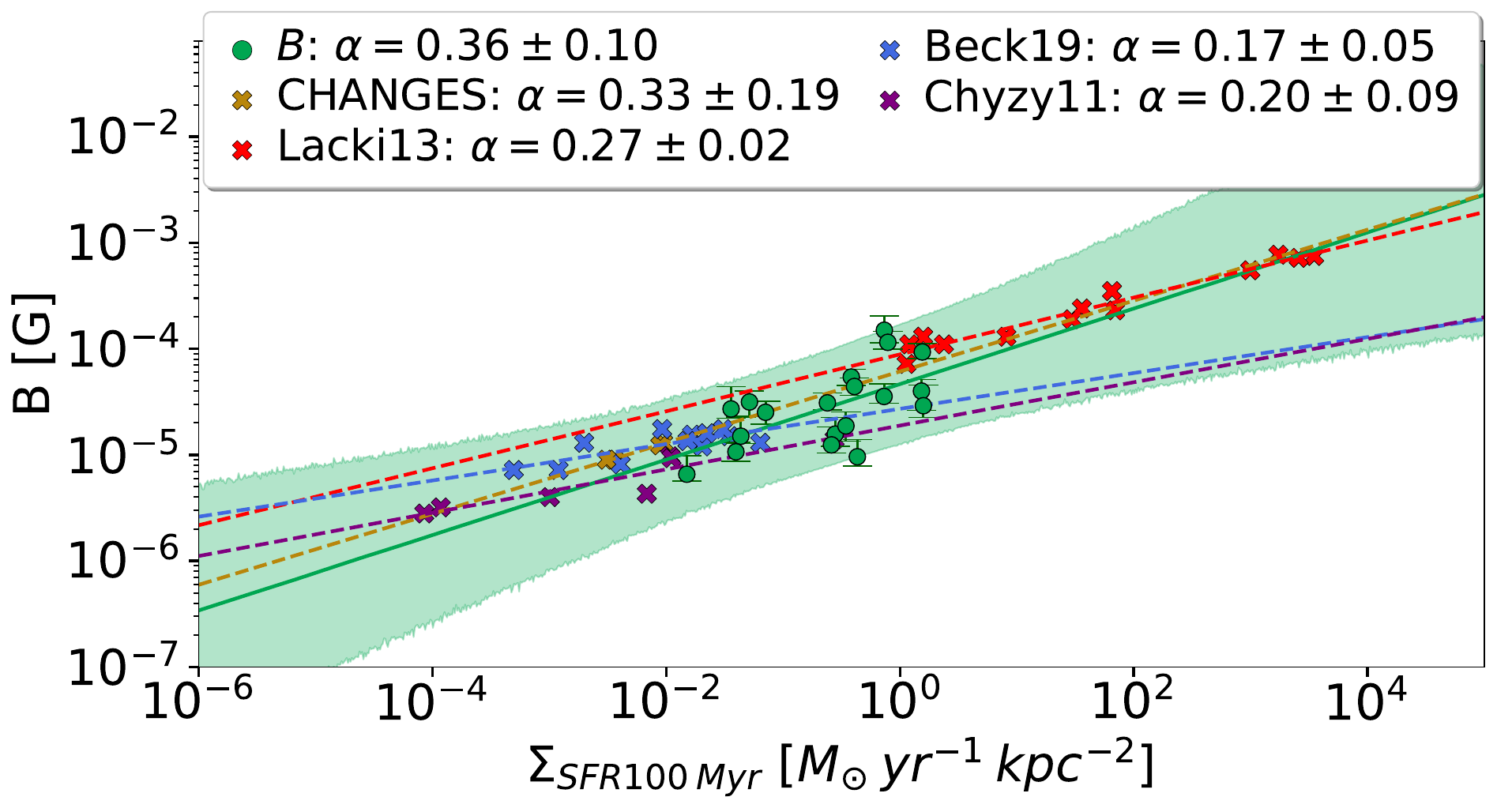}
\includegraphics[width=\linewidth]{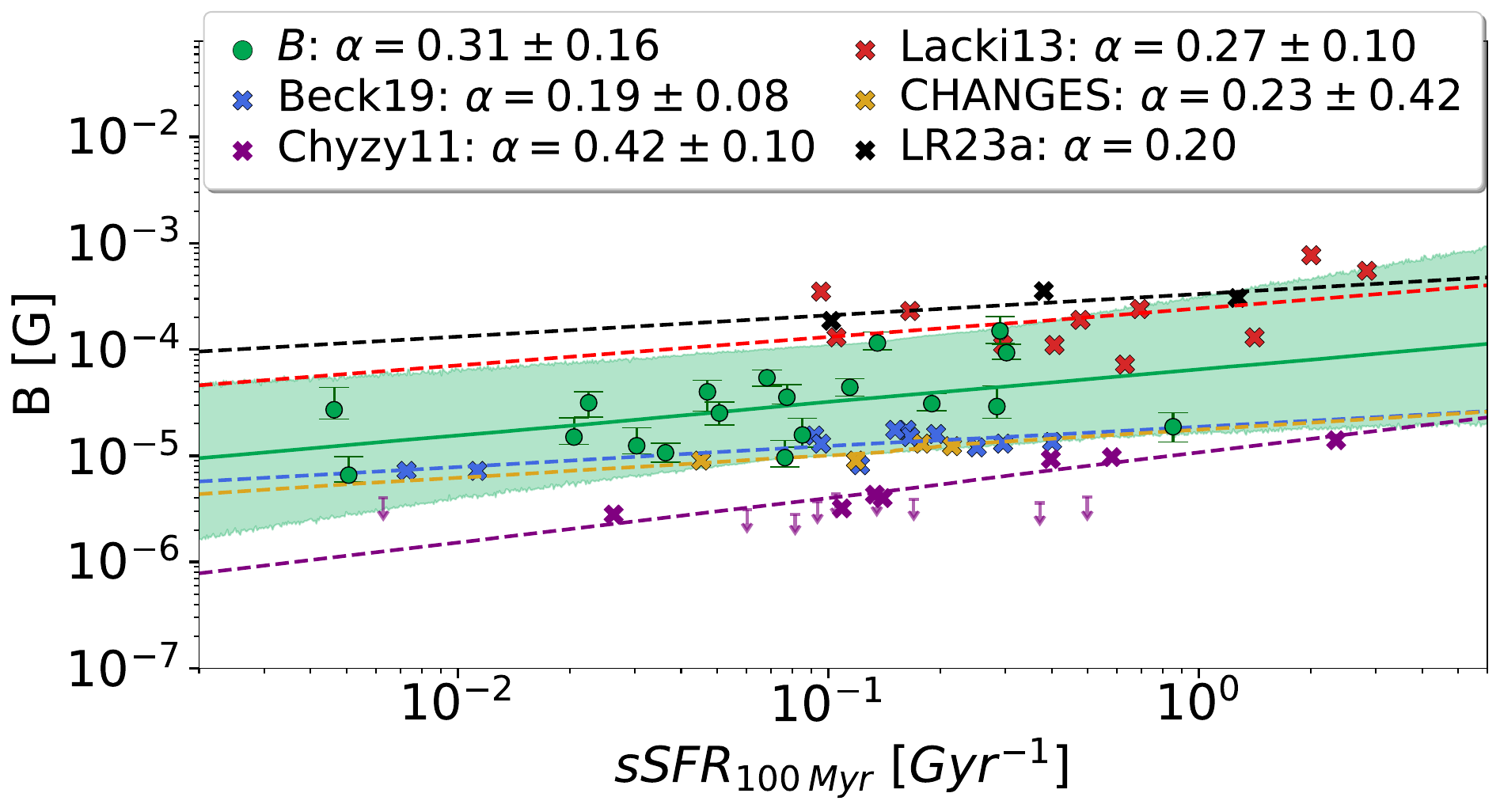}
\caption{{Magnetic field strength as a function of  global star formation and mass indicators.
Each panel shows the median magnetic field strength, $B$, within the disc mask of the simulated sample (solid green circles) measured as described in Section~\ref{The simulated galaxy sample}. Error bars show IQR range. The solid green line shows the best fit, and the shaded region 
indicates the $95\%$ prediction interval, with their fitted slopes and the standard error reported in the legends.
Samples of observed galaxies taken from the literature and their best-fit relations (dashed lines), are overlaid with different color crosses: red \citep{Lacki2013}, black \citep{Lopez-Rodriguez2023}, purple \citep{Chyzy2011}, navy blue \citep{Beck2019}, and goldenrod (\citealp[CHANGES survey]{Heald2022, Stein2020, Stein2019, Mora-Partiarroyo2019}). Their respective best-fit relations are shown as dashed lines. Upper limits from \citet{Chyzy2011} are shown as downward arrows, and are excluded from the fits.
Top: $B$ versus integrated SFR over the last 100\,Myr. 
Central: $B$ versus the SFR surface density.}
Bottom: $B$ versus sSFR.}
\label{fig:Bfield_strength_SFR}
\end{figure}

\subsection{Magnetic fields versus star formation}
\label{subsec:B_vs_SFR}

To review the interrelation between the magnetic field across our population of galaxies in \textsc{Azahar}, we show the relation of their magnetic field strength with the SFR in the top panel of  Figure~\ref{fig:Bfield_strength_SFR}. 
We find a scaling of magnetic field strength
\[
B \propto \mathrm{SFR}^{\,\alpha},\qquad \alpha = 0.28 \pm 0.08.
\]
This relation is shown by the solid green line fit, with its $95\%$ prediction interval as a shaded region.

We compare this result to estimates of the magnetic fields extracted from the literature. We selected the following samples of magnetic field strength estimations:
\begin{itemize}[label=-]
\item  Thirteen starburst cores from \citet{Lacki2013}. The magnetic field strength has been inferred using the revised equipartition formula between electron CRs and magnetic field strength \citep{Beck2005,Lacki2013}. Specifically, they used the integrated GHz synchrotron emission from the galaxy and assumed it to arise mostly from the central $0.1\text{--}1$ kpc regions.
The values are shown as red crosses in the top panel of Figure \ref{fig:Bfield_strength_SFR}.

\item  Seven irregular dwarfs from \citet{Chyzy2011}. These are low-mass systems ($1.05\times10^{6}\,\mathrm{M_\odot} \leq M_\star \leq 1.50\times10^{9}\,\mathrm{M_\odot}$). The magnetic fields strength were computed as integrating fluxes in apertures encompassing all detected radio emission (or over the optical extent for non-detections) via the revised radio–equipartition formula \citep{Beck2005}, which assumes energy equipartition between CRs and magnetic fields. The values characterise the global disc rather than nuclear regions. They are shown as purple crosses in the top panel of Figure~\ref{fig:Bfield_strength_SFR}. 

\item Twelve spiral galaxies from \citet{Beck2019}. The sample consists mainly of moderately to highly inclined systems, with inclinations ranging from $24^\circ$ (M\,51) to $78^\circ$ (NGC\,253). The magnetic field strengths were estimated across the resolved disc using equipartition between electron CRs and magnetic field strength. We estimated the median $B$ across the disc, typically within $\sim$6-8 kpc. 
These points characterise the kiloparsec scale magnetic field over the galaxy disc rather than the nuclear core. They are shown as navy crosses in the top panel of Figure~\ref{fig:Bfield_strength_SFR}.

\item  Four edge-on spirals from the CHANGES survey \citep{Mora-Partiarroyo2019,Stein2019,Stein2020,Heald2022}. 
Magnetic fields were derived from thermally corrected 1.5–6\,GHz synchrotron maps using the revised equipartition formula \citep{Beck2005} with $K_0=100$, assuming geometric path lengths of 6–20\,kpc appropriate for edge-on discs. The quoted values correspond to mean fields within $\pm1$\,kpc of the midplane over the star-forming discs ($\sim$15–20\,kpc in diameter) at $\sim$10–15" resolution ($\approx0.8$–1.5\,kpc). All values assume equipartition between CRs and magnetic fields and are representative of kiloparsec-scale averages across the galactic discs. They are shown as yellow crosses in the top panel of Figure~\ref{fig:Bfield_strength_SFR}.

\item  Three starburst cores (M\,82, NGC\,253, and NGC\,2146) from \citet{Lopez-Rodriguez2023}.
The magnetic fields strength in M82 were measured by the modified Davis–Chandrasekhar–Fermi method (DCF) \citep{Davis1951a,Chandrasekhar1953} accounting for large-scale flows by analytically solving the ideal-MHD equations \citep{Lopez-Rodriguez2021}. Magnetic fields for NGC\,253 and NGC\,2146 were then estimated by scaling from M\,82 using the empirical SFR–B relation 
$B = B_{0}\,\mathrm{SFR}^{0.34\pm0.04}$ with $B_{0}=128~\mu\mathrm{G}\,(\mathrm{M_\odot\,yr^{-1}})^{-0.34}$. We include these three systems as reference points for the magnetic field in the cold phase of the ISM at the core of starburst systems. They are shown as black crosses in the top panel of Figure~\ref{fig:Bfield_strength_SFR}.

\end{itemize}

For each galaxy in the observed sample, we selected an SFR tracer that is sensitive to star formation over roughly the past hundred million years (e.g. FUV+22\,$\mu$m, UV+IR, radio-based calibrations), approximately matching the SFR timescale employed for our simulated sample measurements. 
The magnetic field strengths, SFRs, sSFR, $\Sigma_{\rm SFR}$, SFR tracer, stellar masses, inclination, and references for the compiled observational sample are reported in Table~\ref{tab:Btot_obs_properties}. 

We find the following slopes for the $B \propto \mathrm{SFR}^{\,\alpha}$ relation for the observed samples:
\begin{itemize}[label=-]
\item Starburst cores \citep{Lacki2013}: $\alpha = 0.28 \pm 0.12$.
\item Irregular dwarfs \citep{Chyzy2011}: $\alpha = 0.24 \pm 0.09$.
\item Nearby spirals \citep{Beck2019}: $\alpha = 0.19 \pm 0.07$.
\item Edge-on spirals (CHANGES): $\alpha = 0.21 \pm 1.44$.
\end{itemize}

We measured an approximately universal scaling of magnetic field strength in galaxies with their SFR of $\alpha = 0.2 - 0.3$. This result is invariant across both simulated and observed measurements, as well as across the large variety of galactic environments and stages (e.g. nuclear region, starburst galaxies, dwarf galaxies, and both edge-on and face-on views).

The \citet{Lacki2013} starbursts show a trend highly consistent with our simulations, even with the SFR reaching significantly higher values than our simulated sample. 
The \citet{Chyzy2011}, \citet{Beck2019}, and CHANGES samples all show shallower slopes.

We find a scaling difference between samples. We attribute the different normalisations to the fact that the magnetic field has been measured in different regions of the galaxies. 
Measurements from \citealt{Lacki2013} show the higher normalisation due to their $B$ measurements being taken within compact nuclear measurement regions, where energy density budgets are more elevated. Similarly to our simulated measurements, all taken in the innermost 0.5 - 1~kpc starburst regions.
Conversely, measurements from dwarf irregulars \citep{Chyzy2011}, nearby spirals \citep{Beck2019}, and edge-on spirals \citep[CHANGES;][]{Heald2022, Stein2020, Stein2019, Mora-Partiarroyo2019} all display the lower normalisation. We attribute it to the measurement of the diffuse gas within the entire disc of the galaxy.

To verify that the inferred magnetic field–star formation scaling is not driven by projection effects, we performed a complementary analysis using face-on projections of the simulated galaxies. The resulting B–SFR relations are fully consistent with those obtained from edge-on views, demonstrating that the measured scaling is independent of galaxy inclination. The face-on analysis is presented in Appendix~\ref{sec:faceon_test}, with the corresponding relations shown in Figure~\ref{fig:Bfield_strength_SFR_faceon}.

\subsubsection{Magnetic field strength versus SFR surface density}
\label{subsec:B_vs_Sigma_SFR}

As was discussed in the introduction, the standard interpretation of the $B$–SFR correlation arises from equipartition between SN-driven turbulence and magnetic energy. Under the assumption of the magnetic energy being amplified to some fraction of saturation, $f_\text{sat}$, of the turbulent energy, \citet{Schleicher2013} derives a scaling of $B \propto \Sigma_{\rm SFR}^{1/3}$. 

In the central panel of Figure \ref{fig:Bfield_strength_SFR}, we show our measured $B \propto \Sigma_{\rm SFR}^{\alpha}$ relation for the simulated sample. The star formation surface density is measured in the same innermost region of our galaxies, using the semi-major axes of the disc region and following the approach used in \citealp{Wiegert2015} for their measurements. Our galaxies follow a scaling of $\alpha = 0.36 \pm 0.10$, in good agreement with a SN-driven turbulence assumption.

We compare the results with the observed samples.
In the case of the CHANGES galaxies, we used the $\Sigma_{\rm SFR}$ values directly reported in the same works where the magnetic field strengths were derived \citep{Heald2022, Stein2020, Stein2019, Mora-Partiarroyo2019}.
For the remaining samples, we estimated the SFR surface density, $\Sigma_{\rm SFR}$, by dividing the total SFR of each galaxy by the area of the region where the magnetic field strength was measured. In each case, we used the aperture radius reported in the corresponding study to define this circular area.

This approach is subject to several caveats. The SFR values compiled from the literature represent integrated measurements, which are not necessarily computed over the same physical region as the magnetic field. To the first order, we assume that the SFR is approximately uniform across the disc area encompassing the magnetic field measurement. In practice, spatial variations in the local SFR, particularly between central starburst regions and outer discs, may introduce systematic uncertainties in $\Sigma_{\rm SFR}$. Nevertheless, this method provides a reasonable first-order estimate that enables a meaningful comparison between simulated and observed galaxies.

The observed samples show the following scaling of magnetic field with $\Sigma_{\rm SFR}$:
\begin{itemize}[label=-]
\item Starburst cores \citep{Lacki2013}: $\alpha = 0.22 \pm 0.02$.
\item Irregular dwarfs \citep{Chyzy2011}: $\alpha = 0.20 \pm 0.09$.
\item Nearby spirals \citep{Beck2019}: $\alpha = 0.17 \pm 0.05$.
\item Edge-on spirals (CHANGES): $\alpha = 0.33\pm 0.19$.
\end{itemize}

The \citealp{Lacki2013} starburst cores are consistent with saturation of the small–scale dynamo in dense, high–pressure nuclei.
Edge-on spirals (CHANGES) are consistent with the $1/3$ expectation within large uncertainties ($\alpha \approx 0.33\pm0.19$) due to the limited number of sources in this sample. 
The nearby spirals from \citet{Beck2019} and irregular dwarfs from \citet{Chyzy2011} show a shallower slope similar to their global $B$-$SFR$ relation, likely reflecting their low star formation efficiencies and turbulent energy budgets. All observational trends lie within the $95\%$ prediction interval of our fit, reinforcing a robust $B$–$\Sigma_{\rm SFR}$ coupling across galaxy types and measurement strategies.

\subsubsection{Magnetic field strength versus specific SFR}
\label{subsec:B_vs_sSFR}

\begin{figure}[htbp]
\centering
\includegraphics[width=\linewidth]{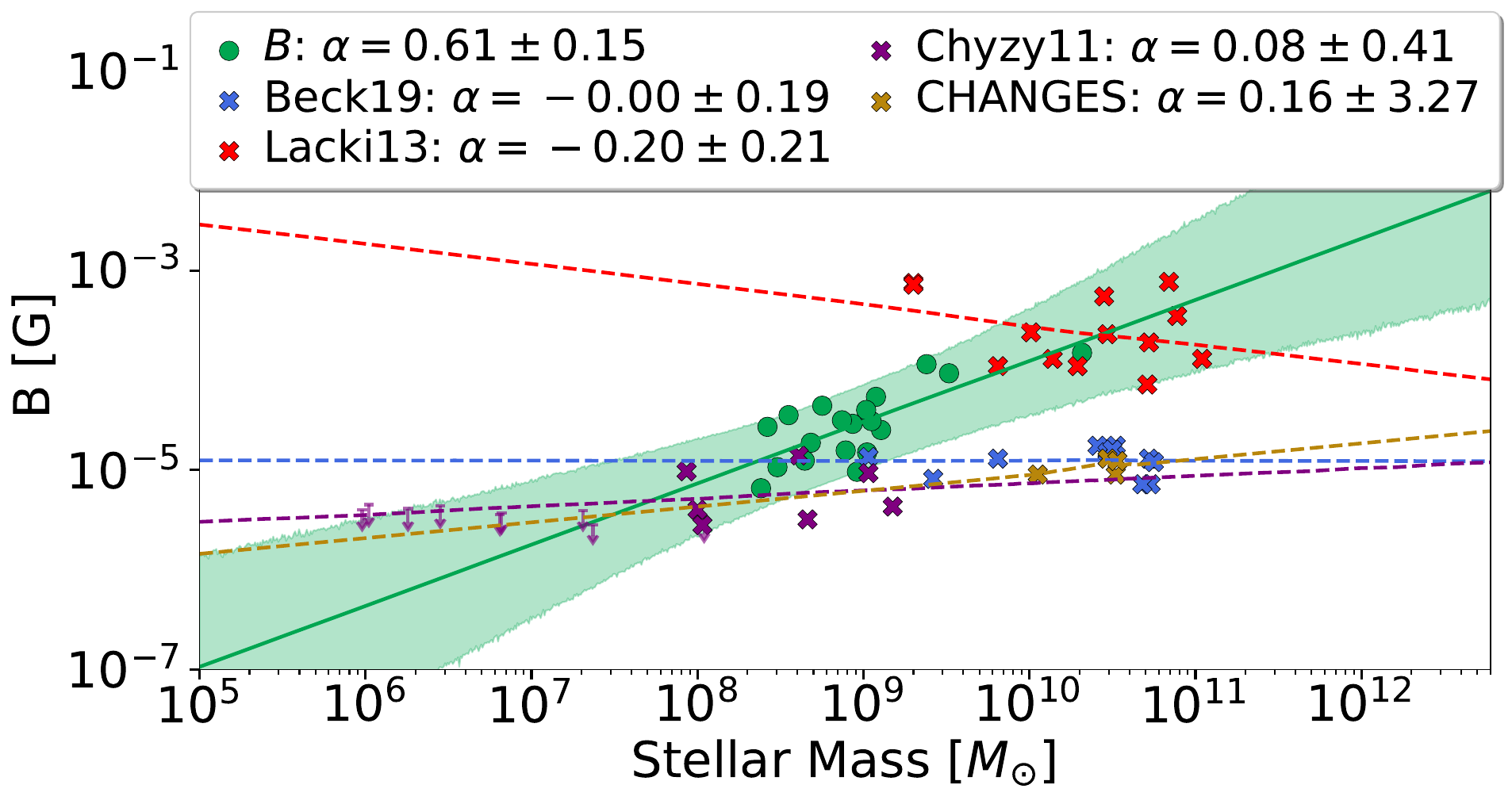}
\caption{Magnetic field strength as a function of stellar mass, $M_{\star}$ (solid green circles). The solid green line shows the best fit, and the shaded region indicates the $95\%$ prediction interval, with the fitted slopes and the standard error reported in the legend. 
Samples of observed galaxies taken from the literature and their best-fit relations (dashed lines) are overlaid with different colour crosses: red \citep{Lacki2013}, black \citep{Lopez-Rodriguez2023}, purple \citep{Chyzy2011}, navy blue \citep{Beck2019}, and goldenrod (\citealp[CHANGES survey]{Stein2019, Mora-Partiarroyo2019, Stein2020, Heald2022}). Their respective best-fit relations are shown as dashed lines. Upper limits from \citet{Chyzy2011} are shown as downward arrows, and are excluded from the fits.}
\label{fig:Bfield_stellarmass}
\end{figure}

To understand the effect of the mass on the $B$ and SFR relation, we examine this dependence across in our simulated galaxies in Figure~\ref{fig:Bfield_stellarmass}. The sample follows a tight correlation between the magnetic field strength, $B$, and stellar mass, $M_\star$,
\[
B \propto M_\star^{\,0.61 \pm 0.15},
\] reflecting how part of the scaling measured for the $B$--SFR plane is driven by galaxy stellar mass.

A simple energetic argument shows that this is plausible. If turbulent kinetic and magnetic energies remain in rough equipartition and scale with the potential depth, then at a fixed radius, $r$, one expects
\begin{equation}
B\,(r)\ \propto\ \sqrt{\frac{G\,M\,(<r)}{r}\ \rho_\mathrm{gas}}\ \sim \sqrt{\rho_\mathrm{gas}}\, v_c (r),
\label{eq:equipart_mass}
\end{equation}
so that, for modest variations in $\rho_\mathrm{gas}$ at fixed $r$, $B$ should scale approximately as $M_\star^{1/2}$, consistent with our best‑fit slope. This scaling can also be rewritten as a function of the circular velocity ($v_c$) of such systems. This is qualitatively aligned with observations that more massive, faster‑rotating discs host stronger large‑scale fields (via the Tully–Fisher connection between $v_\mathrm{rot}$, dynamical mass, and $M_\star$; \citealt{Reyes2011, Tabatabaei2016}) and with galaxy‑formation simulations where central $B$ grows with bulge mass and central gas density.

We tested if the observed samples show this correlation, using stellar masses estimates taken from various sources from the literature, as listed in Table \ref{tab:Btot_obs_properties}.
We found:

\begin{itemize}[label=-]
    \item \citet{Beck2019}: $\alpha = -0.00 \pm 0.19$
    \item \citet{Lacki2013}: $\alpha = -0.20 \pm 0.21$
    \item \citet{Chyzy2011}: $\alpha = 0.08 \pm 0.41$
    \item CHANGES sample: $\alpha = 0.16 \pm 3.27$
\end{itemize}

Within uncertainties, all observed samples are consistent with a flat $B$–$M_\star$ relation, with little dependence of the magnetic field strength on the total stellar mass.
However, these trends can be explained by the methodology employed to measure the magnetic field strength in the observed samples.
In all observational compilations, the stellar masses were integrated over the entire galaxy, while magnetic field strengths were derived from radio synchrotron emission measured within specific apertures of regions with considerably different sizes and physical conditions:
\begin{itemize}[label=-]
\item Central (starburst-core) measurements. In the starburst sample of \citet{Lacki2013}, the radio flux is explicitly restricted to compact nuclear regions, intentionally weighting the densest, most CRs rich gas and yielding the highest inferred magnetic field strength values.
One extreme case is Arp220, an ultra-luminous infrared galaxy with two merger nuclei ~200 pc in size separated by ~300 pc \citet{BarcosMunoz2015}. Due to the merger, the strong star formation activity, and molecular outflows, this galaxy has been measured to have one of the strongest, microgauss, magnetic field strength \citet{McBride2014}. This outlier may be biasing the fitting to show a negative correlation using the \citealp{Lacki2013} sample. However, note that this sample is composed of starburst galaxies, which agree with the trend found in the simulated galaxies.

\item Galaxy disc averages. In nearby spirals, the magnetic field strength is estimated from total or ring-averaged synchrotron intensities using the revised equipartition formula \citep{Beck2005}, which mixes the relatively small area corresponding to bright centres with extended and fainter outer discs. This measuring procedure leads to a lower mean $B$ than the central measurements, and flattens trends with $M_\star$ due to increasing $r$ and a non-spherical configuration. For a~dex increase in stellar mass of disc galaxies, the expected variation in $B$ is approximately a factor of two, consistent with the observed shallow scalings.  CHANGES edge-on systems likewise report kiloparsec-scale, disc-averaged fields from thermally corrected 1.5–6\,GHz maps under revised equipartition. 
\item Local Group dwarfs. \citet{Chyzy2011} derive magnetic field strength from single-dish maps, integrating over each galaxy’s optical extent. For compact starburst dwarfs such as IC~10 the flux is core-dominated; for more extended dwarfs, the estimate approximates a global mean field. 
\end{itemize}

Because $M_\star$ is always a global, integrated quantity, while the typical measurement of magnetic field strength is effectively density‑weighted within the radio aperture, the two quantities are not directly comparable. Variations in the $B$-measurement aperture (core‑only vs whole disc) and the mass aperture (whole galaxy) weaken or cancel the intrinsic $B$–mass coupling and produce flat slopes when samples are mixed or contain bulge-disc component variations. In practice, central core measurements (\citealp{Lacki2013}; and our synthetic maps) yield higher normalisations at a given $M_\star$, whereas whole‑disc averages (e.g. nearby spirals) yield lower normalisations and flatter trends. We note that for dwarf galaxies, the same trend may still take place with the HI mass (and/or the velocity dispersion, $\sigma_v)$ of the galaxies, possibly modulated by galaxy outflows \citep{Chyzy2011}. This poses measuring magnetic fields in dwarf galaxies as the most promising avenue to fully characterise whether the scaling proposed by Eq.~\ref{eq:equipart_mass} holds.

To determine the relative importance of star formation and remove mass-driven effects in its relation with $B$, we reviewed the magnetic field strength as a function of specific SFR (sSFR $\equiv{\rm SFR}/M_\star$). This is shown in the bottom panel of Figure~\ref{fig:Bfield_strength_SFR}.

Our simulated galaxies show the relation \[
B \propto \mathrm{sSFR}^{\,\alpha},\qquad \alpha = 0.31 \pm 0.16.
\] 
The \citet{Lacki2013} measurements in the core regions of starburst galaxies exhibit a very similar slope with a tighter correlation ($\alpha = 0.27 \pm 0.10$), and stronger magnetisations. Both nearby spirals ($\alpha = 0.19 \pm 0.08$) studied by \citet{Beck2019} and by the CHANGES collaboration (\citealp{Stein2019, Mora-Partiarroyo2019, Stein2020, Heald2022}; $\alpha = 0.23 \pm 0.42$) exhibit somewhat shallower scaling, and a weaker norm for their magnetic field strength. Conversely, the dwarf galaxies measured by \citet{Chyzy2011} have the steepest correlation ($\alpha = 0.42 \pm 0.10$), although with the weakest field strength. All of the observed samples and our simulated galaxies, however, are consistent within 1$\sigma$. Interestingly, expressing the magnetic field strength interrelation with star formation in specific terms yields the largest separation between datasets, both in terms of power-law index variation and normalisations.

\subsection{Specific energy budgets versus SFR}
\label{subsec:energies_vs_sfr}

In the previous sections, we showed how the star formation activity in the discs of galaxies is tightly correlated with magnetic field strength, suggesting a scenario in which star formation is driving magnetic field amplification in galaxies. Since our MHD simulations make no a priori equipartition assumptions between turbulent and magnetic energy, the correlation arises self-consistently from SN-driven turbulence and MHD amplification. This indicates a coupling between turbulent kinetic and magnetic energies across the phases of the ISM that needs to be further characterised.

\begin{figure}[htbp]
\centering
\includegraphics[width=\linewidth]{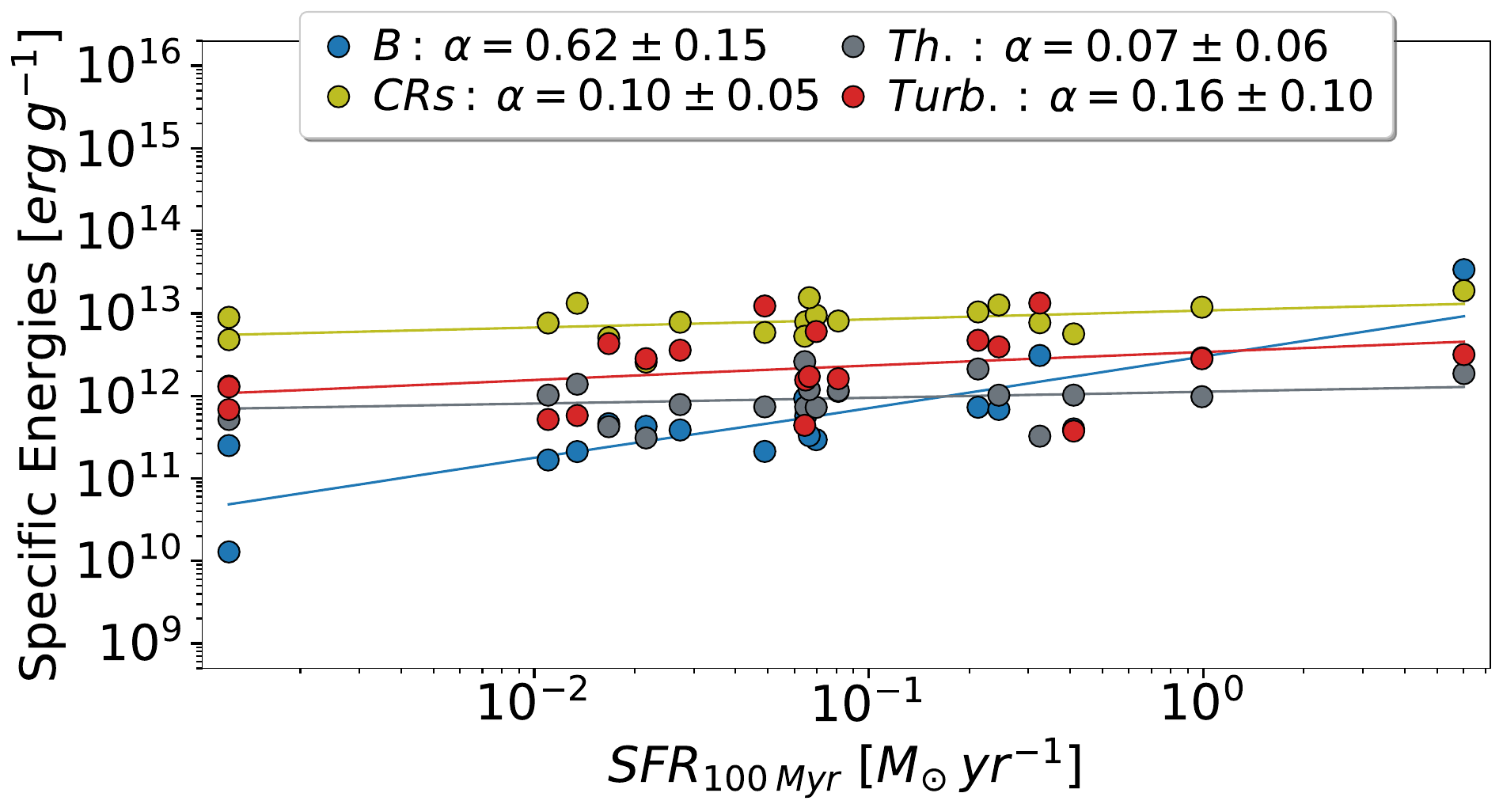}
\includegraphics[width=\linewidth]{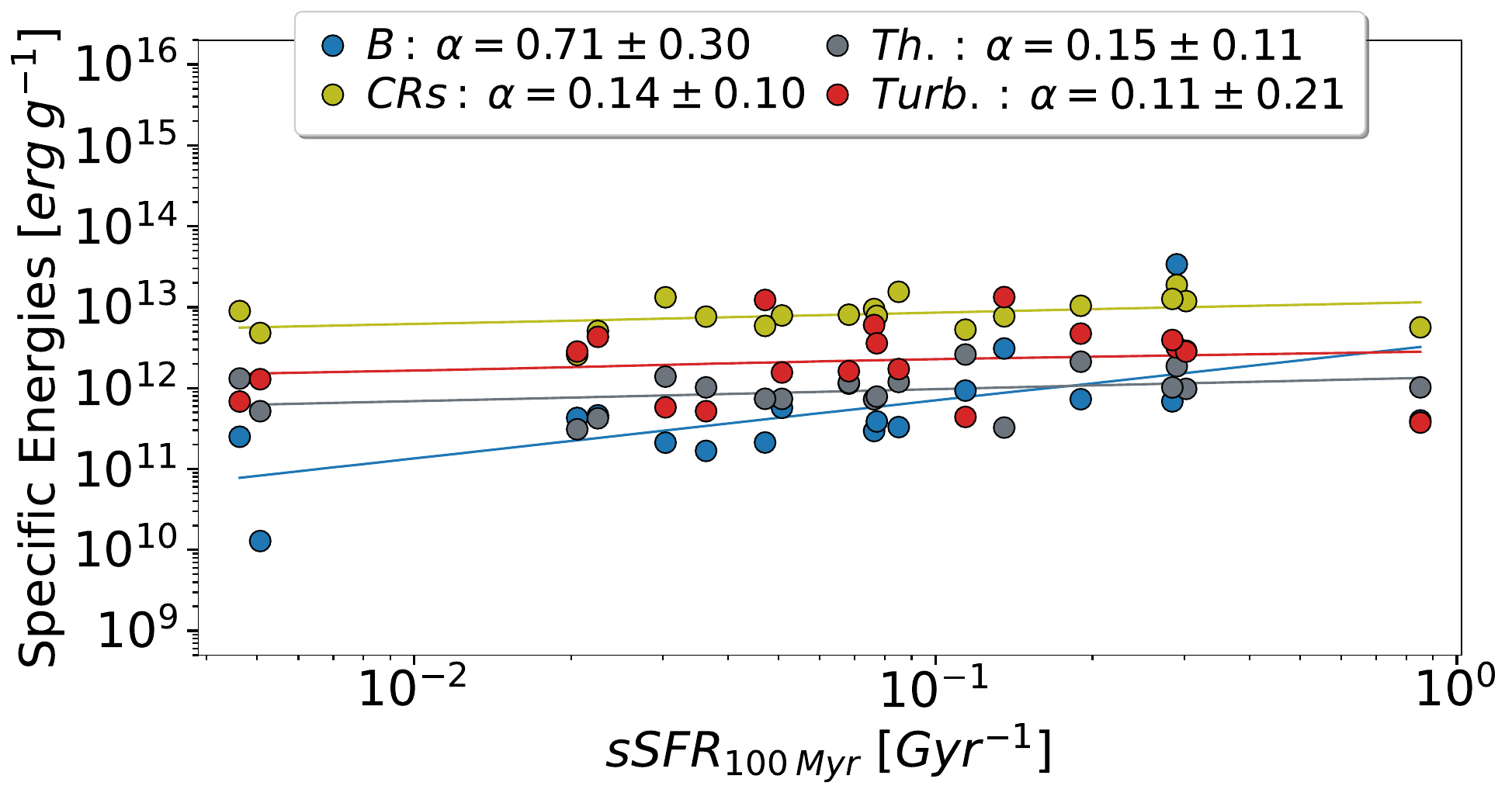}
\caption{Specific energies as a function of the SFR (top panel) and sSFR (bottom panel). Each point represents the specific energy associated with the magnetic (blue), thermal (grey), turbulent (red), and CR (yellow) component within the disc of each galaxy. The best fits for each energy component are shown as coloured lines, with their corresponding slopes and standard errors provided in the legends.}
\label{fig:Energies_SFR_fullgas}
\end{figure}

We examined the relationship between the total SFR and the specific energy (i.e. the energy per unit gas mass) for the full gas content of the simulated galaxies.
Figure~\ref{fig:Energies_SFR_fullgas} shows these relationships, where different colour points indicate the specific energy, $\mu_{\!}\equiv U/M_{\rm gas}$, of the magnetic (blue), thermal (grey), turbulent (red), and CR (yellow) components. Specific energies were computed as the total energy in each component integrated over the galactic disc, and divided by the corresponding total gas mass in the same region. The dotted lines show bootstrap-based best-fit power-law relations for each energy component. The specific energy values for each galaxy of the sample are shown in Table \ref{tab:combined_mu}.

All energy components exhibit a positive correlation with star formation, both for absolute and sSFR. This indicates that more actively star-forming galaxies tend to have higher specific energies. The thermal ($\alpha_{\rm therm} = 0.15 \pm 0.11$), CR ($\alpha_{\rm CR} = 0.14 \pm 0.10$), and turbulent ($\alpha_{\rm turb} = 0.11 \pm 0.21$) energy components show similar scalings with sSFR. These slopes indicate a self-regulated energy content, moderately sensitive to star formation, and consistent with a scenario in which CRs, turbulent energy, and thermal energy are replenished by SN feedback during secular evolutions. Note that the selected galaxies are not undergoing mergers.

The magnetic specific energy, $\mu_{\rm mag}\!\equiv\!U_{\rm mag}/M_{\rm gas}$, scales most steeply with both the SFR ($\alpha_{\rm mag}=0.62\pm0.15$) and the sSFR ($\alpha_{\rm mag}=0.71\pm0.30$). 
This scaling leads to sub-equipartition magnetisations in galaxies with low star-formation activity and approximate equipartition in those with ongoing star formation, with the transition occurring around sSFR $\sim 0.1\,\mathrm{Gyr}^{-1}$ and SFR $\sim 1\, \mathrm{M_{\odot}} \, \mathrm{yr}^{-1}$.
This steep scaling reinforces our findings for the $B$–SFR relations: since \(\mu_{\rm mag}\propto B^2/\rho_{\rm eff}\), where $\rho_{\rm eff}$ denotes the effective gas density associated with the region over which the magnetic energy is averaged, and \(B\propto{\rm SFR}^{\,0.28\pm0.08}\), one expects \(\mu_{\rm mag}\propto{\rm SFR}^{\,0.56 \pm 0.16}\rho_{\rm eff}^{-1}\). The measured \(\alpha_{\rm mag}\) suggests a scenario in which turbulence driven by SN feedback couples to magnetic fields through small-scale dynamo processes \citep{Kulsrud1999, Schleicher2013, Federrath2016, Martin-Alvarez2018, Pakmor2023}.

These trends are consistent with a self-regulated ISM picture.
In the inner, star-forming disc of neutral gas, the turbulent energy fraction remains approximately constant, as is expected for self-regulated star formation with nearly constant gas velocity dispersions (e.g. \citealt{Ostriker2011, Krumholz2016, Varidel2020}).
The CR energy fraction increases sub-linearly with sSFR, since injection scales proportionally to SFR but it is reduced by short CR residence times (e.g. \citealt{Lacki2013a}).
The thermal energy fraction increases weakly with sSFR, likely due to efficient radiative cooling maintaining thermal balance in the denser ISM (e.g. \citealt{Wolfire2003}).
Finally, the magnetic energy fraction increases most rapidly with sSFR, through increased coupling with the kinematic components \citep{Schleicher2013, Martin-Alvarez2022, Pakmor2024}), and eventually reaches approximate equipartition at high sSFR values. 
This magnetic pressures is comparable to the thermal and turbulent terms, implying a dynamically important magnetic component, with important consequences for vertical support, feedback coupling, and wind launching (e.g. \citealt{Heesen2011, Pillepich2018a, Su2020, Martin-Alvarez2020}).

It is worth emphasising that the linear fits reported here should not be extrapolated to high SFRs ($>100$ M$_{\odot}$ yr$^{-1}$). The small–scale dynamo saturates once Lorentz forces back–react on the flow, limiting further field amplification and field line stretching (e.g. \citealt{Schleicher2013,Martin-Alvarez2022,Pakmor2024}). Beyond this point, the magnetic energy is expected to approach equipartition with the turbulent component, and the magnetic slope should gradually converge on that of the turbulent energy, corresponding to a state in which magnetic, turbulent, and thermal pressures remain in near balance.

While our simulations favour a scenario in which magnetic field amplification is primarily regulated by SN-driven turbulence through small-scale dynamo processes, it is worth highlighting that the resulting B field configurations are not restricted to purely tangled, small-scale structures.
Across the full sample of simulated galaxies included in our analysis, several systems exhibit ordered, spiral-like magnetic field patterns on kiloparsec scales in their face-on projections (see Figure \ref{fig:Bfieldsmaps_all_faceon}). This indicates the presence of a large-scale magnetic field component, driven by differential rotation and shear.
In the corresponding edge-on projections (see Figure~\ref{fig:Bfieldsmaps_all_edgeon}), the B field appears predominantly aligned with the disc midplane.

Nevertheless, the presence of ordered magnetic field structures should not be interpreted as evidence for the operation of a dominant large-scale mean-field dynamo. The field morphology does not exhibit a well-defined vertical parity or symmetry with respect to the disc midplane, which would be expected for a fully developed, mature large-scale mean-field $\alpha$--$\Omega$ dynamo \citep{Ntormousi2020}.
When selecting only gas cells associated with galactic outflows, defined by an escaping velocity threshold of $v_{\rm out} > 50\,{\rm km\,s^{-1}}$, the edge-on magnetic field streamlines appear more open and vertically extended. These features still do not display a coherent symmetry or polarity pattern that would point to a dominant large-scale dynamo operating in the disc.

Furthermore, we find that the mean toroidal magnetic field in the discs of our sample of simulated galaxies is weaker than the spatial fluctuations of the magnetic field strength, with typical ordered-to-turbulent ratios of $\sim$5\% up to $\sim$20\%, indicating dominance of a turbulent magnetic component.
While this suggests that a large-scale dynamo does not dominate in these highly star-forming systems, our findings do not address its importance during stages of lower relative star formation activity.

\subsection{The dependence of gas phase magnetisations and energy fractions on star formation}
\label{subsec:phase_B_vs_SFR}

\begin{figure*}
\centering
\includegraphics[width=0.49\linewidth]{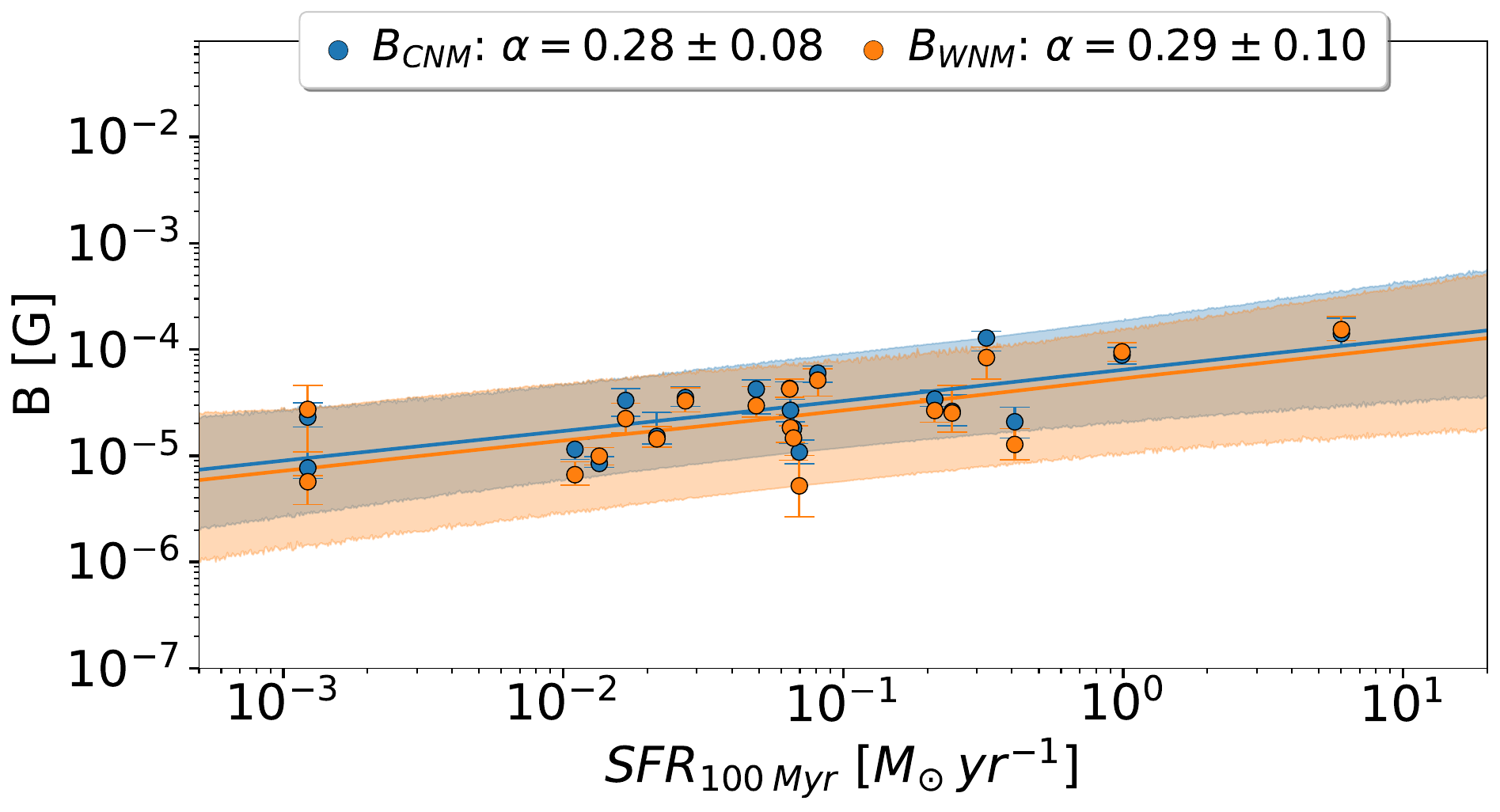}
\includegraphics[width=0.49\linewidth]{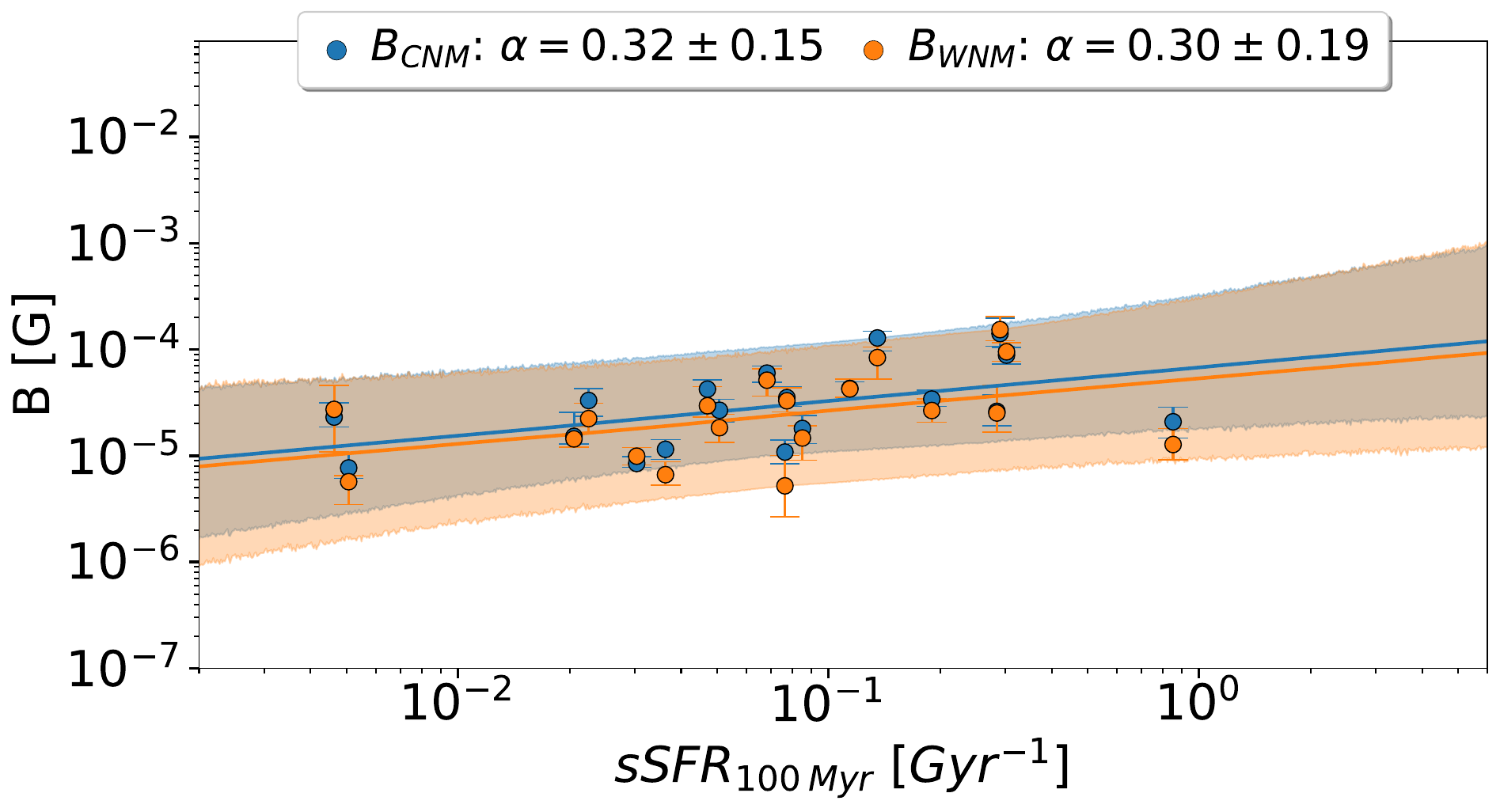}
\caption{Magnetic field strength versus global star formation for neutral ISM phases.
Panels show the median magnetic field strength, $B$, within the disc mask of the simulated sample (solid circles) in the CNM (blue) and WNM (orange) phases. Error bars for each data point show the IQR for each galaxy. Solid lines indicate best-fit power laws, and the shaded region indicates the 95\% prediction interval; fitted slopes are reported in the legends.
Left: $B$ versus integrated SFR. 
Right: $B$ versus specific SFR (sSFR).}
\label{fig:Bfield_strength_SFR_gas_phase}
\end{figure*}

In Section~\ref{subsec:energies_vs_sfr}, we showed that the magnetic field strength in the discs of spiral galaxies increases with star formation, reaching approximate magnetic equipartition with the thermal and turbulent components for systems with significant ongoing star formation (sSFR $\gtrsim 0.1\,\mathrm{Gyr}^{-1}$, SFR $\gtrsim 1\, \mathrm{M_{\odot}} \, \mathrm{yr}^{-1}$; Section~\ref{subsec:energies_vs_sfr}). A global analysis of this magnetisation provides only a limited picture, as colder ISM phases are expected to have higher relative magnetisations \citep{Ferriere2001, Draine2011, Crutcher2012}. To understand this increased relevance of magnetic fields across ISM phases, we now extend this analysis to the CNM and WNM phases. The left panel of Figure~\ref{fig:Bfield_strength_SFR_gas_phase} shows the same B–SFR relation previously shown in Figure~\ref{fig:Bfield_strength_SFR} (top left), now for the CNM values in blue and WNM values in orange.  The best-fit power laws are shown as solid colour-coded lines, with the shaded areas indicating a $95\%$ prediction interval.
Both phases show significant positive and comparable trends:
\[
\begin{aligned}
\alpha_{\rm CNM} &= 0.28\pm0.08,\\
\alpha_{\rm WNM} &= 0.29\pm0.10.
\end{aligned}
\]
These broadly similar slopes suggest that the same underlying mechanism operates across phases in the disc, especially when compared with the scaling obtained for the entire gas distribution ($\alpha = 0.28\pm0.08$).

The right panel of Figure~\ref{fig:Bfield_strength_SFR_gas_phase} shows the $B$–sSFR relation for the cold and warm neutral media (CNM in blue; WNM in orange). The fitted power-law indices are
\[
\begin{aligned}
\alpha_{\rm CNM} &=0.32\pm 0.15,\\
\alpha_{\rm WNM} &=0.30\pm 0.19.
\end{aligned}
\]
Within uncertainties, the two phases are consistent with each other, and once again in agreement with the global scaling ($\alpha = 0.31\pm0.16$).
The approximate phase-independence of the $B$--SFR/sSFR scalings suggests that the same processes regulating magnetic field strength operate across the CNM and WNM neutral phases.

\begin{figure}[htbp]
\includegraphics[width=0.49\textwidth]{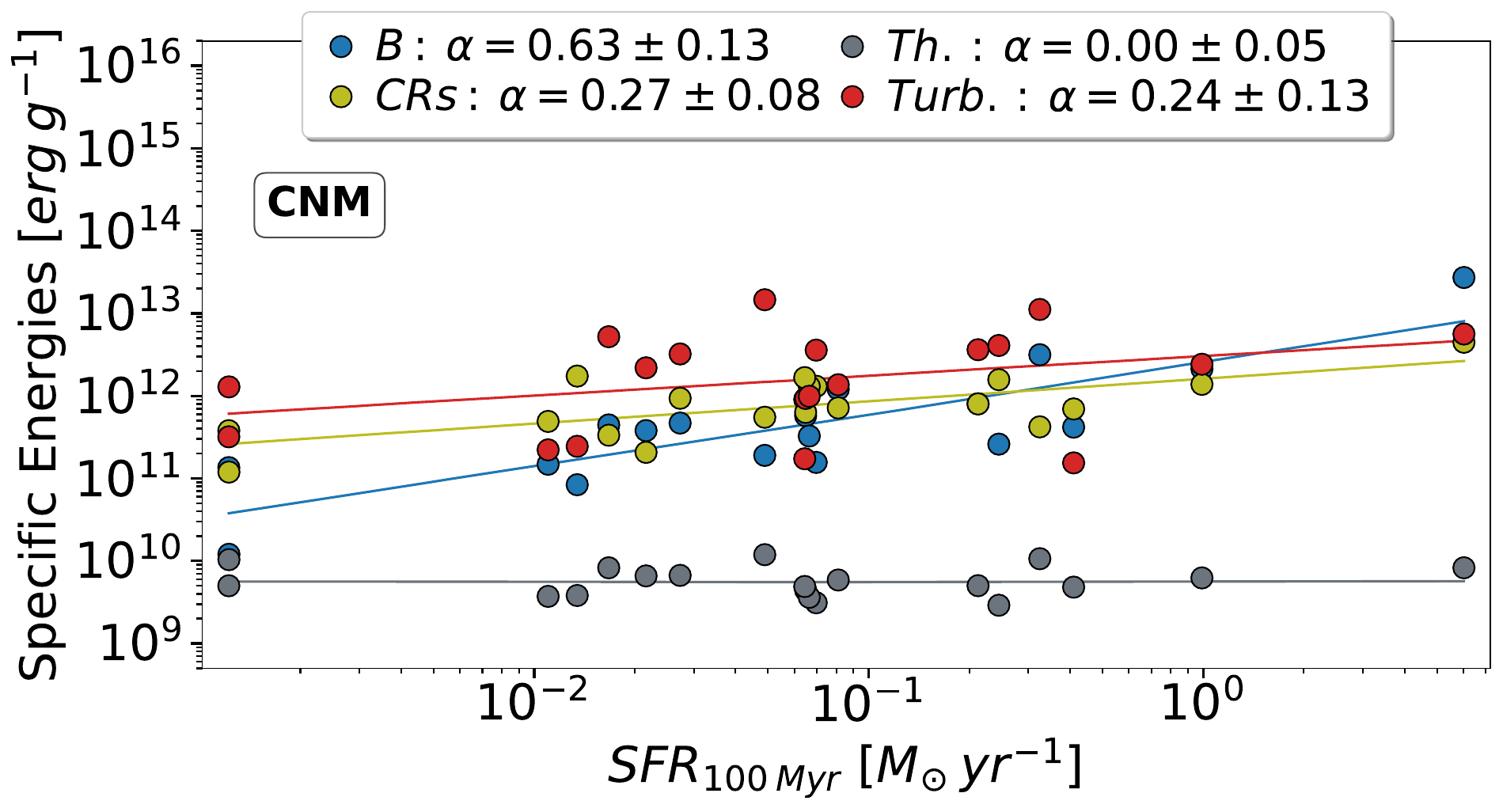}
\includegraphics[width=0.49\textwidth]{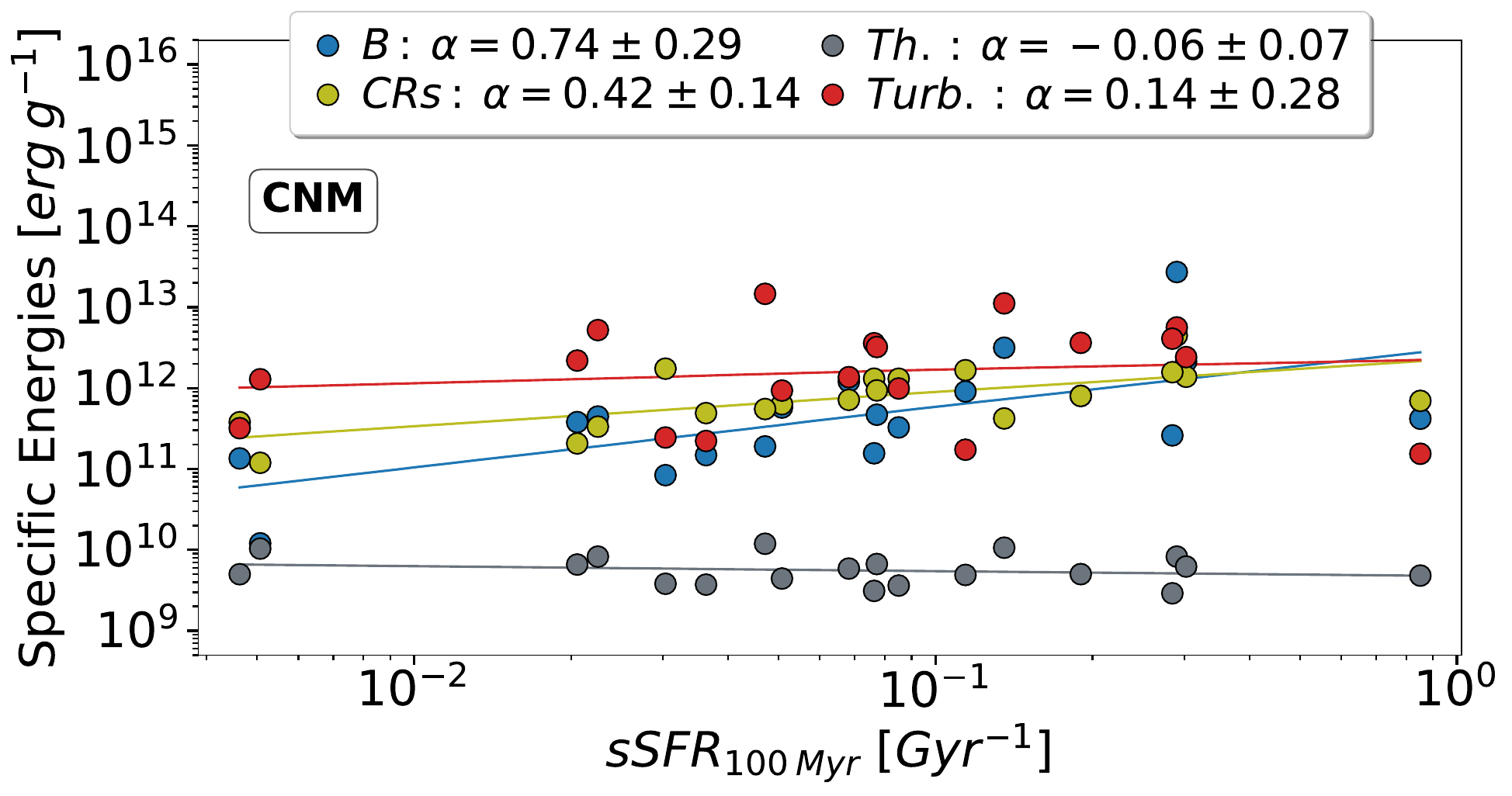}
\caption{
Specific energies as a function of the SFR over the last 100 Myr (top panel) and over the specific SFR (bottom panel) in the CNM phase. Each point represents the specific energy associated with the magnetic (blue), thermal (grey), turbulent (red), and CR (yellow) component within the disc mask of each galaxy. The best fit for each energy component associated with each component is shown as a dotted line, with the slope and Pearson index shown on the legend at the top.}
\label{fig:Energies_SFR_gasphase_CNM}
\end{figure}

\paragraph{CNM.}
We show in Figure~\ref{fig:Energies_SFR_gasphase_CNM} the scaling of the specific energies in the CNM as a function of SFR (top panel) and sSFR (bottom panel). Here we focus our analysis on the scalings between the different energy components and the specific star formation. The power-law indices between each energy component and the integrated SFR follow comparable trends and are shown in the figure. 
The specific magnetic energy has the steepest scaling with sSFR ($\alpha_{\rm mag,CNM}=0.74\pm0.29$), followed by the CR component ($\alpha_{\rm CR,CNM}=0.42\pm0.14$) and the turbulent component ($\alpha_{\rm turb,CNM}=0.14\pm0.28$). The thermal specific energy ($\alpha_{\rm therm,CNM}=-0.06\pm0.07$) follows a slight decrease with increasing sSFR.

Energetically, most of our galaxies are turbulence-dominated in the CNM, with non-negligible contributions from the CRs component. The magnetic component is only significant on the high sSFR end, reaching approximate equipartition with the other energy forms for sSFR $\gtrsim 0.1\,\mathrm{Gyr}^{-1}$ and for SFR $\gtrsim 1\, \mathrm{M_{\odot}} \, \mathrm{yr}^{-1}$. As star formation in our simulations takes place almost exclusively in the CNM gas, this reflects a scenario in which higher sSFR values can increase the magnetic contribution up to equipartition. However, this energy component cannot significantly exceed the energy budget of the other components, as this would suppress star formation. This behaviour is observed in galaxies with unrealistically strong magnetic field strengths, $B_{10} = 10^{-10}$ G, in the \textsc{nut} simulations \citep{Martin-Alvarez2018,Martin-Alvarez2020}.
The slight decrease in thermal energy with an increasing sSFR can be similarly interpreted as colder, denser gas clouds that feature more efficient star formation. Successful star-forming regions will in turn significantly increase gas temperatures through either photoheating ($\sim10^4\,\K$) or SN feedback ($\gtrsim10^6\,\K$), and their gas will no longer be classified as CNM \citep{Martin-Alvarez2023}.

\begin{figure}[htbp]
\includegraphics[width=0.49\textwidth]{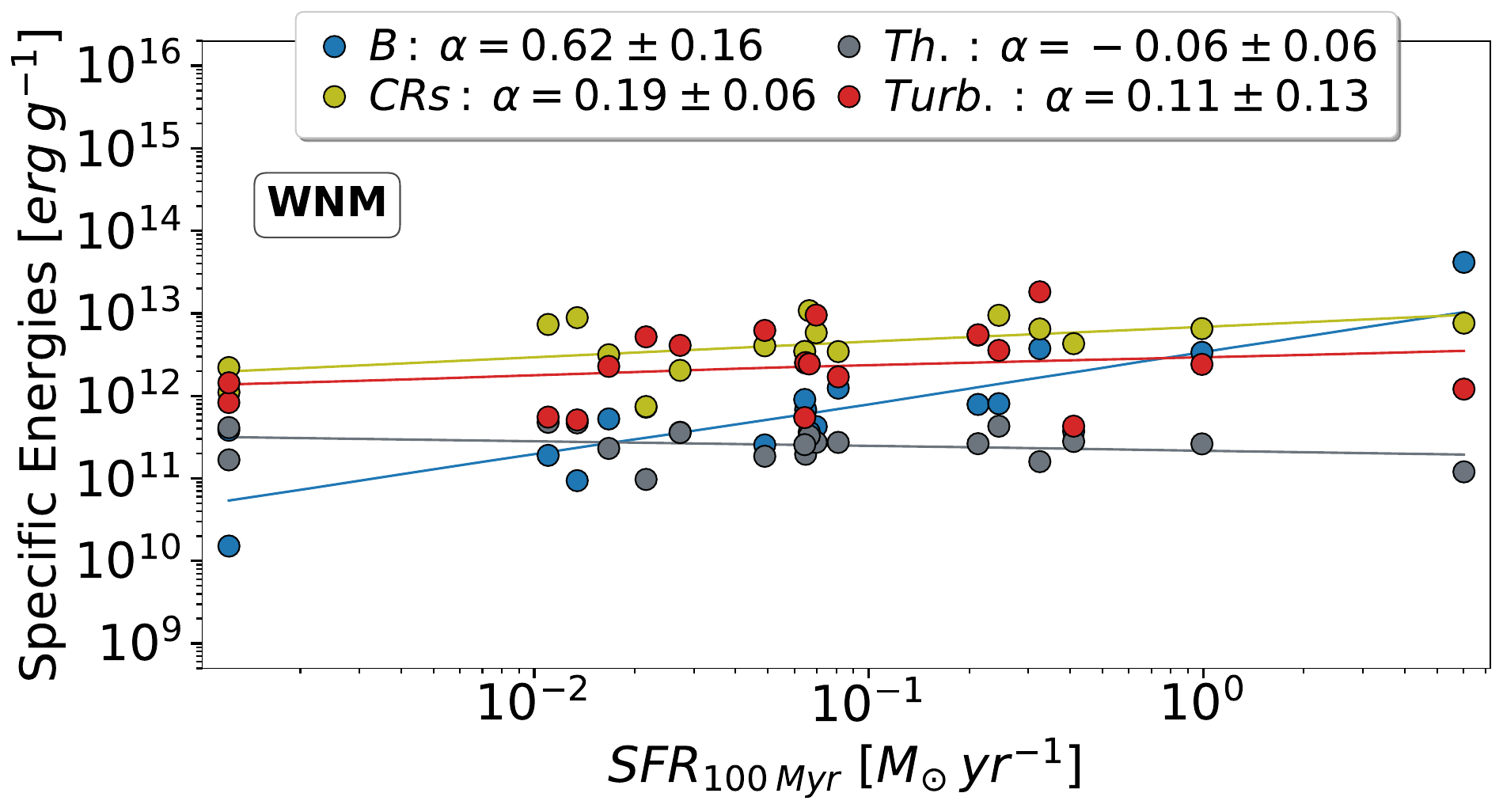}
\includegraphics[width=0.49\textwidth]{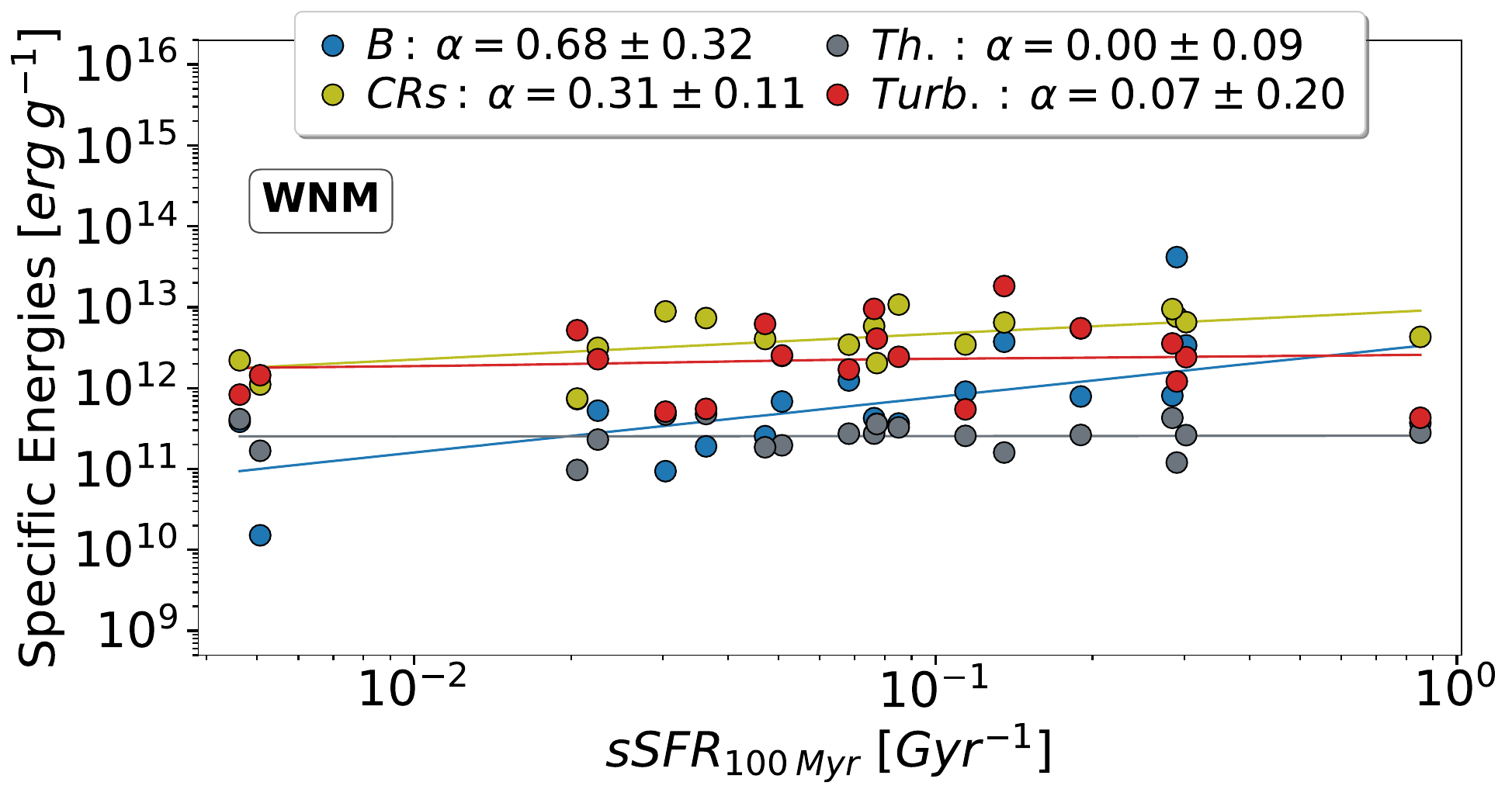}
\caption{Same as Figure~\ref{fig:Energies_SFR_gasphase_CNM}, but for the WNM phase.}
\label{fig:Energies_SFR_gasphase_WNM}
\end{figure}

\paragraph{WNM.}
Figure~\ref{fig:Energies_SFR_gasphase_WNM} shows again the scaling of the specific energies, now for the WNM. The top panel shows these quantities as a function of the SFR, whereas the bottom panel shows them as a function of the sSFR.
While the scalings of the specific energies still follow similar approximate trends, they are broadly shallower for the WNM phase. The magnetic component ($\alpha_{\rm mag,WNM}=0.68\pm0.32$) is approximately unchanged. The CR component is now shallower ($\alpha_{\rm CR,WNM}=0.31\pm0.11$), likely as a result of higher hadronic energy losses for CRs in the denser CNM gas. The turbulent component ($\alpha_{\rm turb,WNM}=0.07\pm0.20$) features only a very shallow correlation with sSFR, in agreement with SN-driven energy injection taking place at $\sim 100$~pc scales, and in the WIM and hot phases; and subsequently cascading down into the WNM \citep{Korpi1999, Joung2006, Martin-Alvarez2022}. The thermal specific energy ($\alpha_{\rm therm,WNM}=0.0\pm0.09$) is approximately independent of the sSFR across the inner star-forming disc of the sampled galaxies.

Unlike for the CNM, the WNM thermal energy provides a significant contribution to the total energy: at a low SFR it can be near or above the magnetic energy (e.g. in 886xp574558438, g8, 719xp463552479, and 1478xp591488450), and in rough equipartition with the turbulent and CR components. However, for ${\rm sSFR}\gtrsim 0.1\,{\rm Gyr^{-1}}$, the non-thermal components strengthen relative to the thermal one, and the magnetic term overtakes the thermal term.
Across both phases, the combination of a steep \(\mu_{\rm mag}\) rise with SFR, a weak trend for \(\mu_{\rm turb}\), and approximately flat \(\mu_{\rm thermal}\) leads to galaxies that become increasingly magnetically dominated once SFR $\gtrsim 1\, \mathrm{M_{\odot}} \, \mathrm{yr}^{-1}$ and sSFR $\gtrsim 0.1\,\mathrm{Gyr}^{-1}$, with this trend being more pronounced in the CNM.

\subsection{Model dependence of the $B$–SFR relation: Seeds versus SN injection}
\label{sec:bsfr_model_dependence}

\begin{figure}[htbp]
\centering
\includegraphics[width=0.49\textwidth]{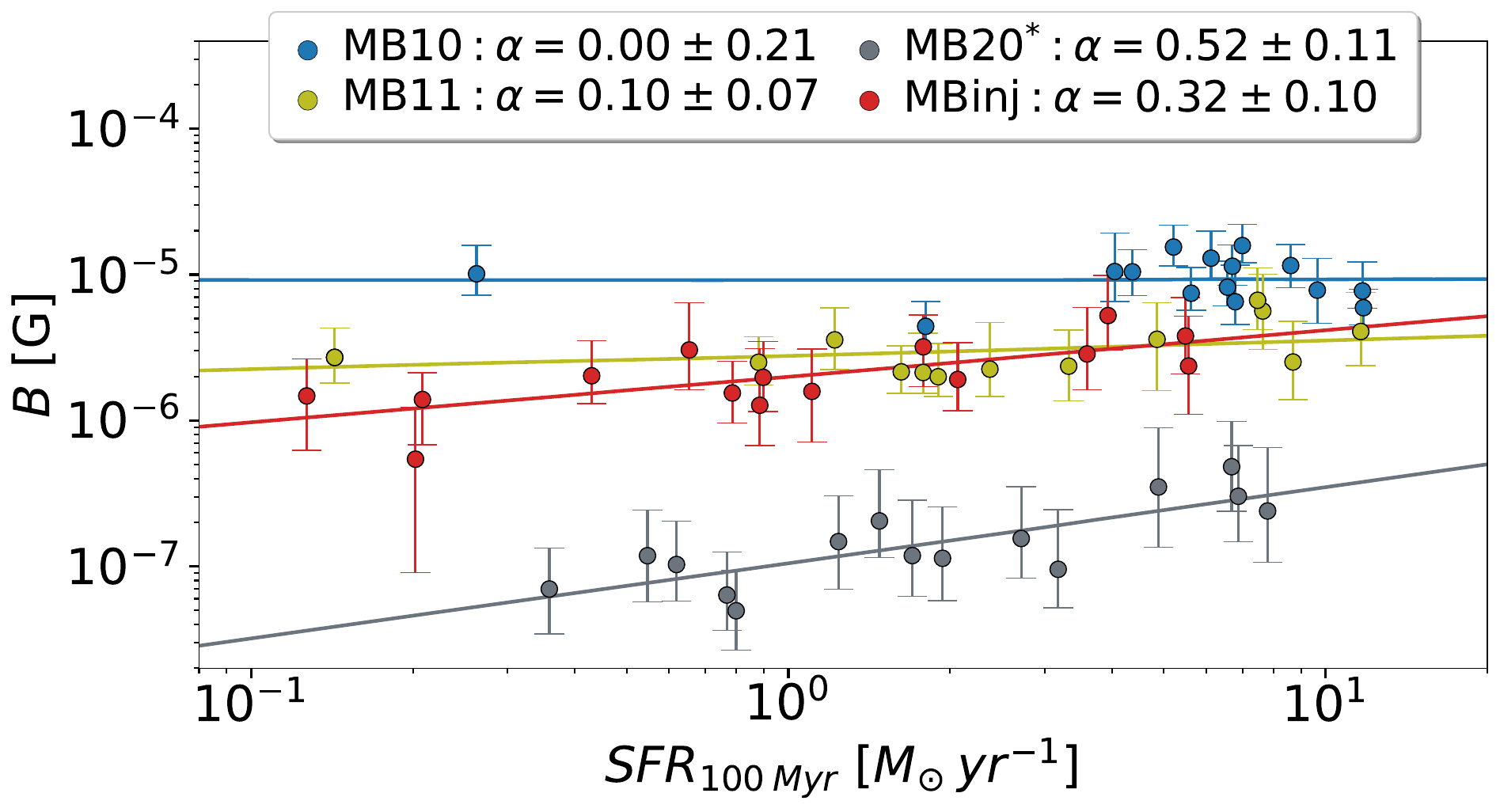}
\includegraphics[width=0.49\textwidth]{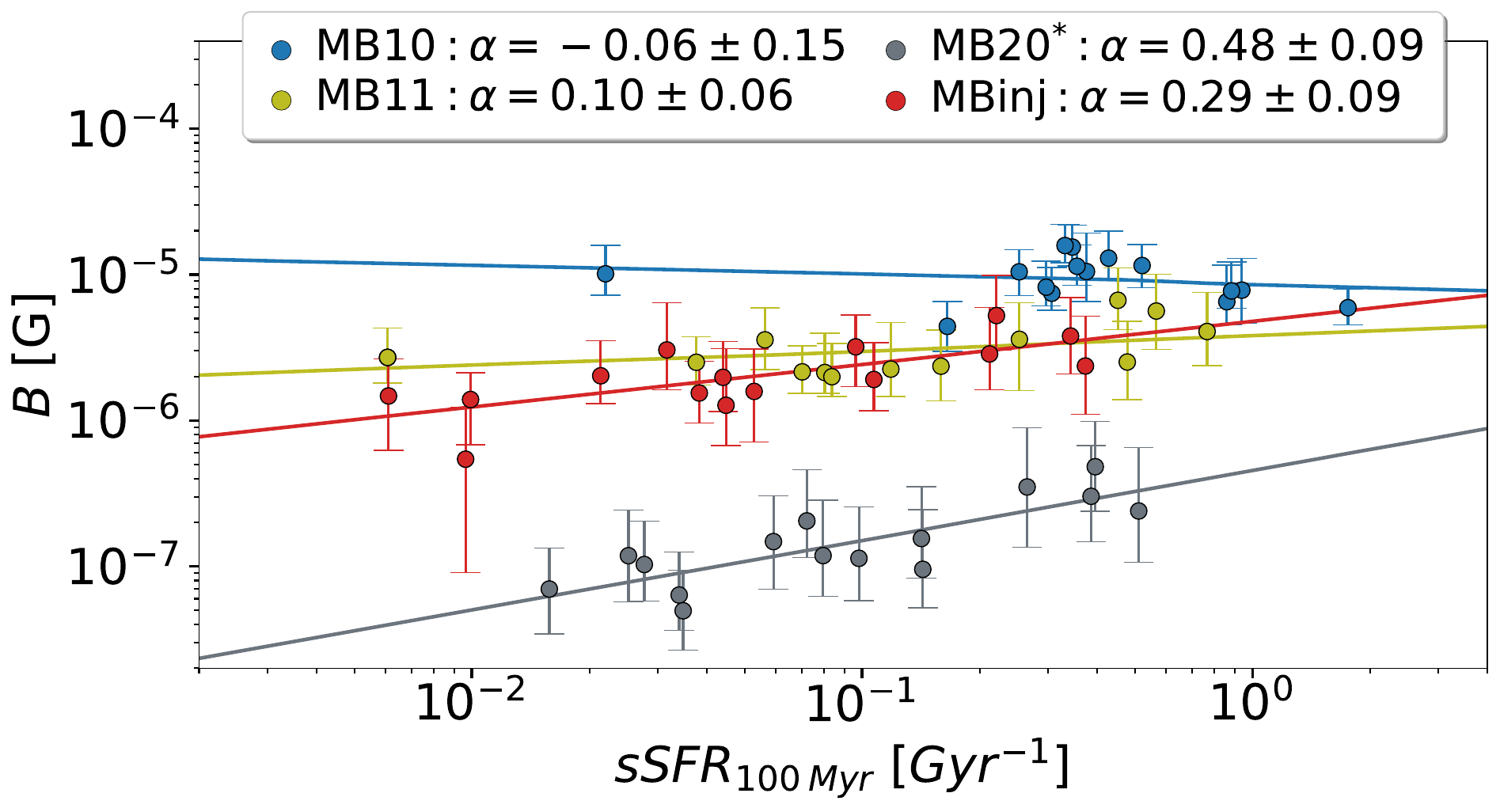}
\caption{Magnetic field strength, $B$, versus SFR (top panel) and specific SFR (bottom panel) for a NUT galaxy with different magnetisation prescriptions. MB$X$ runs adopt primordial seed fields of $B_0=3\times10^{-X}$\,G, while MBinj includes SN magnetic injection. Solid lines show best-fit power laws. Error bars represent IQR computed for each galaxy. \\
$^*$The magnetic field estimates of the MB$20$ model have been multiplied by $10^7$ for better visualisation.}
\label{fig:Bfield_strength_SFR_NUTS}
\end{figure}

To assess how the $B$--SFR relation depends on seed magnetisation and on the injection prescription, we analysed a single Milky-Way like \textsc{nut} galaxy simulated with different initial magnetic configurations (see Section \ref{sec:NUTsetup} for the details of the \textsc{nut} simulation set-up). Each point along a given run corresponds to a different epoch (thus different SFR), so trends reflect the joint evolution in $z$ and activity. We compared three runs with primordial seed fields of amplitude $B_0 = 3\times10^{-X}\,\mathrm{G}$ (hereafter MB$X$, with $X=10\ ,\ 11,\ 20$) and one run with SN magnetic injection (MBinj), which adopts the same prescription used in our main sample.
The data and power-law fits are shown in Fig.~\ref{fig:Bfield_strength_SFR_NUTS} for both the SFR (top panel) and sSFR (bottom panel).

We find the following slopes for the different magnetisation models:
\begin{itemize}[label=-]
    \item MB10: $\alpha_{\rm SFR}=0.00\pm0.21$, \quad $\alpha_{\rm sSFR}=-0.06\pm0.15$,
    \item MB11: $\alpha_{\rm SFR}=0.10\pm0.07$, \quad $\alpha_{\rm sSFR}=0.10\pm0.06$,
    \item MB20: $\alpha_{\rm SFR}=0.52\pm0.11$, \quad $\alpha_{\rm sSFR}=0.48\pm0.09$,
    \item MBinj: $\alpha_{\rm SFR}=0.32\pm0.10$, \quad $\alpha_{\rm sSFR}=0.29\pm0.09$.
\end{itemize}

These slopes reveal differences in the magnetic response to star formation among the various seed configurations.
Models with high primordial seeds produce shallow slopes (consistent with $0$--$0.1$), as expected once the dynamo saturates early: after saturation, $B$ varies only weakly with instantaneous SFR, since magnetic energy has already reached near-equipartition with the turbulent component.

The MB20 model has the larger scaling with star formation activity with SFR, but with very low absolute values for the magnetic field strength ($< 10^{-14} \: G$). 
The steep MB20 slope is consistent with a pre-saturation small-scale dynamo: $B$ is still amplified efficiently by turbulent motions, so it responds more strongly to changes in SFR and sSFR than it would in the saturated/equipartition regime, where the slope is expected to flatten.
The SN-injected run (MBinj) is fully consistent with our main spirals and with saturated small-scale dynamo expectations, supporting the physical fidelity of the injection model.

Overall, these results suggest that within a single galaxy the $B$--SFR slope is set by the state of the small-scale dynamo. Strong seeds or continuous SN injection bring the system to saturation early and yield shallow $\alpha$ (MB10/MB11) or the canonical $\alpha\!\approx\!0.3$ (MBinj). Extremely weak seeds (MB20) stay kinematic for longer, producing steep slopes but negligible absolute field strengths.

\section{Summary and conclusions}
\label{sec:conclusions}

In this work, we have investigated the interrelation of magnetic fields scale with star-formation activity in spiral galaxies. We used the \textsc{Azahar} simulation suite (Martin-Alvarez et al. in prep) full-physics models, incorporating RT, CRs, and magnetism with state-of-the-art star formation and SN feedback models. We selected a sample of secularly evolving disc galaxies with $10^{-3} \, \mathrm{M_\odot\,yr^{-1}}\lesssim \mathrm{SFR}_{100} \lesssim 10 \, \mathrm{M_\odot\,yr^{-1}}$ and stellar masses of $10^8 \, \msun \lesssim M_* \lesssim  10^{10} \, \msun$, corresponding to $z \sim 3$ analogues of local starburst galaxies. For these systems, we generated edge-on measurements of various quantities, focusing on their inner star-forming disc (with typical major axis of $\sim 0.5$ kpc). We compared our results with observed samples \citep{Chyzy2011, Lacki2013, Beck2019, Stein2019, Mora-Partiarroyo2019, Stein2020, Heald2022}, across neutral ISM phases (CNM and WNM). We tested the robustness of our results by reviewing this trends with an additional simulation suite of a Milky Way-like spanning multiple degrees and channels of magnetisation \citep{Martin-Alvarez2021}.
Our main findings are:

\begin{itemize}[label=-]
    \item We find an approximately universal scaling of magnetic field strength with SFR in galaxies ($\alpha \sim 0.2 - 0.3$). This scaling is preserved across galaxy masses (from dwarf to spiral galaxies), inclinations, neutral phases (CNM and WNM), and galactic environments. 
    
    \item The star formation surface density follows the expectation for SN-driven ISM turbulence, and turbulence-regulated magnetic field strength ($\alpha_\Sigma \sim 1/3$; \citealt{Schleicher2013}). Remarkably, when expressed in terms of $\Sigma_{\mathrm{SFR}}$, all systems align along the common sequence defined by our simulated galaxies, independently of their individual slopes.

    \item The $B$-sSFR relation provides the largest segregation between environments and galaxy types, both in terms of magnetic field strength normalisation and its power-law index. All systems, however, remain broadly consistent with a common scaling of $\alpha \sim 0.3$.

    \item All energy components (i.e. turbulent, magnetic, CRs, and thermal) in our simulated galaxies increase with star formation, with the magnetic energies displaying the steepest dependence. While these trends are both present in both the CNM and WNM phases, we measure a steeper increase in the CRs and turbulent components with sSFR in the CNM phase.

    \item Overall, our CNM and WNM gas phases are primarily dominated by the turbulent energy component, followed by the CR energy, and with a more negligible contribution of the thermal energy in the WNM. Following the steepness of the magnetic energy with sSFR, we find the magnetic energy to range from sub-dominant at low star formation (sSFR $< 0.1\,\mathrm{Gyr}^{-1}$, SFR $< 1 \, \mathrm{M_{\odot}} \, \mathrm{yr}^{-1}$) to approximate equipartition with turbulent and CR at high star formation activity (sSFR $\gtrsim 0.1\,\mathrm{Gyr}^{-1}$, SFR $\gtrsim 1\,\mathrm{M_{\odot}} \, \mathrm{yr}^{-1}$).

    \item Varying the initial magnetisation and the seeding channel in a single Milky-Way-like \textsc{nut} galaxy suggests that the $B$–SFR slope is controlled by the dynamo’s saturation state. Strong magnetic seeds ($B_0 = 10^{-10}\,G$, $B_0 = 10^{-11}\,G$) saturate early and yield very shallow scalings with SFR/sSFR ($\alpha\sim0$–0.1), whereas SN magnetic injection (MBinj) reproduces the canonical $\alpha\simeq0.3$ with realistic normalisation. Extremely weak seeds ($B_0 = 10^{-20}\,G$) produce steep apparent slopes ($\alpha\simeq0.5$) but remain at negligible absolute field strengths ($B\lesssim10^{-14}\,$G) for simulations with a resolution of $\sim1{-}10$~pc. Thus, the injection channel sets the time to saturation, whereas the normalisation is determined by the saturation balance with turbulence. Once SN-driven turbulence maintains a saturated small-scale dynamo, a near-universal $B$–star formation relation emerges.

\end{itemize}

Our results emphasise that magnetic fields are dynamically important agents in shaping the structure and evolution of galaxies.  
Their tight correlation with star formation activity across mass, phase, and environment demonstrates that magnetism is intimately linked to SN-driven turbulence and feedback. Magnetic energy becomes dynamically relevant, often approaching equipartition with turbulent and cosmic-ray components, for systems with ongoing star formation (${\rm sSFR} \gtrsim 0.1~{\rm Gyr^{-1}}$).  
In the neutral ISM, both the CNM and WNM exhibit this coupling, highlighting the role of magnetic pressure in regulating cloud support and disc stability.

Recent ALMA FIR polarimetric observations of gravitationally lensed dusty star-forming galaxies at $z=2.6$ (9io9) and $z=5.6$ (SPT0346$-$52) reveal kiloparsec-scale ordered magnetic fields, with SFRs of $\gtrsim10^3\,\mathrm{M_\odot\,yr^{-1}}$ and inferred field strengths of $B\!\lesssim\!514~\mu\mathrm{G}$ and $\lesssim\!450~\mu\mathrm{G}$, respectively \citep{Geach2023,Chen2024}. Under these extreme conditions, magnetic pressure and tension are likely to play a direct role in regulating star formation and shaping the dynamics of the ISM from the very early stages of galaxy formation.
The qualitative agreement between these recent high-redshift observations and the regular magnetic field morphology in our simulations suggests that signatures of large-scale magnetic ordering emerge rapidly even in highly star-forming systems, where the magnetic energy budget is still dominated by small-scale, turbulence-driven field components.

More broadly, these findings underline the importance of including multi-phase ISM physics and magnetic fields in cosmological galaxy-formation simulations, especially as the resolution in these models reaches into the sub-$100$~pc regime.  
Such models, coupled with realistic cosmic-ray transport and radiative-transfer treatments, will enable a more physically complete picture of how magnetic energy builds up and interacts with the multiphase gas.  
Progress on this front will require synergy with multi-wavelength polarimetric observations, spanning from the FIR (e.g. the mission concept Probe far-infrared Mission for Astrophysics PRIMA, \citealp{Glenn2025}) to the radio: FIR dust polarisation tracing the dense, small-scale field structure of molecular clouds, and radio synchrotron and Faraday-rotation diagnostics probing the diffuse, large-scale magnetised halos.

\begin{acknowledgements}

We sincerely thank the referee for their constructive and insightful comments, which have helped us to improve the quality and clarity of the manuscript.

D.B. thanks support from the Marco Polo Programme of the University of Bologna, which funded his research period abroad at the University of South Carolina, and thanks Enrique Lopez-Rodriguez for supervision and support during this period. D.B. also thanks Sergio Martín-Alvarez for sponsoring access to the Sherlock high-performance computing cluster at Stanford University and for guidance in its use.

E.L.-R. thanks support by the NASA Astrophysics Decadal Survey Precursor Science (ADSPS) Program (NNH22ZDA001NADSPS) with ID 22-ADSPS22-0009 and agreement No. 80NSSC23K1585. 

S.M.A. is supported by the Kavli Institute for Particle Astrophysics and Cosmology. 

\end{acknowledgements}

\bibliographystyle{aa} % style aa.bst
\bibliography{bibliography.bib} % your references Yourfile.bib

\begin{appendix}

\twocolumn[
\section{Simulated galaxy maps}
\label{sec:gal_maps}

\noindent

\begin{figure*}[!b]
\centering
\includegraphics[width=0.20\linewidth]{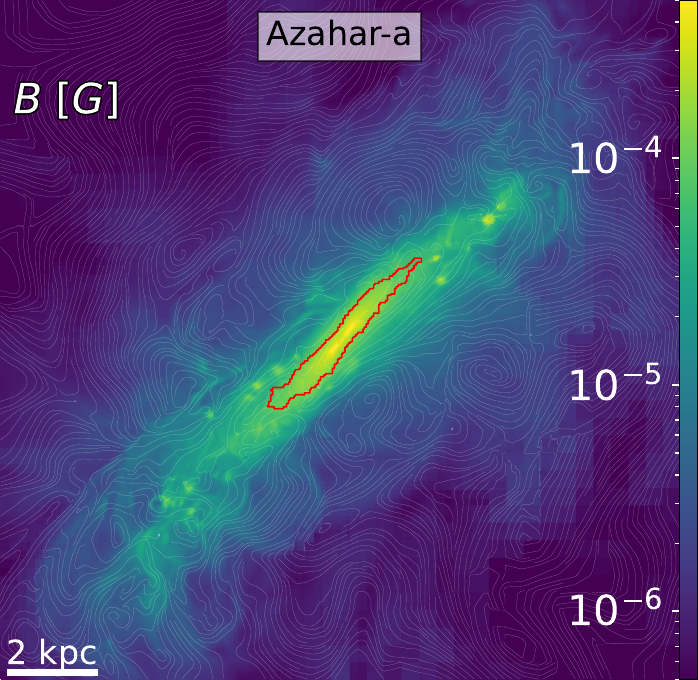}
\includegraphics[width=0.20\linewidth]{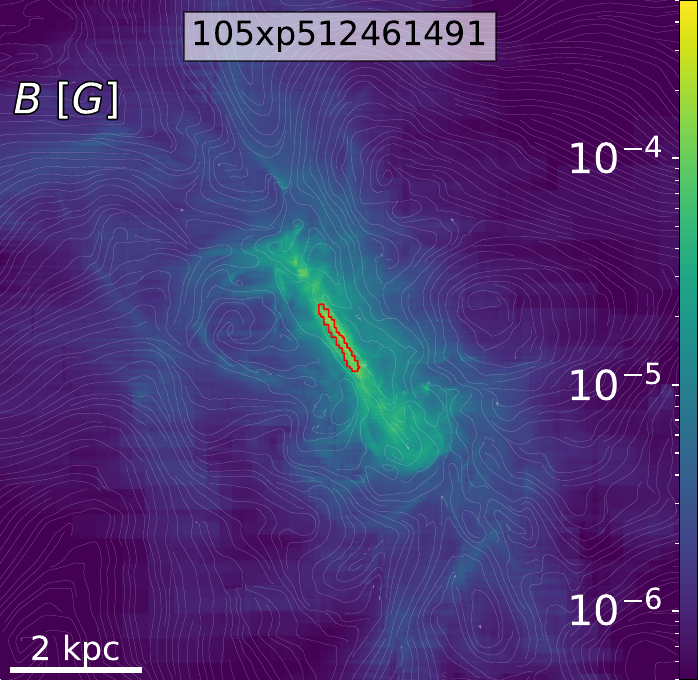}
\includegraphics[width=0.20\linewidth]{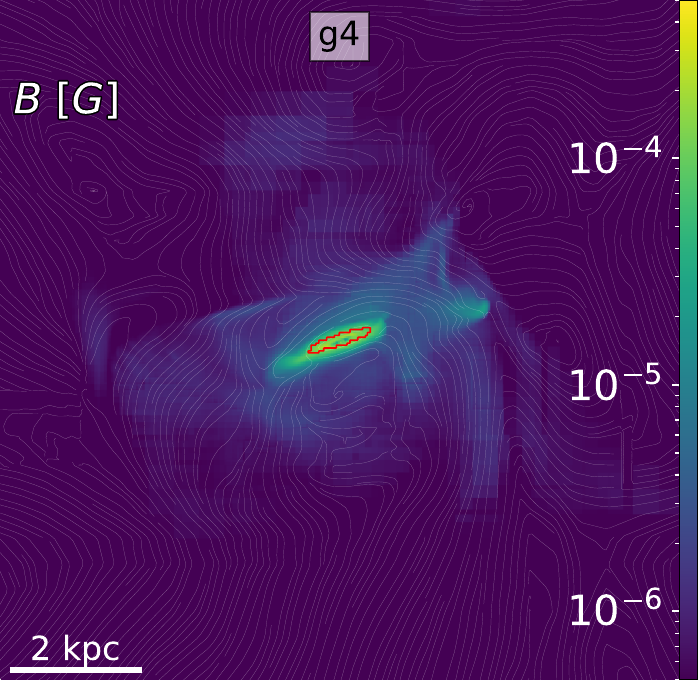}
\includegraphics[width=0.20\linewidth]{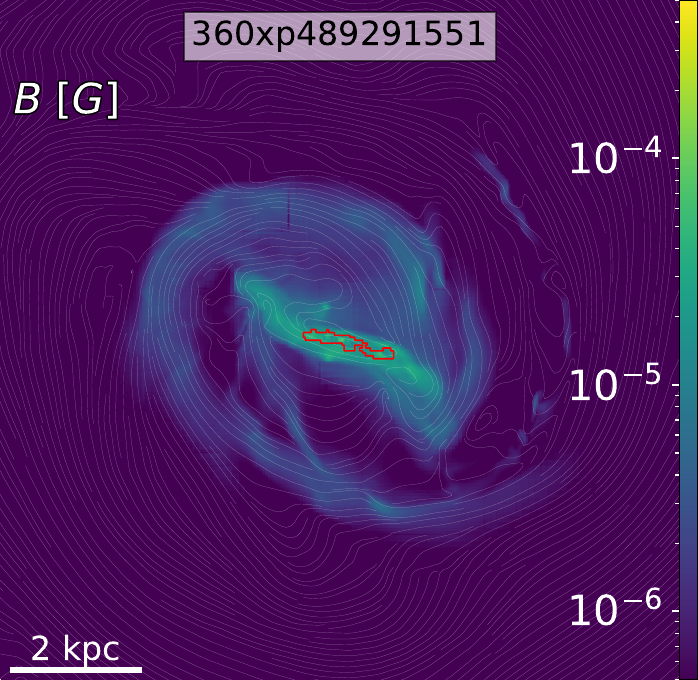}
\includegraphics[width=0.20\linewidth]{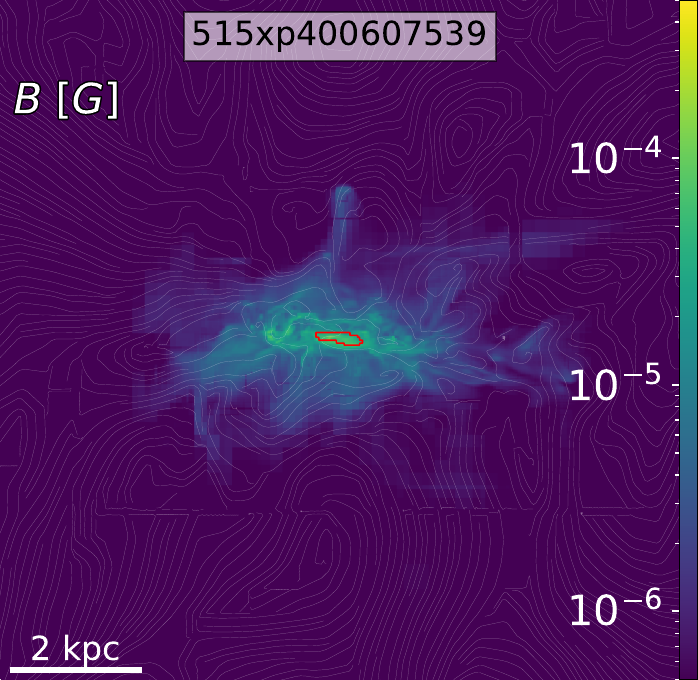}
\includegraphics[width=0.20\linewidth]{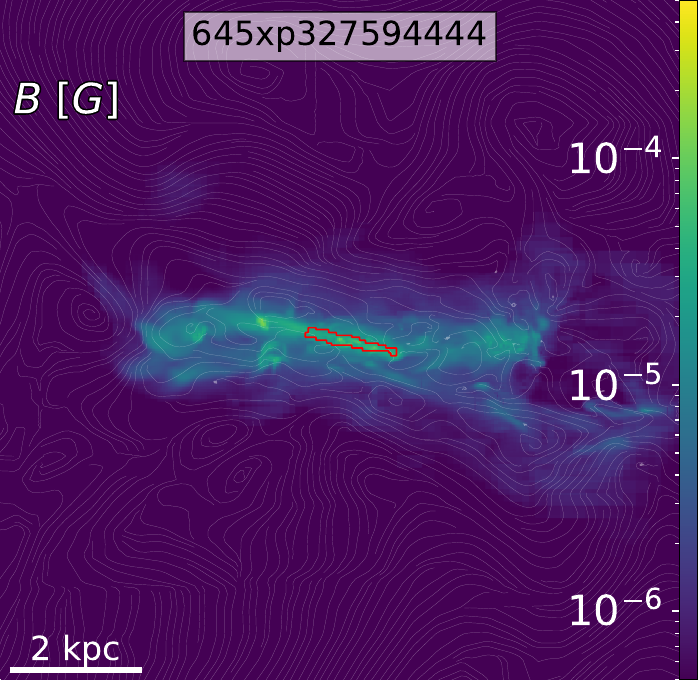}
\includegraphics[width=0.20\linewidth]{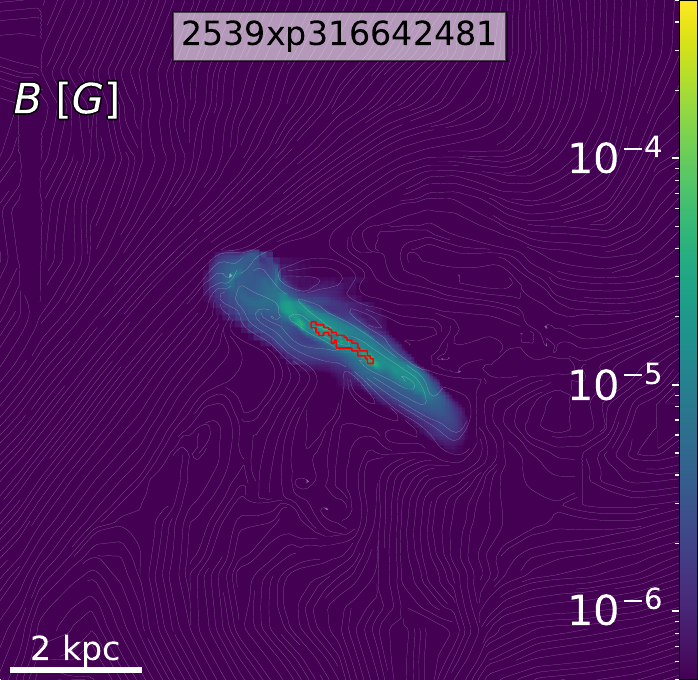}
\includegraphics[width=0.20\linewidth]{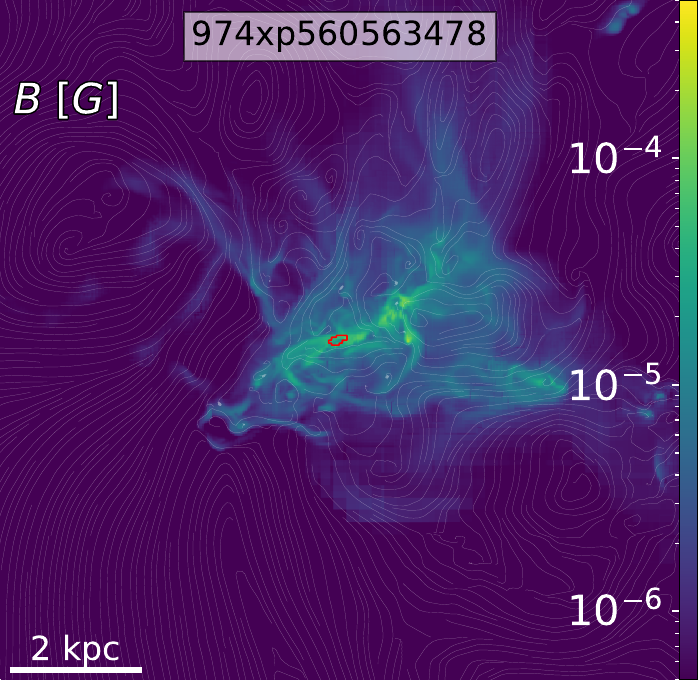}
\includegraphics[width=0.20\linewidth]{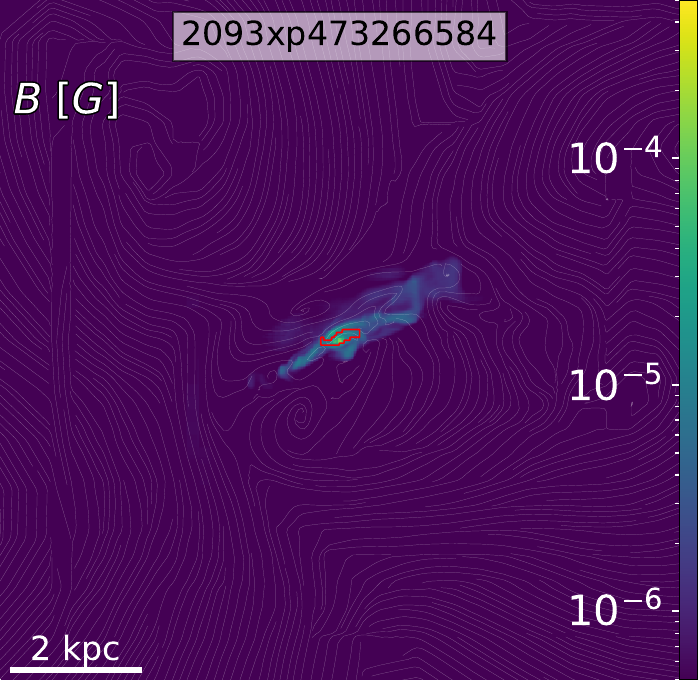}
\includegraphics[width=0.20\linewidth]{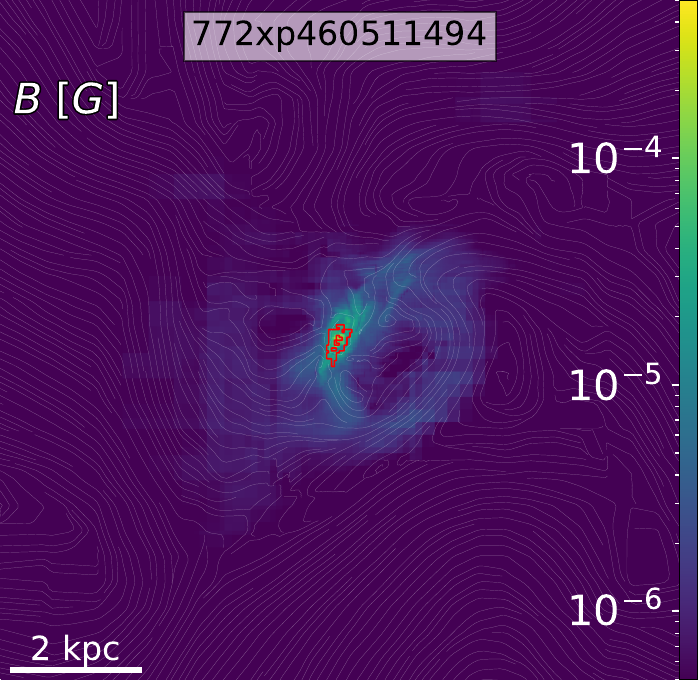}
\includegraphics[width=0.20\linewidth]{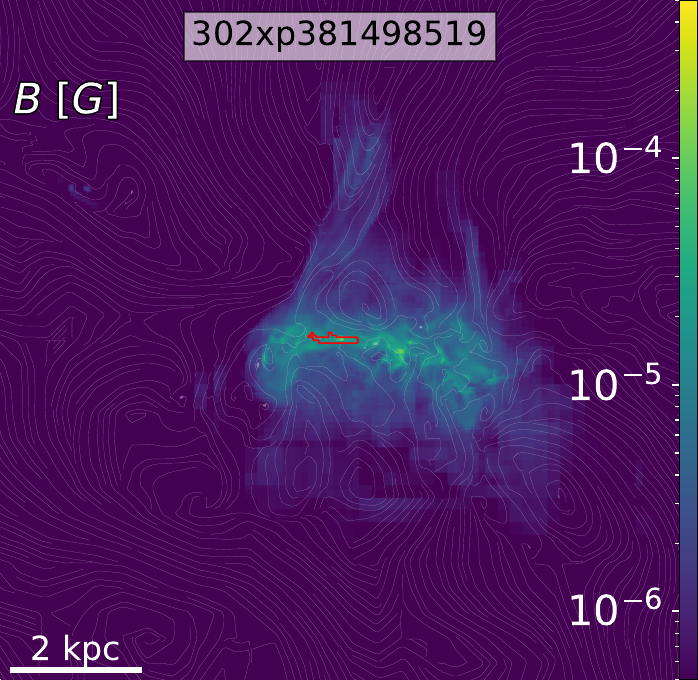}
\includegraphics[width=0.20\linewidth]{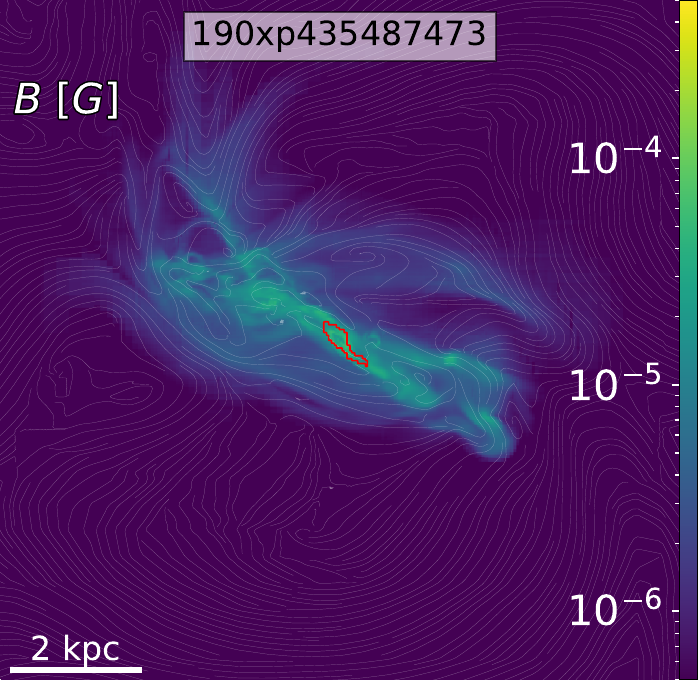}
\includegraphics[width=0.20\linewidth]{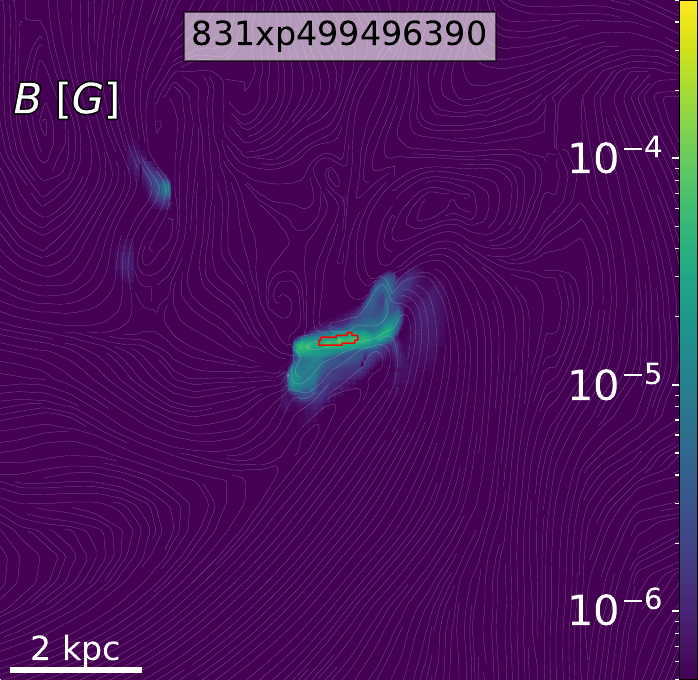}
\includegraphics[width=0.20\linewidth]{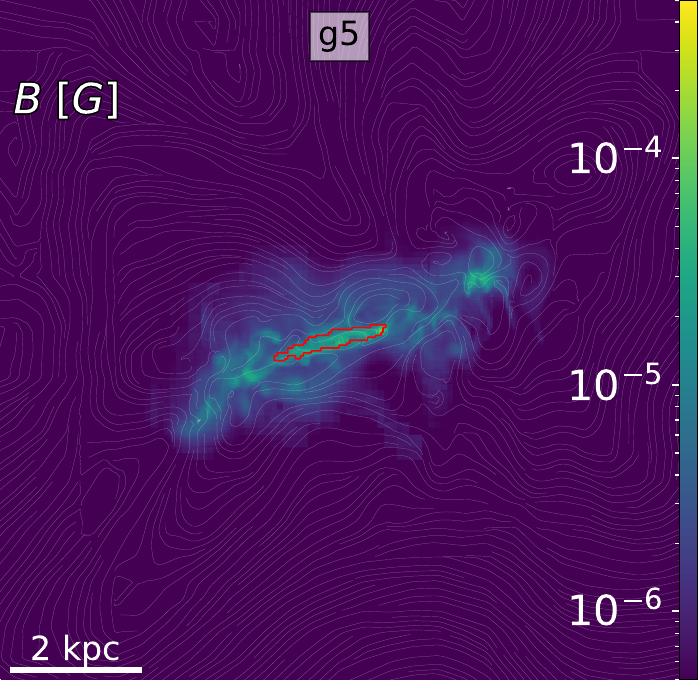}
\includegraphics[width=0.20\linewidth]{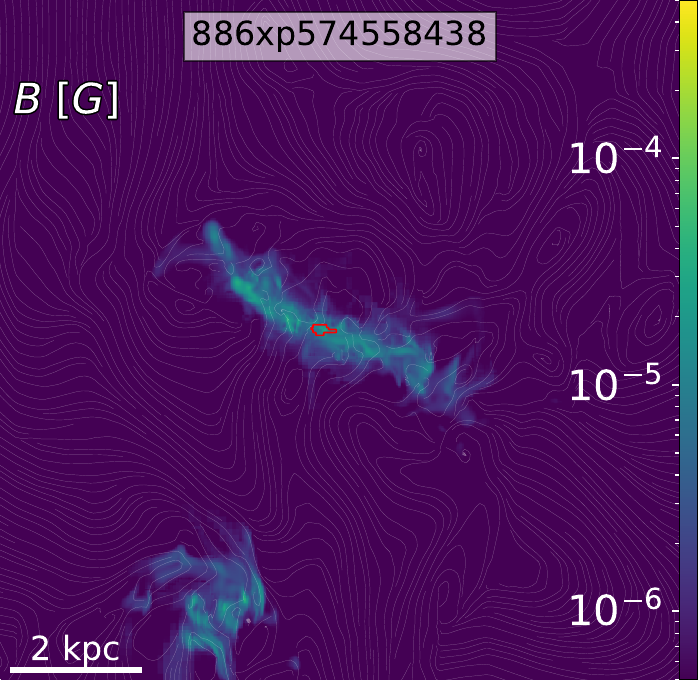}
\includegraphics[width=0.20\linewidth]{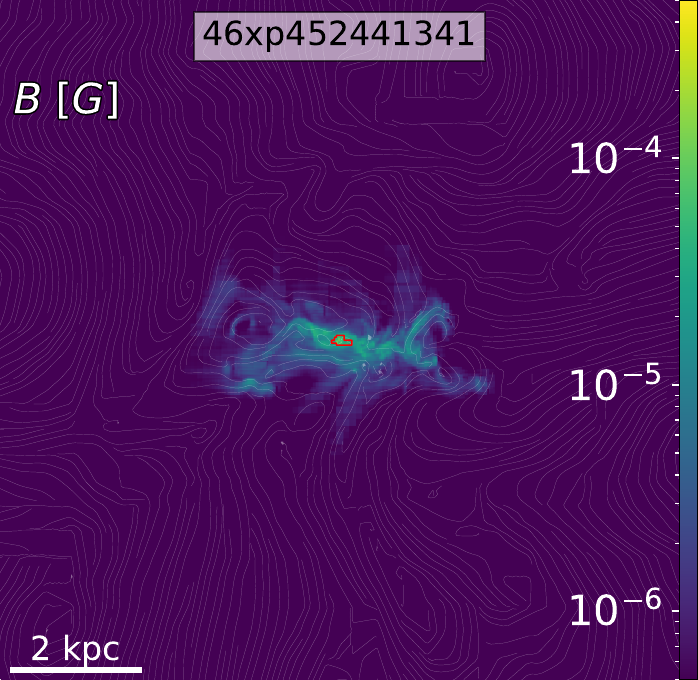}
\includegraphics[width=0.20\linewidth]{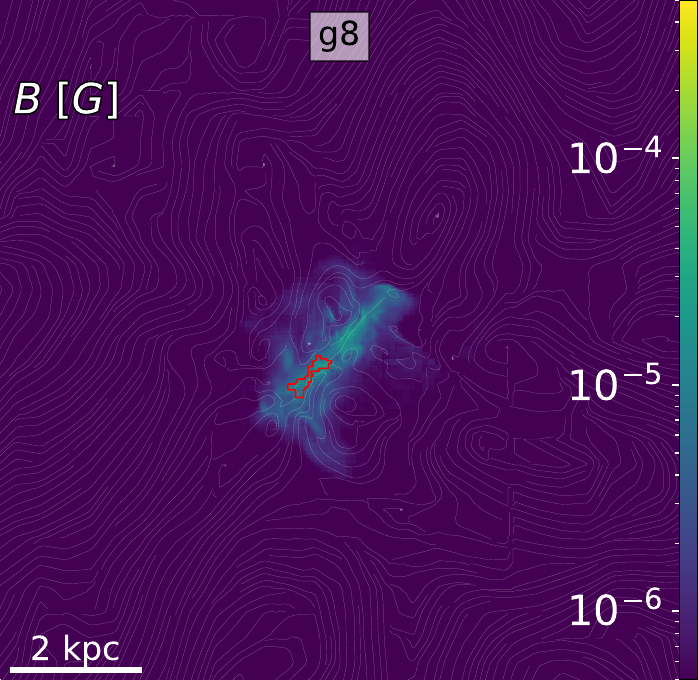}
\includegraphics[width=0.20\linewidth]{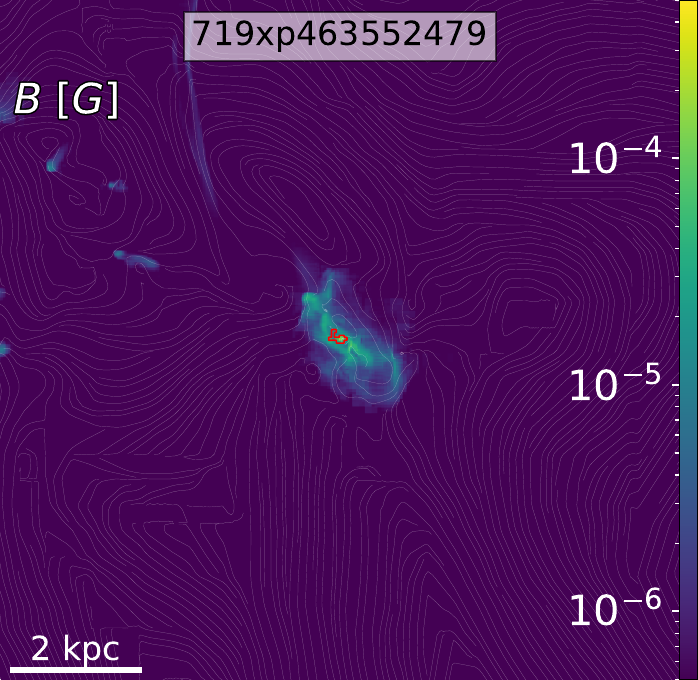}
\includegraphics[width=0.20\linewidth]{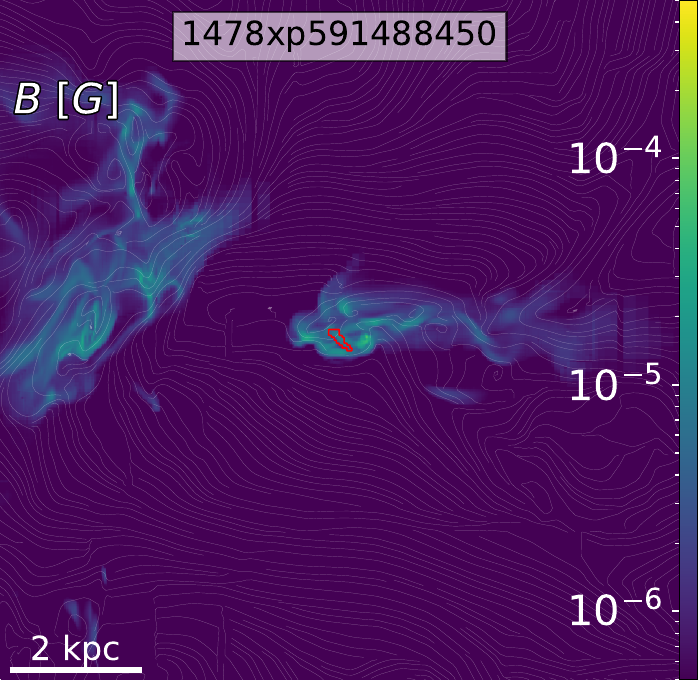}
\caption{Magnetic field strength maps of the simulated galaxies shown face-on. White streamlines trace the projected magnetic field orientation, while the red contours delineate the region of the galactic disc used in the analysis.}
\label{fig:Bfieldsmaps_all_edgeon}
\end{figure*}

\begin{figure*}[!b]
\centering
\includegraphics[width=0.20\linewidth]{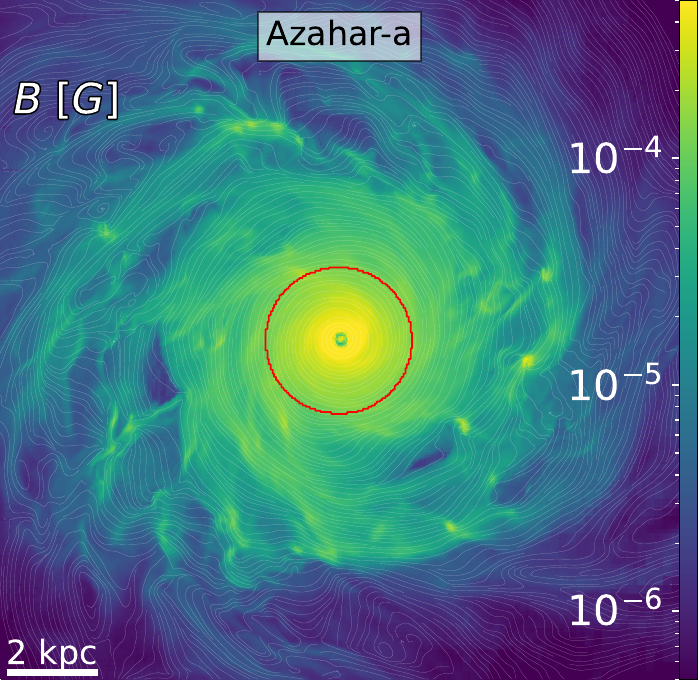}
\includegraphics[width=0.20\linewidth]{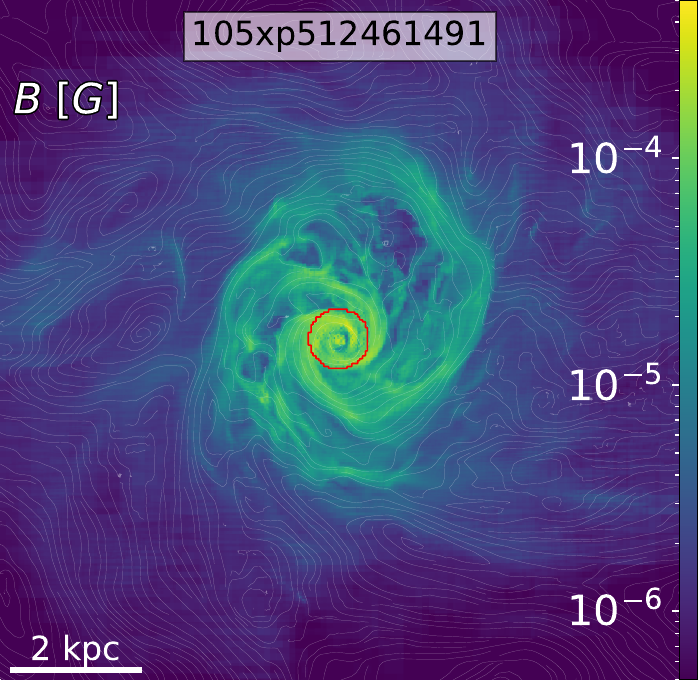}
\includegraphics[width=0.20\linewidth]{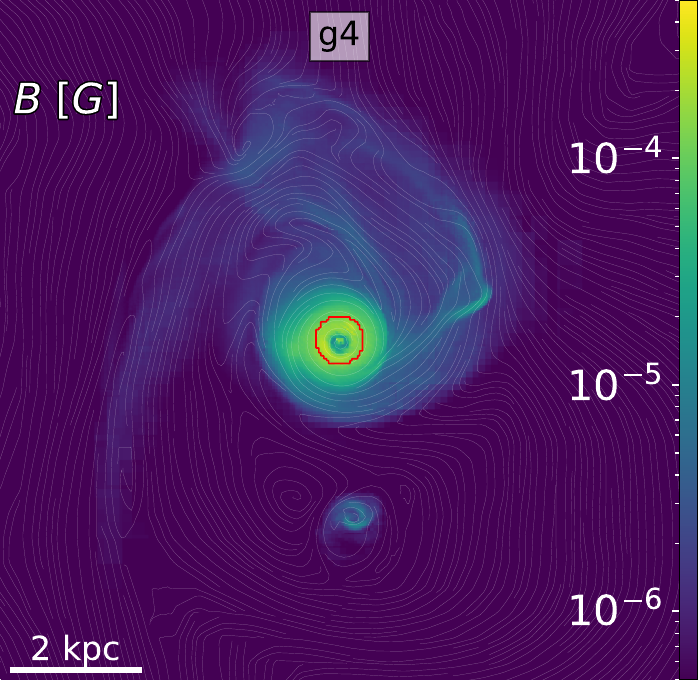}
\includegraphics[width=0.20\linewidth]{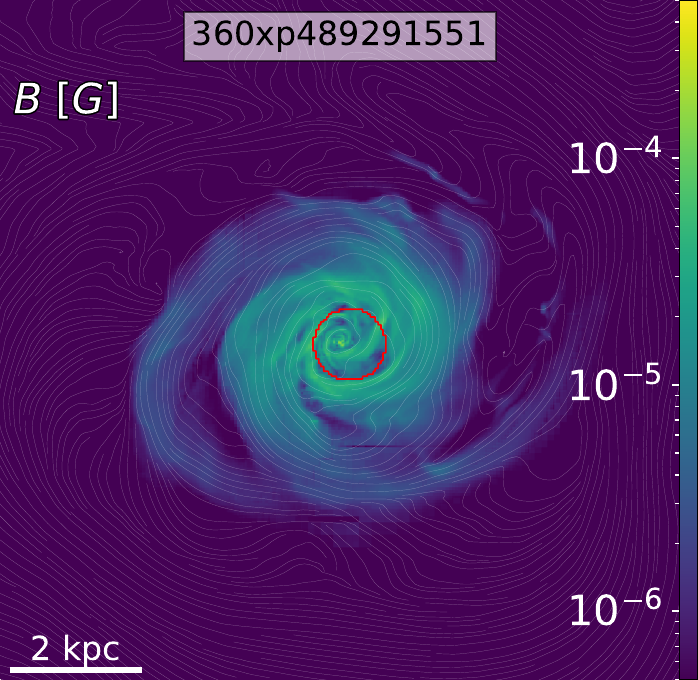}
\includegraphics[width=0.20\linewidth]{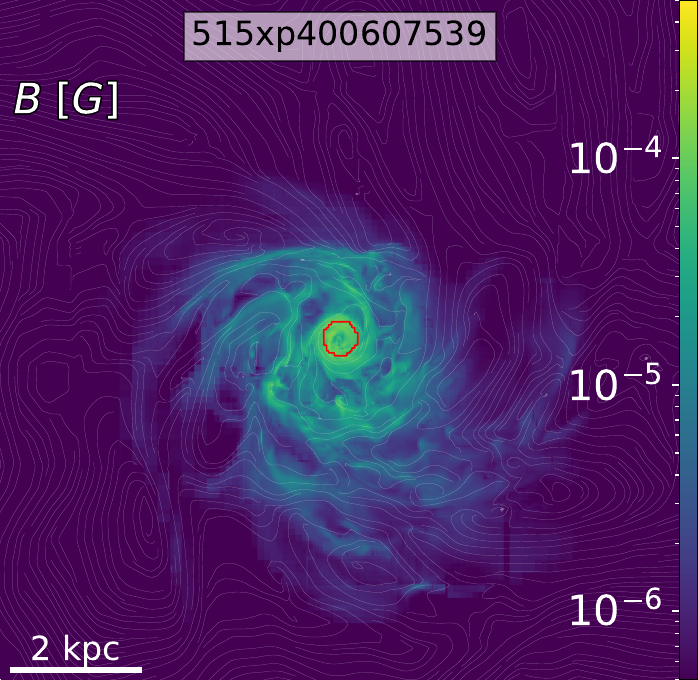}
\includegraphics[width=0.20\linewidth]{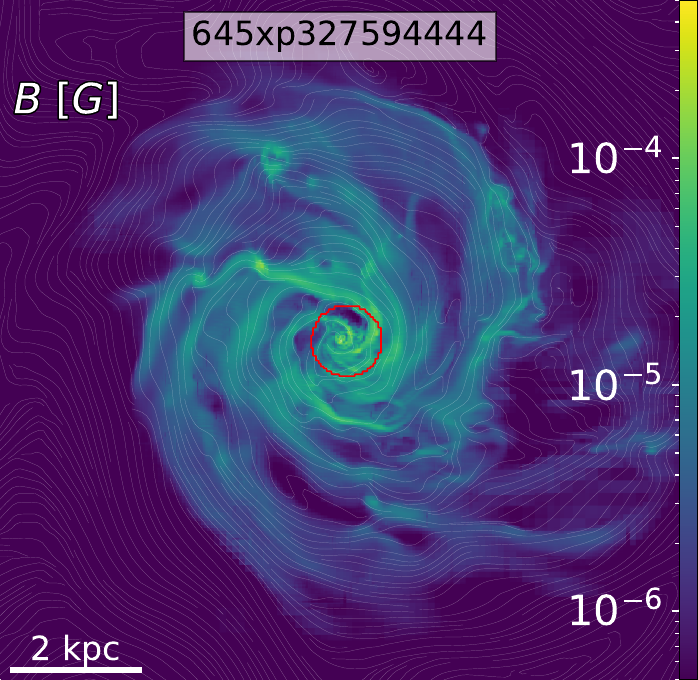}
\includegraphics[width=0.20\linewidth]{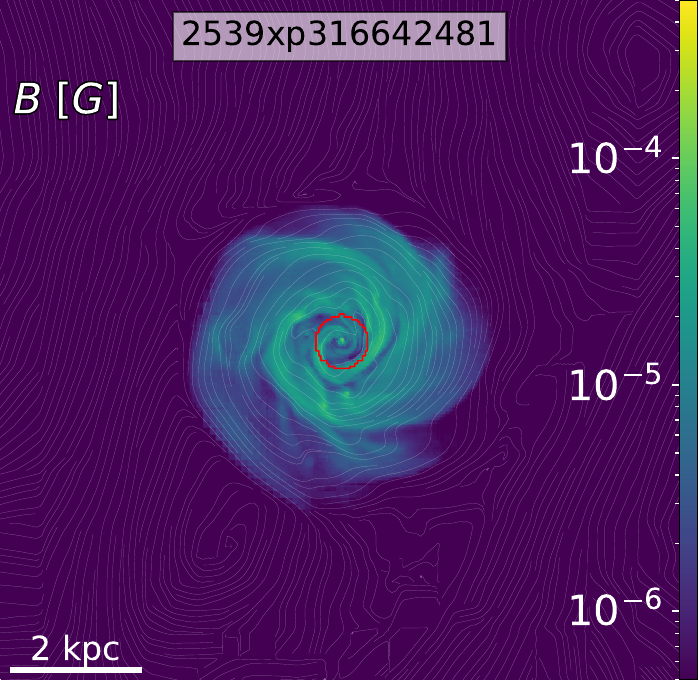}
\includegraphics[width=0.20\linewidth]{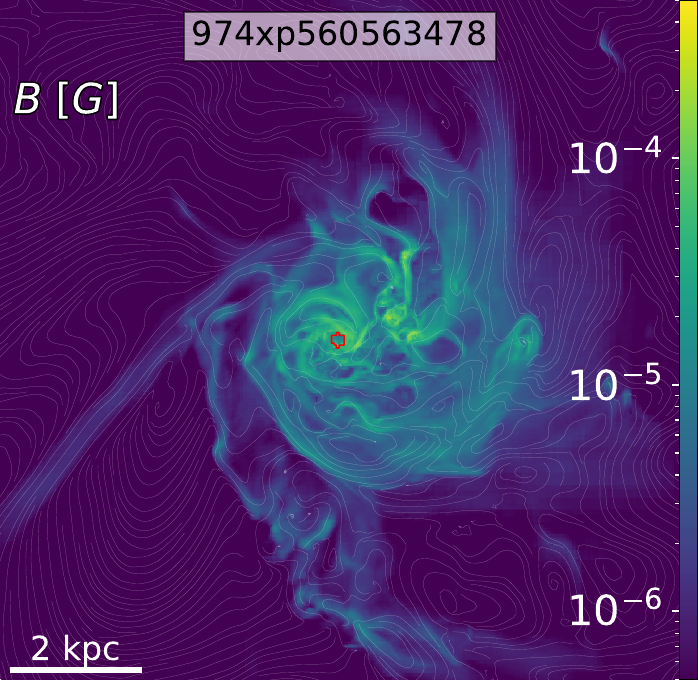}
\includegraphics[width=0.20\linewidth]{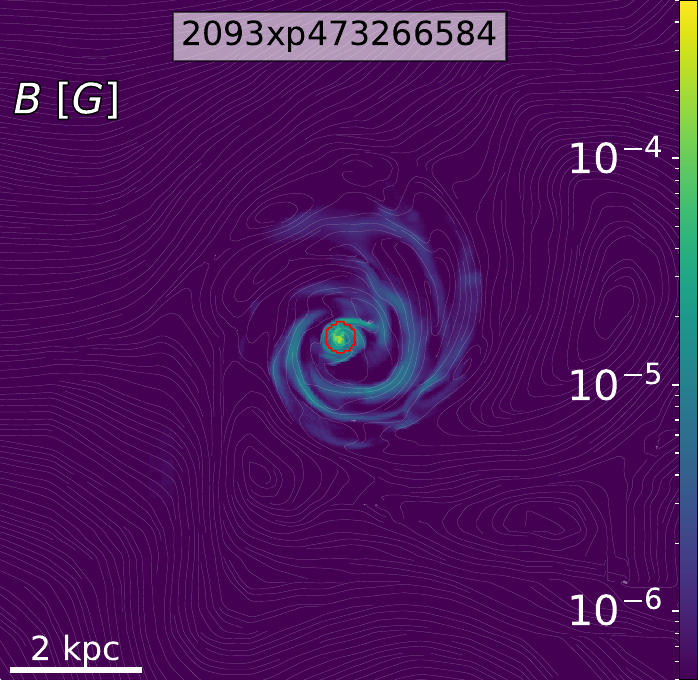}
\includegraphics[width=0.20\linewidth]{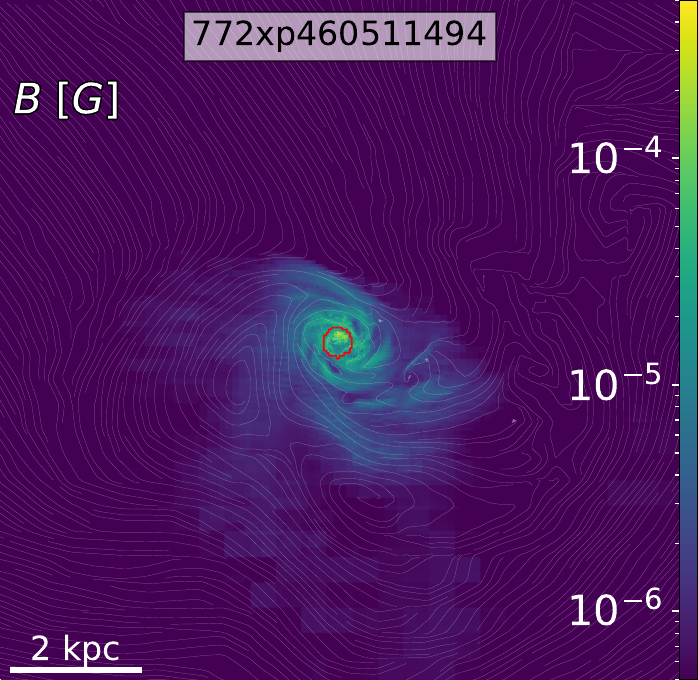}
\includegraphics[width=0.20\linewidth]{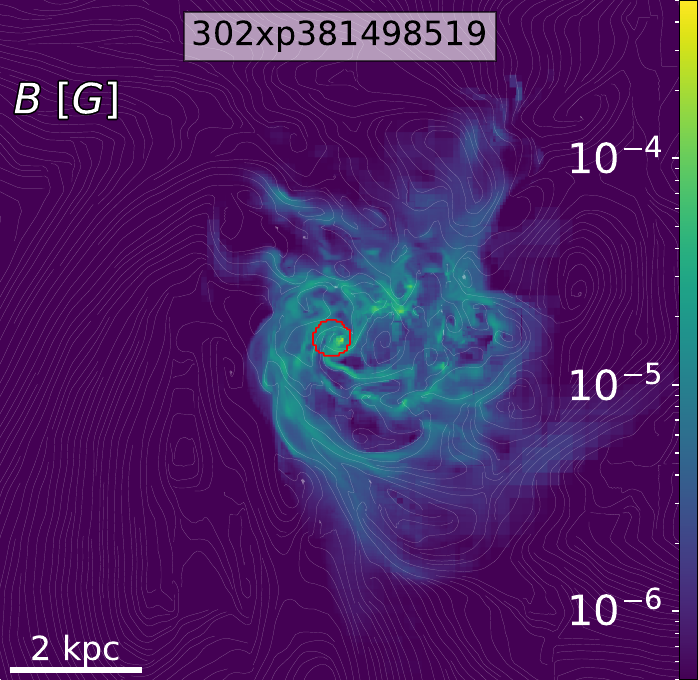}
\includegraphics[width=0.20\linewidth]{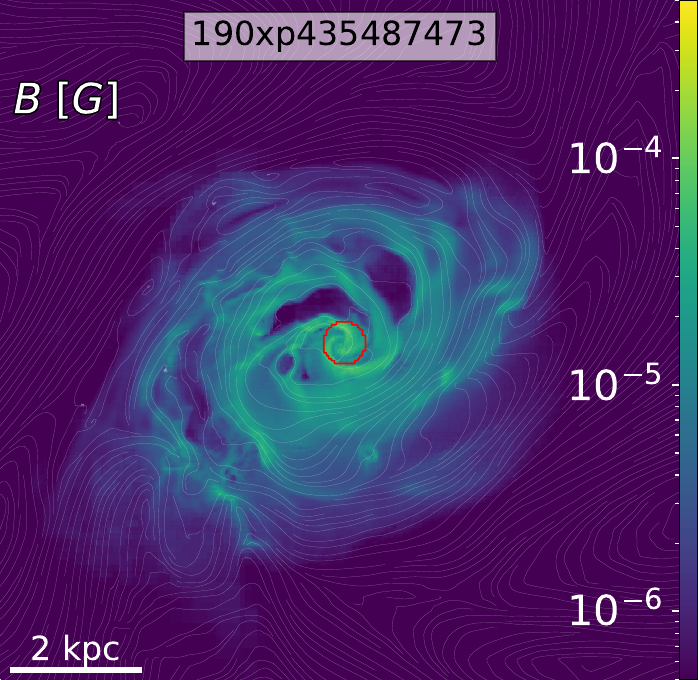}
\includegraphics[width=0.20\linewidth]{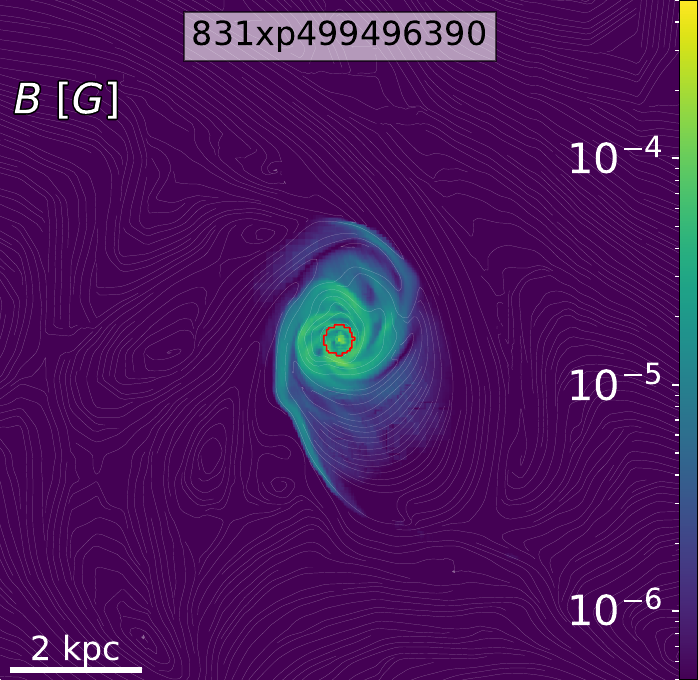}
\includegraphics[width=0.20\linewidth]{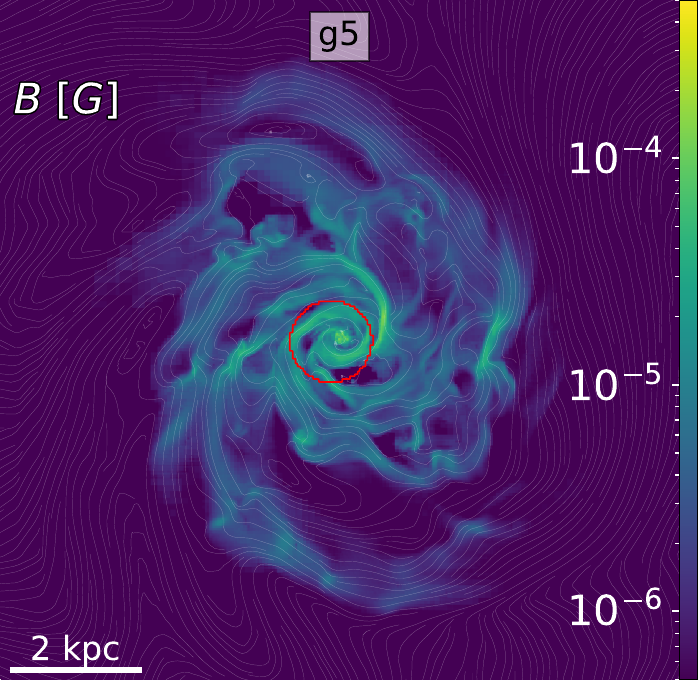}
\includegraphics[width=0.20\linewidth]{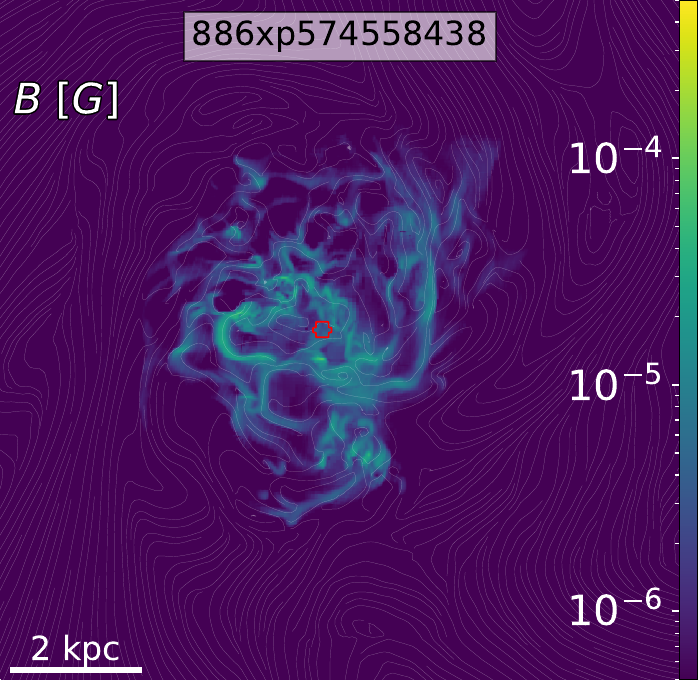}
\includegraphics[width=0.20\linewidth]{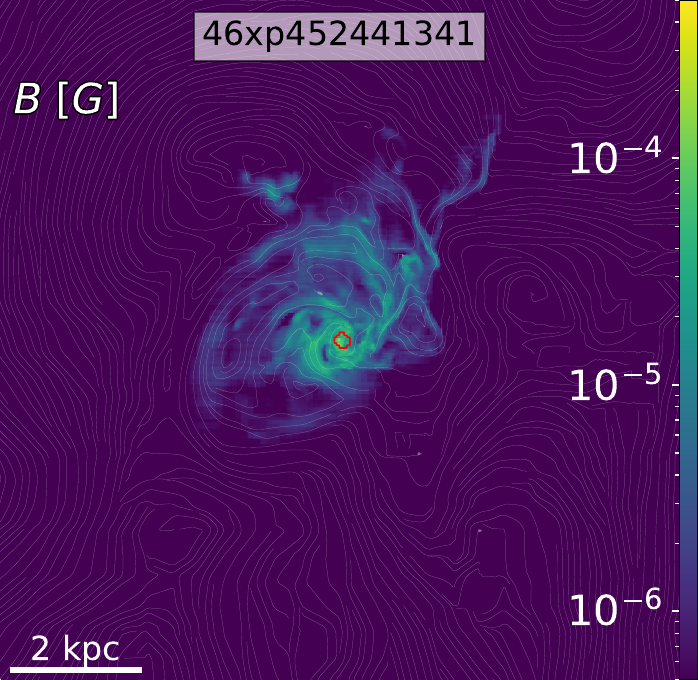}
\includegraphics[width=0.20\linewidth]{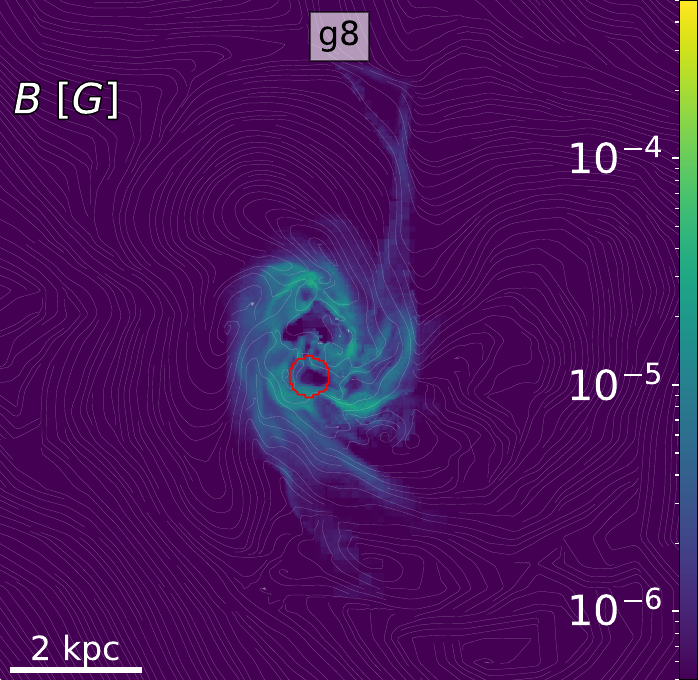}
\includegraphics[width=0.20\linewidth]{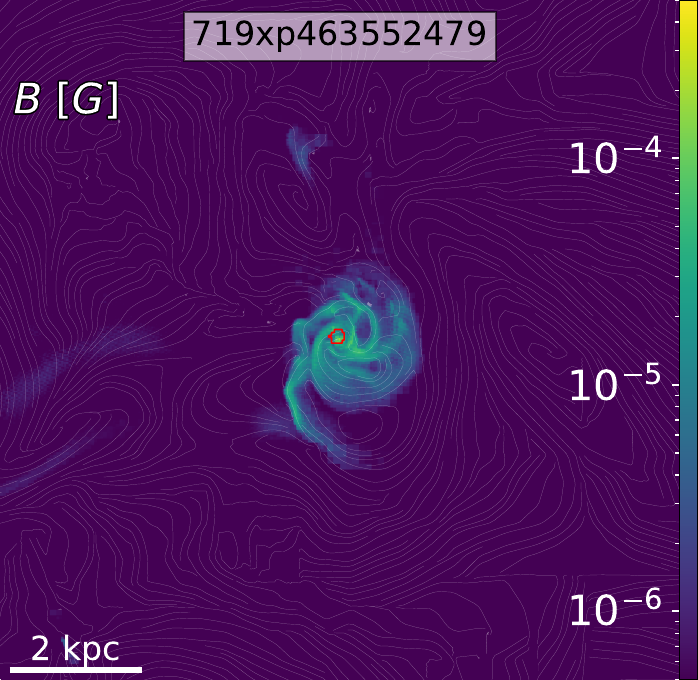}
\includegraphics[width=0.20\linewidth]{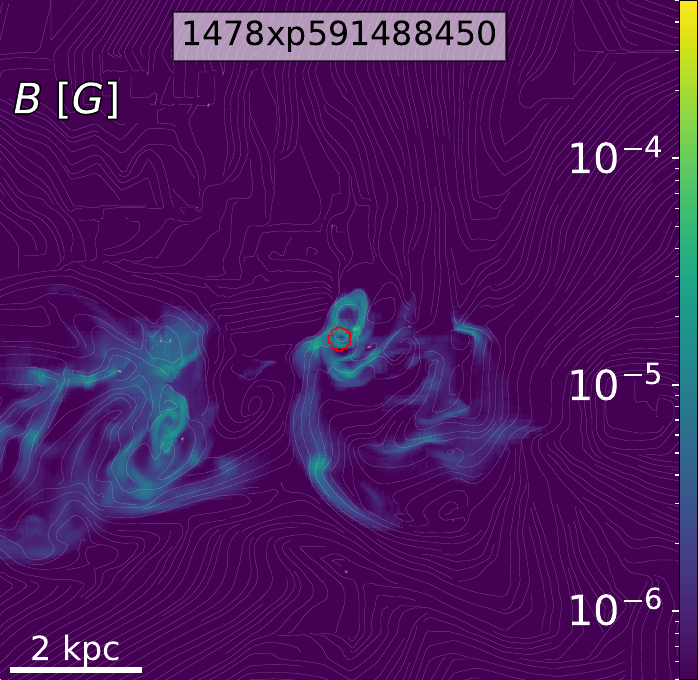}
\caption{Magnetic field strength maps of the simulated galaxies shown faceon. White streamlines trace the projected magnetic field orientation. The red circles indicate the measurement apertures, defined for each galaxy as circular regions with radius equal to the semi-major axis of the disc region used in the corresponding edge-on projection shown in Figure~\ref{fig:Bfieldsmaps_all_edgeon}
}
\label{fig:Bfieldsmaps_all_faceon}
\end{figure*}

Figure~\ref{fig:Bfieldsmaps_all_edgeon} presents the edge-on magnetic field strength maps of the simulated galaxies used in our analysis, with the measurement region indicated by the red contours. For completeness, the corresponding face-on maps are shown in Figure~\ref{fig:Bfieldsmaps_all_faceon}, where the measurement region is defined as a circular aperture with radius set by the semi-major axis of the edge-on disc region.
The B field morphology is also shown on both figures through white streamlines calculated using the density-weighted
magnetic field for the entire column displayed along the line-of-sight.
]

\twocolumn[

\section{Tables}
\label{sec:tables}

In this section we summarise the properties of the simulated and observed samples used throughout the paper. Table~\ref{tab:gal_properties} lists, for each simulated galaxy, the stellar mass $M_\star$, the SFR averaged over the pas $100$~Myr (SFR$_{100\:Myr}$), the corresponding specific SFR (sSFR$_{100\:Myr}$), and the star formation surface density $\Sigma_{\rm SFR}$ measured over the aperture adopted for the magnetic field estimates. Table~\ref{tab:Btot_disk} reports the median total magnetic field strength in the disc for the full neutral gas, and separately for the cold (CNM) and warm (WNM) phases, with IQR percentile ranges indicating the scatter. Table~\ref{tab:combined_mu} compiles the mass–specific magnetic, thermal, and turbulent energies ($\mu_{\rm mag}$, $\mu_{\rm th}$, $\mu_{\rm turb}$; in $\mathrm{erg,g^{-1}}$) for the same regions and phases, enabling a direct comparison of energy budgets across the sample. For reference, Table~\ref{tab:Btot_obs_properties} presents the observational compilation used for comparison—SFRs (with tracer noted), total magnetic field strengths, inclinations, stellar masses, and sSFRs—where duplicate entries correspond to distinct regions within a galaxy.

]

\renewcommand{\arraystretch}{1.5}
\begin{table*}[!b]
\centering
\caption{Unique galaxy ID (first column), stellar mass $M_*$ (second column), SFR integrated over the last $100 \: Myr$ (third column), specific SFR (fourth column)}
\begin{tabular}{lcccccccccc}
\hline
Galaxy & Stellar mass [$M_\odot$] & SFR$_{100\:Myr}$ [$M_\odot\,\mathrm{yr}^{-1}$] & sSFR$_{100\:Myr}$ [$\mathrm{Gyr}^{-1}$] & $\Sigma_{\mathrm{SFR}}$ [$M_\odot\,\mathrm{yr}^{-1}\,\mathrm{kpc}^{-2}$] \\
\hline
\textsc{Azahar-a} & $2.07\times 10^{10}$ & $6.02\times 10^{0}$ & $2.9\times 10^{-1}$ & $7.35\times 10^{-1}$ \\
105xp512461491 & $3.28\times 10^{9}$ & $9.91\times 10^{-1}$ & $3.02\times 10^{-1}$ & $1.55\times 10^{0}$ \\
g4 & $2.4\times 10^{9}$ & $3.25\times 10^{-1}$ & $1.35\times 10^{-1}$ & $7.87\times 10^{-1}$ \\
360xp489291551 & $1.28\times 10^{9}$ & $6.48\times 10^{-2}$ & $5.08\times 10^{-2}$ & $7.08\times 10^{-2}$ \\
515xp400607539 & $1.19\times 10^{9}$ & $8.11\times 10^{-2}$ & $6.82\times 10^{-2}$ & $3.83\times 10^{-1}$ \\
645xp327594444 & $1.12\times 10^{9}$ & $2.12\times 10^{-1}$ & $1.9\times 10^{-1}$ & $2.39\times 10^{-1}$ \\
2539xp316642481 & $1.05\times 10^{9}$ & $2.16\times 10^{-2}$ & $2.06\times 10^{-2}$ & $4.35\times 10^{-2}$ \\
974xp560563478 & $1.04\times 10^{9}$ & $4.89\times 10^{-2}$ & $4.71\times 10^{-2}$ & $1.53\times 10^{0}$ \\
2093xp473266584 & $9.14\times 10^{8}$ & $6.97\times 10^{-2}$ & $7.63\times 10^{-2}$ & $4.32\times 10^{-1}$ \\
772xp460511494 & $8.61\times 10^{8}$ & $2.45\times 10^{-1}$ & $2.85\times 10^{-1}$ & $1.59\times 10^{0}$ \\
302xp381498519 & $7.82\times 10^{8}$ & $6.64\times 10^{-2}$ & $8.5\times 10^{-2}$ & $2.8\times 10^{-1}$ \\
190xp435487473 & $7.42\times 10^{8}$ & $1.67\times 10^{-2}$ & $2.25\times 10^{-2}$ & $5.16\times 10^{-2}$ \\
831xp499496390 & $5.64\times 10^{8}$ & $6.44\times 10^{-2}$ & $1.14\times 10^{-1}$ & $4.08\times 10^{-1}$ \\
g5 & $4.82\times 10^{8}$ & $4.1\times 10^{-1}$ & $8.51\times 10^{-1}$ & $3.45\times 10^{-1}$ \\
886xp574558438 & $4.43\times 10^{8}$ & $1.34\times 10^{-2}$ & $3.04\times 10^{-2}$ & $2.59\times 10^{-1}$ \\
46xp452441341 & $3.54\times 10^{8}$ & $2.73\times 10^{-2}$ & $7.72\times 10^{-2}$ & $7.31\times 10^{-1}$ \\
g8 & $3.03\times 10^{8}$ & $1.1\times 10^{-2}$ & $3.63\times 10^{-2}$ & $3.95\times 10^{-2}$ \\
719xp463552479 & $2.64\times 10^{8}$ & $1.22\times 10^{-3}$ & $4.63\times 10^{-3}$ & $3.59\times 10^{-2}$ \\
1478xp591488450 & $2.41\times 10^{8}$ & $1.22\times 10^{-3}$ & $5.07\times 10^{-3}$ & $1.5\times 10^{-2}$ \\
\hline
\end{tabular}
\label{tab:gal_properties}
\end{table*}

\FloatBarrier

\begin{table*}[htbp]
\centering
\caption{Median magnetic field strength in the disc for the full gas phase, CNM, and WNM (columns 2–4).}
\begin{tabular}{l|ccc}
\hline
Galaxy & $B$ [G] & $B_{\mathrm{CNM}}$ [G] & $B_{\mathrm{WNM}}$ [G] \\
\hline
\textsc{Azahar-a}        & $1.5^{+0.55}_{-0.36}\times 10^{-4}$  & $1.41^{+0.56}_{-0.34}\times 10^{-4}$ & $1.54^{+0.5}_{-0.33}\times 10^{-4}$ \\
105xp512461491  & $9.33^{+1.9}_{-1.25}\times 10^{-5}$  & $8.85^{+1.56}_{-1.56}\times 10^{-5}$ & $9.5^{+2.12}_{-1.78}\times 10^{-5}$ \\
g4              & $1.15^{+0.3}_{-0.16}\times 10^{-4}$  & $1.28^{+0.2}_{-0.32}\times 10^{-4}$  & $8.39^{+2.11}_{-3.13}\times 10^{-5}$ \\
360xp489291551  & $2.52^{+0.69}_{-0.58}\times 10^{-5}$ & $2.68^{+0.73}_{-0.6}\times 10^{-5}$   & $1.84^{+0.62}_{-0.5}\times 10^{-5}$ \\
515xp400607539  & $5.41^{+1.02}_{-0.89}\times 10^{-5}$ & $5.99^{+0.97}_{-1.07}\times 10^{-5}$ & $5.14^{+1.46}_{-1.51}\times 10^{-5}$ \\
645xp327594444  & $3.09^{+0.75}_{-0.44}\times 10^{-5}$ & $3.41^{+0.74}_{-0.51}\times 10^{-5}$ & $2.66^{+0.75}_{-0.59}\times 10^{-5}$ \\
2539xp316642481 & $1.5^{+0.77}_{-0.22}\times 10^{-5}$  & $1.52^{+1.04}_{-0.23}\times 10^{-5}$ & $1.44^{+0.44}_{-0.24}\times 10^{-5}$ \\
974xp560563478  & $4^{+1.16}_{-1.37}\times 10^{-5}$    & $4.23^{+0.93}_{-1.75}\times 10^{-5}$ & $2.96^{+1.52}_{-0.65}\times 10^{-5}$ \\
2093xp473266584 & $9.6^{+4.3}_{-1.72}\times 10^{-6}$    & $1.08^{+0.33}_{-0.24}\times 10^{-5}$ & $5.21^{+4.88}_{-2.54}\times 10^{-6}$ \\
772xp460511494  & $2.91^{+1.66}_{-0.65}\times 10^{-5}$ & $2.62^{+1.12}_{-0.71}\times 10^{-5}$ & $2.53^{+2.06}_{-0.86}\times 10^{-5}$ \\
302xp381498519  & $1.57^{+0.68}_{-0.37}\times 10^{-5}$ & $1.81^{+0.58}_{-0.51}\times 10^{-5}$ & $1.47^{+0.45}_{-0.56}\times 10^{-5}$ \\
190xp435487473  & $3.15^{+0.86}_{-0.85}\times 10^{-5}$ & $3.31^{+0.97}_{-0.97}\times 10^{-5}$ & $2.24^{+0.9}_{-0.59}\times 10^{-5}$ \\
831xp499496390  & $4.41^{+0.92}_{-0.76}\times 10^{-5}$ & $4.3^{+0.68}_{-0.76}\times 10^{-5}$  & $4.25^{+1.03}_{-0.72}\times 10^{-5}$ \\
g5              & $1.88^{+0.67}_{-0.53}\times 10^{-5}$ & $2.09^{+0.77}_{-0.63}\times 10^{-5}$ & $1.28^{+0.52}_{-0.36}\times 10^{-5}$ \\
886xp574558438  & $1.24^{+0.58}_{-0.2}\times 10^{-5}$  & $8.46^{+1.37}_{-0.67}\times 10^{-6}$ & $9.91^{+2.02}_{-1.72}\times 10^{-6}$ \\
46xp452441341   & $3.55^{+1.13}_{-0.48}\times 10^{-5}$ & $3.53^{+0.92}_{-0.62}\times 10^{-5}$ & $3.29^{+1.07}_{-0.72}\times 10^{-5}$ \\
g8              & $1.07^{+0.25}_{-0.2}\times 10^{-5}$  & $1.15^{+0.27}_{-0.22}\times 10^{-5}$ & $6.63^{+2.11}_{-1.36}\times 10^{-6}$ \\
719xp463552479  & $2.71^{+1.69}_{-0.49}\times 10^{-5}$ & $2.31^{+0.84}_{-0.44}\times 10^{-5}$ & $2.73^{+1.86}_{-1.64}\times 10^{-5}$ \\
1478xp591488450 & $6.57^{+3.26}_{-0.89}\times 10^{-6}$ & $7.68^{+3.08}_{-1.59}\times 10^{-6}$ & $5.7^{+0.77}_{-2.24}\times 10^{-6}$ \\
\hline
\end{tabular}
\label{tab:Btot_disk}

\vspace{4mm}

\captionsetup{type=table}
\captionof{table}{Mass--specific magnetic ($\mu_{\mathrm{mag}}$), thermal ($\mu_{\mathrm{th}}$), and turbulent ($\mu_{\mathrm{turb}}$) energies (in $\mathrm{erg\,g^{-1}}$) for each galaxy, measured in the disc region.}
\setlength{\tabcolsep}{6pt}
\footnotesize
\begin{adjustbox}{max width=\textwidth}
\begin{tabular}{l|ccc|ccc|ccc}
\hline
 & \multicolumn{3}{c|}{Full gas} & \multicolumn{3}{c|}{CNM} & \multicolumn{3}{c}{WNM} \\
\cline{2-4} \cline{5-7} \cline{8-10}
Galaxy &
\shortstack{${\mu_{\mathrm{mag}}}$ \\ $[{\rm erg}\,{\rm g}^{-1}]$} & 
\shortstack{${\mu_{\mathrm{th}}}$  \\ $[{\rm erg}\,{\rm g}^{-1}]$} & 
\shortstack{${\mu_{\mathrm{turb}}}$ \\ $[{\rm erg}\,{\rm g}^{-1}]$} &
\shortstack{${\mu_{\mathrm{mag}}}$ \\ $[{\rm erg}\,{\rm g}^{-1}]$} & 
\shortstack{${\mu_{\mathrm{th}}}$  \\ $[{\rm erg}\,{\rm g}^{-1}]$} & 
\shortstack{${\mu_{\mathrm{turb}}}$ \\ $[{\rm erg}\,{\rm g}^{-1}]$} &
\shortstack{${\mu_{\mathrm{mag}}}$ \\ $[{\rm erg}\,{\rm g}^{-1}]$} & 
\shortstack{${\mu_{\mathrm{th}}}$  \\ $[{\rm erg}\,{\rm g}^{-1}]$} & 
\shortstack{${\mu_{\mathrm{turb}}}$ \\ $[{\rm erg}\,{\rm g}^{-1}]$} \\
\hline
\textsc{Azahar-a} & 3.38e+13 & 1.86e+12 & 3.17e+12 & 2.72e+13 & 8.26e+09 & 5.62e+12 & 4.15e+13 & 1.20e+11 & 1.21e+12 \\
105xp512461491 & 2.89e+12 & 9.78e+11 & 2.82e+12 & 2.12e+12 & 6.23e+09 & 2.42e+12 & 3.37e+12 & 2.62e+11 & 2.41e+12 \\
g4 & 3.09e+12 & 3.25e+11 & 1.33e+13 & 3.15e+12 & 1.06e+10 & 1.11e+13 & 3.74e+12 & 1.60e+11 & 1.82e+13 \\
360xp489291551 & 5.71e+11 & 7.35e+11 & 1.56e+12 & 5.75e+11 & 4.43e+09 & 9.31e+11 & 6.82e+11 & 1.96e+11 & 2.53e+12 \\
515xp400607539 & 1.13e+12 & 1.16e+12 & 1.62e+12 & 1.20e+12 & 5.88e+09 & 1.37e+12 & 1.24e+12 & 2.73e+11 & 1.70e+12 \\
645xp327594444 & 7.28e+11 & 2.12e+12 & 4.71e+12 & 8.02e+11 & 5.02e+09 & 3.63e+12 & 7.87e+11 & 2.63e+11 & 5.50e+12 \\
2539xp316642481 & 4.28e+11 & 3.09e+11 & 2.83e+12 & 3.80e+11 & 6.59e+09 & 2.19e+12 & 7.30e+11 & 9.72e+10 & 5.19e+12 \\
974xp560563478 & 2.12e+11 & 7.37e+11 & 1.23e+13 & 1.90e+11 & 1.19e+10 & 1.46e+13 & 2.55e+11 & 1.85e+11 & 6.21e+12 \\
2093xp473266584 & 2.94e+11 & 7.26e+11 & 6.02e+12 & 1.56e+11 & 3.11e+09 & 3.58e+12 & 4.25e+11 & 2.74e+11 & 9.53e+12 \\
772xp460511494 & 6.85e+11 & 1.02e+12 & 3.93e+12 & 2.60e+11 & 2.91e+09 & 4.08e+12 & 8.05e+11 & 4.29e+11 & 3.56e+12 \\
302xp381498519 & 3.30e+11 & 1.19e+12 & 1.72e+12 & 3.27e+11 & 3.62e+09 & 9.89e+11 & 3.65e+11 & 3.27e+11 & 2.44e+12 \\
190xp435487473 & 4.60e+11 & 4.23e+11 & 4.30e+12 & 4.45e+11 & 8.26e+09 & 5.21e+12 & 5.26e+11 & 2.31e+11 & 2.28e+12 \\
831xp499496390 & 9.31e+11 & 2.60e+12 & 4.40e+11 & 9.08e+11 & 4.90e+09 & 1.73e+11 & 9.04e+11 & 2.57e+11 & 5.46e+11 \\
g5 & 3.98e+11 & 1.02e+12 & 3.75e+11 & 4.18e+11 & 4.81e+09 & 1.54e+11 & 3.72e+11 & 2.80e+11 & 4.30e+11 \\
886xp574558438 & 2.11e+11 & 1.39e+12 & 5.78e+11 & 8.39e+10 & 3.81e+09 & 2.45e+11 & 9.39e+10 & 4.71e+11 & 5.11e+11 \\
46xp452441341 & 3.86e+11 & 7.84e+11 & 3.58e+12 & 4.68e+11 & 6.69e+09 & 3.22e+12 & 3.62e+11 & 3.59e+11 & 4.09e+12 \\
g8 & 1.67e+11 & 1.02e+12 & 5.18e+11 & 1.48e+11 & 3.72e+09 & 2.22e+11 & 1.90e+11 & 4.78e+11 & 5.54e+11 \\
719xp463552479 & 2.50e+11 & 1.31e+12 & 6.82e+11 & 1.35e+11 & 5.00e+09 & 3.19e+11 & 3.83e+11 & 4.13e+11 & 8.31e+11 \\
1478xp591488450 & 1.29e+10 & 5.18e+11 & 1.29e+12 & 1.20e+10 & 1.04e+10 & 1.29e+12 & 1.51e+10 & 1.68e+11 & 1.44e+12 \\
\hline
\end{tabular}
\end{adjustbox}
\label{tab:combined_mu}
\end{table*}

\begingroup
\onecolumn

\begingroup
\setlength{\tabcolsep}{5.5pt}
\renewcommand{\arraystretch}{1.25}

\begin{longtable}{@{}lcccccc@{}}
\caption{Star formation rates (SFRs) and total magnetic field strengths for the galaxy sample. Duplicate entries indicate different regions.}
\label{tab:Btot_obs_properties}\\
\toprule
Galaxy & SFR [$M_\odot\,\mathrm{yr}^{-1}$] & $B$ [$\mu$G] & Inclination [°] & $\log M_\star$ [$M_\odot$] & sSFR [$\mathrm{Gyr}^{-1}$] & $\Sigma_{\mathrm{SFR}} \,[\mathrm{M_\odot\,yr^{-1}\,kpc^{-2}}]$ \\
\midrule
\endfirsthead

\caption[]{Star formation rates (SFRs) and total magnetic field strengths for the galaxy sample (continued).}\\
\toprule
Galaxy & SFR [$M_\odot\,\mathrm{yr}^{-1}$] & $B$ [$\mu$G] & Inclination [°] & $\log M_\star$ [$M_\odot$] & sSFR [$\mathrm{Gyr}^{-1}$] & $\Sigma_{\mathrm{SFR}} \,[\mathrm{M_\odot\,yr^{-1}\,kpc^{-2}}]$ \\
\midrule
\endhead

\midrule
\multicolumn{7}{r}{Continued on next page}\\
\endfoot

\bottomrule
\endlastfoot

M\,82\tsB{*}  & $1.30\times10^{1}$\tsS{a} & $305$ & $81$\tsI{a} & 10.01\tsM{a} & $1.27$ & -- \\  
NGC\,253      & $3.00$\tsS{q}             & $186$ & $78$\tsI{b} & 10.47\tsM{a} & $0.10$ & -- \\
NGC\,2146     & $2.00\times10^{1}$\tsS{r} & $354$ & $63$\tsI{c} & 10.72\tsM{a} & $0.38$ & -- \\

\midrule
\multicolumn{7}{c}{Equipartition magnetic field from \citealp{Beck2019}}\\
\hline 
M\,31                   & $3.90\times10^{-1}$\tsS{b} & $7.2$  & $77$\tsI{d}   & 10.73\tsM{a} & $0.01$ & $4.97\times10^{-4}$ \\
M\,33                   & $3.20\times10^{-1}$\tsS{b} & $8.15$ & $52$\tsI{e}   & 9.42\tsM{a}  & $0.12$ & $4.08\times10^{-3}$ \\
M\,51                   & $2.70$\tsS{c}              & $15.5$ & $22$\tsI{f}   & 10.47\tsM{b} & $0.09$ & $1.66\times10^{-2}$ \\
M\,81                   & $5.40\times10^{-1}$\tsS{b} & $7.2$  & $59$\tsI{g}   & 10.68\tsM{a} & $0.01$ & $1.20\times10^{-3}$ \\
M\,83                   & $4.17$\tsS{b}              & $17.5$ & $24$\tsI{h}   & 10.41\tsM{a} & $0.16$ & $9.23\times10^{-3}$ \\
NGC\,253 (disc)         & $4.90$\tsS{b}              & $15.0$ & $78$\tsI{b}   & 10.47\tsM{a} & $0.17$ & $3.47\times10^{-2}$ \\
NGC\,1097               & $5.01$\tsS{b}              & $13.0$ & $48.6$\tsI{i} & 10.72\tsM{a} & $0.10$ & $6.38\times10^{-2}$ \\
NGC\,1365               & $1.41\times10^{1}$\tsS{b}  & $12.0$ & $40$\tsI{j}   & 10.75\tsM{a} & $0.25$ & $2.03\times10^{-2}$ \\
NGC\,4254               & $5.01$\tsS{b}              & $17.5$ & $42$\tsI{k}   & 10.52\tsM{a} & $0.15$ & $3.07\times10^{-2}$ \\
NGC\,4449 (global disc) & $4.30\times10^{-1}$\tsS{b} & $13.5$ & $56.2$\tsI{l} & 9.03\tsM{a}  & $0.40$ & $1.52\times10^{-2}$ \\
NGC\,6946               & $6.17$\tsS{b}              & $16.0$ & $38$\tsI{m}   & 10.50\tsM{a} & $0.20$ & $2.23\times10^{-2}$ \\
IC\,342 (disc)          & $1.91$\tsS{b}              & $13.0$ & $31$\tsI{n}   & 9.81\tsM{a}  & $0.30$ & $1.99\times10^{-3}$ \\

\midrule
\multicolumn{7}{c}{Equipartition magnetic field from \citealp{Lacki2013}}\\
\hline
M\,82                       & $7.08$\tsS{b}             & $240$ & $81$\tsI{a}   & 10.01\tsM{a} & $0.69$  & $3.60\times10^{1}$ \\
NGC\,253 (core/starburst)   & $4.90$\tsS{b}             & $230$ & $78$\tsI{b}   & 10.47\tsM{a} & $0.17$  & $6.94\times10^{1}$ \\
NGC\,4945                   & $1.45$\tsS{b}             & $130$ & $75$\tsI{o}   & 10.14\tsM{a} & $0.10$  & $1.58\times10^{0}$ \\
NGC\,1068                   & $3.24\times10^{1}$\tsS{b} & $72$  & $40$\tsI{p}   & 10.71\tsM{a} & $0.63$  & $1.14\times10^{0}$ \\
IC\,342 (central starburst) & $1.91$\tsS{b}             & $110$ & $31$\tsI{n}   & 9.81\tsM{a}  & $0.30$  & $1.20\times10^{0}$ \\
NGC\,2146                   & $2.51\times10^{1}$\tsS{b} & $190$ & $63$\tsI{c}   & 10.72\tsM{a} & $0.48$  & $2.96\times10^{1}$ \\
NGC\,3690                   & $1.55\times10^{2}$\tsS{b} & $130$ & --             & 11.04\tsM{a} & $1.41$  & $8.14\times10^{0}$ \\
NGC\,1808                   & $7.94$\tsS{b}             & $110$ & $57$\tsI{z}   & 10.29\tsM{a} & $0.41$  & $2.38\times10^{0}$ \\
NGC\,3079                   & $7.41$\tsS{b}             & $350$ & $84$\tsI{aa}  & 10.89\tsM{a} & $0.10$  & $6.52\times10^{1}$ \\
Arp\,220\,West              & $1.10\times10^{2}$\tsS{d} & $750$ & $>52$\tsI{q}  & 9.30\tsM{d}  & $55.0$  & $3.50\times10^{3}$ \\
Arp\,220\,East              & $8.10\times10^{1}$\tsS{d} & $720$ & $31$\tsI{r}   & 9.30\tsM{d}  & $40.5$  & $2.58\times10^{3}$ \\
Arp\,193                    & $8.00\times10^{1}$\tsS{l} & $550$ & --             & 10.45\tsM{e} & $2.84$  & $9.94\times10^{2}$ \\
Mrk\,273                    & $1.39\times10^{2}$\tsS{k} & $770$ & $78$\tsI{s}   & 10.84\tsM{e} & $2.01$  & $1.73\times10^{3}$ \\

\midrule
\multicolumn{7}{c}{Equipartition magnetic field from \citealp{Chyzy2011}}\\ 
\hline
IC\,1613   & $2.82\times10^{-3}$\tsS{b} & $2.8$   & $58$\tsI{t}    & 8.03\tsM{a}   & $0.03$ & $8.51\times10^{-5}$ \\
NGC\,6822  & $1.40\times10^{-2}$\tsS{e} & $4.0$   & --              & 8.00\tsM{c}   & $0.14$ & $1.01\times10^{-3}$ \\
IC\,10     & $5.00\times10^{-2}$\tsS{f} & $9.7$   & --              & 7.93\tsM{c}   & $0.11$ & $1.08\times10^{-2}$ \\
LMC        & $2.00\times10^{-1}$\tsS{g} & $4.3$   & $35.8$\tsI{u}  & 9.18\tsM{c}   & $0.13$ & $6.82\times10^{-3}$ \\
SMC        & $5.00\times10^{-2}$\tsS{h} & $3.2$   & --              & 8.66\tsM{c}   & $0.58$ & $1.18\times10^{-4}$ \\
NGC\,4449  & $4.27\times10^{-1}$\tsS{b} & $9.3$   & $56.2$\tsI{l}  & 9.03\tsM{a}   & $2.34$ & $1.11\times10^{-2}$ \\
NGC\,1569  & $9.55\times10^{-1}$\tsS{b} & $14$    & $63.2$\tsI{v}  & 8.61\tsM{a}   & $0.40$ & $2.76\times10^{-1}$ \\
Aquarius   & $1.41\times10^{-4}$\tsS{n} & $<4.5$  & --              & 6.02\tsM{a}   & $0.14$ & $2.46\times10^{-4}$ \\
GR\,8      & $2.40\times10^{-3}$\tsS{m} & $<3.6$  & --              & 6.81\tsM{c}   & $0.37$ & $1.70\times10^{-3}$ \\
WLM        & $3.47\times10^{-3}$\tsS{b} & $<3.9$  & $74$\tsI{w}    & 7.31\tsM{a}   & $0.17$ & $3.34\times10^{-4}$ \\
LGS\,3     & $6.00\times10^{-6}$\tsS{p} & $<4.0$  & --              & 5.98\tsM{c}   & $0.01$ & $1.47\times10^{-5}$ \\
SagDIG     & $9.00\times10^{-4}$\tsS{o} & $<4.1$  & --              & 6.26\tsM{f}   & $0.50$ & $5.10\times10^{-4}$ \\
Sextans\,A & $6.61\times10^{-3}$\tsS{b} & $<3.1$  & --              & 8.04\tsM{a}   & $0.06$ & $5.01\times10^{-4}$ \\
Sextans\,B & $1.91\times10^{-3}$\tsS{b} & $<2.8$  & --              & 7.37\tsM{a}   & $0.08$ & $2.18\times10^{-4}$ \\
Leo\,A     & $2.95\times10^{-4}$\tsS{b} & $<4.4$  & --              & 6.45\tsM{a}   & $0.10$ & $1.44\times10^{-4}$ \\
Pegasus    & $6.17\times10^{-4}$\tsS{i} & $<3.7$  & --              & 6.82\tsM{c}   & $0.09$ & $2.96\times10^{-4}$ \\

\midrule
\multicolumn{7}{c}{Equipartition magnetic field from CHANGES survey}\\
\hline
NGC\,5775 & $5.28$\tsS{j}  & $13$\tsB{e}   & $85.8$\tsI{x} & 10.47\tsM{g} & $0.26$ & $9.41\times10^{-3}$ \\
NGC\,4217 & $1.50$\tsS{j}  & $9$\tsB{f}    & $89$\tsI{ab}  & 10.52\tsM{g} & $0.046$& $3.60\times10^{-3}$ \\
NGC\,4666 & $7.30$\tsS{j}  & $12.3$\tsB{g} & $85$\tsI{ac}  & 10.53\tsM{g} & $0.21$ & $8.90\times10^{-3}$ \\
NGC\,4631 & $1.33$\tsS{j}  & $9$\tsB{h}    & $89$\tsI{y}   & 10.05\tsM{g} & $0.12$ & $3.10\times10^{-3}$ \\

\end{longtable}
\endgroup

\begin{minipage}{\textwidth}
\footnotesize
\textbf{Notes.} SFR tracers used: FUV+WISE4 (hybrid, $\sim$100 Myr), UV+IR (hybrid), GALEX FUV ($\sim$100 Myr), CMD–SFH (resolved stars), FIR (10–100 Myr), IR/TIR (8–1000 $\mu$m), radio 1.4 GHz (non-thermal, $\sim$100 Myr). Magnetic field strengths are synchrotron equipartition values. Inclinations are in degrees; “—” = not available/meaningful.

\textbf{SFR methods by superscript:}
\tsS{a}\,\citep{Schreiber2003} IR/TIR (8–1000\,$\mu$m);\;
\tsS{b}\,\citep{Leroy2019} FUV+WISE4 hybrid;\;
\tsS{c}\,\citep{Eufrasio2017} UV+IR hybrid;\;
\tsS{d}\,\citep{Varenius2016} radio 1.4\,GHz;\;
\tsS{e}\,\citep{Efremova2011} GALEX FUV;\;
\tsS{f}\,\citep{Yin2010} IR/TIR;\;
\tsS{g}\,\citep{Harris2009} CMD-based SFH (HST);\;
\tsS{h}\,\citep{Wilke2004} FIR;\;
\tsS{i}\,\citep{Hunter2012} GALEX FUV;\;
\tsS{j}\,\citep{Wiegert2015} FUV+WISE4 hybrid;\;
\tsS{k}\,\citep{Cicone2014} IR/TIR;\;
\tsS{l}\,\citep{Modica2012} Radio+IR composite;\;
\tsS{m}\,\citep{Ott2012} GALEX FUV;\;
\tsS{n}\,\citep{Leroy2021} FUV+WISE4 hybrid;\;
\tsS{o}\,\citep{Tang2016} CMD-based SFH (HST);\;
\tsS{p}\,\citep{Miller2001} CMD-based SFH (HST/WFPC2);\;
\tsS{q}\,\citep{Bolatto2013} IR/TIR;\;
\tsS{r}\,\citep{Gorski2018} IR/TIR.

\textbf{$B_{\mathrm{tot}}$ superscripts:}
\tsB{a}\,\citep{Lopez-Rodriguez2023};\;
\tsB{b}\,\citep{Beck2019};\;
\tsB{c}\,\citep{Lacki2013};\;
\tsB{d}\,\citep{Chyzy2011};\;
\tsB{e}\,\citep{Heald2022};\;
\tsB{f}\,\citep{Stein2020};\;
\tsB{g}\,\citep{Stein2019};\;
\tsB{h}\,\citep{Mora-Partiarroyo2019}.

\textbf{Inclination superscripts:}
\tsI{a}\,\citep{Achtermann1995};
\tsI{b}\,\citep{Pence1980};
\tsI{c}\,\citep{DellaCeca1999};
\tsI{d}\,\citep{McConnachie2005};
\tsI{e}\,\citep{Kam2015};
\tsI{f}\,\citep{Colombo2014};
\tsI{g}\,\citep{Karachentsev2013};
\tsI{h}\,\citep{Tilanus1993};
\tsI{i}\,\citep{Lang2020};
\tsI{j}\,\citep{Jorsater1995};
\tsI{k}\,\citep{Phookun1993};
\tsI{l}\,\citep{Bajaja1994};
\tsI{m}\,\citep{Boomsma2008};
\tsI{n}\,\citep{Crosthwaite2000};
\tsI{o}\,\citep{Henkel2018};
\tsI{p}\,\citep{Schinnerer2000};
\tsI{q}\,\citep{Sakamoto2021};
\tsI{r}\,\citep{Scoville2015};
\tsI{s}\,\citep{Schmelz1987};
\tsI{t}\,\citep{Lake1989};
\tsI{u}\,\citep{Olsen2002};
\tsI{v}\,\citep{Sanchez-Cruces2022};
\tsI{w}\,\citep{Read2016};
\tsI{x}\,\citep{Irwin1994};
\tsI{y}\,\citep{Mora-Partiarroyo2019};
\tsI{z}\,\citep{Dahlem1990};
\tsI{aa}\,\citep{Israel1998};
\tsI{ab}\,\citep{Stein2020};
\tsI{ac}\,\citep{Stein2019}.

\textbf{$\log M_\star$ superscripts:}
\tsM{a}\,\citep{Leroy2021};
\tsM{b}\,\citep{Wei2021};
\tsM{c}\,\citep{McConnachie2012};
\tsM{d}\,\citep{Sakamoto1999};
\tsM{e}\,\citep{Vivian2012};
\tsM{f}\,\citep{Kirby2017};
\tsM{g}\,\citep{Leroy2019}.
\end{minipage}

\twocolumn
\endgroup

\FloatBarrier

\section{Magnetic fields versus star formation for face-on simulated galaxies}
\label{sec:faceon_test}

In this section, we assess the impact of projection effects on the magnetic field–star formation relation by repeating our measurements in face-on projections of the simulated \textsc{Azahar} galaxies (shown in Figure~\ref{fig:Bfieldsmaps_all_faceon}). This test is designed to directly evaluate whether the relations inferred from the edge-on analysis depend on galaxy inclination.

For each galaxy, the measurement region in the face-on view is defined as a circular aperture whose radius is set equal to the semi-major axis of the disc region used in the corresponding edge-on projection, as illustrated in Figure~\ref{fig:Bfieldsmaps_all_faceon}. This choice ensures that the same physical extent of the star-forming disc is probed in both orientations, allowing for a consistent comparison between face-on and edge-on measurements while avoiding inclination-dependent biases in the sampled area.

We find the following scalings of magnetic field strength with star formation:
\begin{equation*}
\begin{aligned}
B &\propto \mathrm{SFR}^{\,\alpha},        & \alpha &= 0.35 \pm 0.10. \\
\end{aligned}
\end{equation*}
Within uncertainties, all slopes are fully consistent with those obtained for the edge-on projections. This indicates that the inferred magnetic field–star formation relations in the \textsc{Azahar} galaxies are not significantly affected by projection effects, and the magnetic field scaling with star formation is largely independent of galaxy inclination.

We emphasise that the magnetic field strengths used in this analysis are measured directly from the intrinsic magnetic field in the simulations. In contrast, observational estimates are typically inferred indirectly and may be affected by additional systematics such as line-of-sight integration, cosmic-ray electron distributions, and RT effects (e.g. \citealp{Dacunha2025}).

We find that face-on projections exhibit a modestly increased scatter with respect to the edge-on case, which further improves the agreement between the simulated relations and the observational samples spanning a range of inclinations.

We interpret the increased scatter observed in the face-on projections as a consequence of the heterogeneous spatial distribution of star formation within the disc. When galaxies are viewed face-on, the SFR surface density appears more clumpy and spatially intermittent.
In this configuration, different galaxies with similar SFRs within the selected aperture can exhibit different degrees of star formation clustering. Such large local star formation activity variations will lead to significantly different local levels of turbulence and magnetic amplification across the spatial region, and therefore to a larger scatter.
In contrast, edge-on projections collapse one of the large disc dimensions along the line of sight, averaging over differently star-forming and magnetised. This, in turn, leads to tighter correlations.
The increased scatter in the face-on case therefore reflects projection-driven differences in how spatial variations within the star-forming disc are sampled, rather than a change in the underlying physical coupling between magnetic fields and star formation.

\begin{figure}[htbp]
\centering
\includegraphics[width=0.44\linewidth]{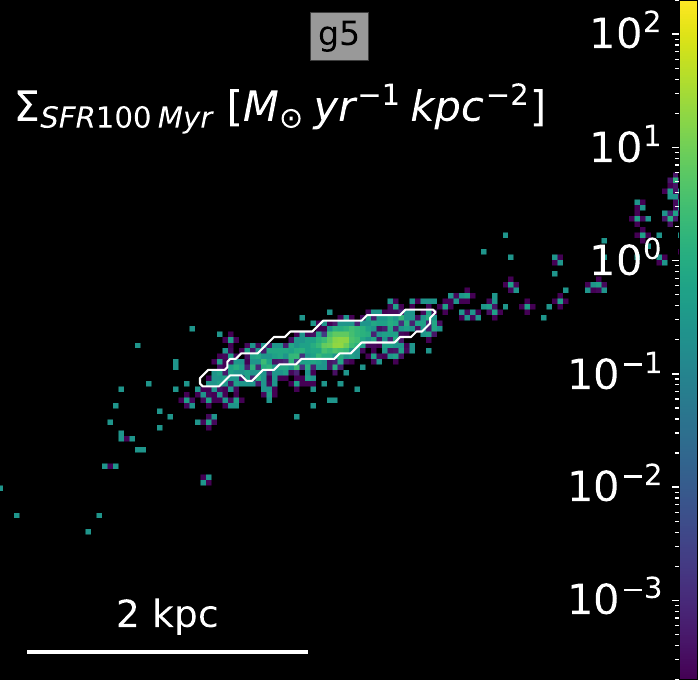}
\includegraphics[width=0.44\linewidth]{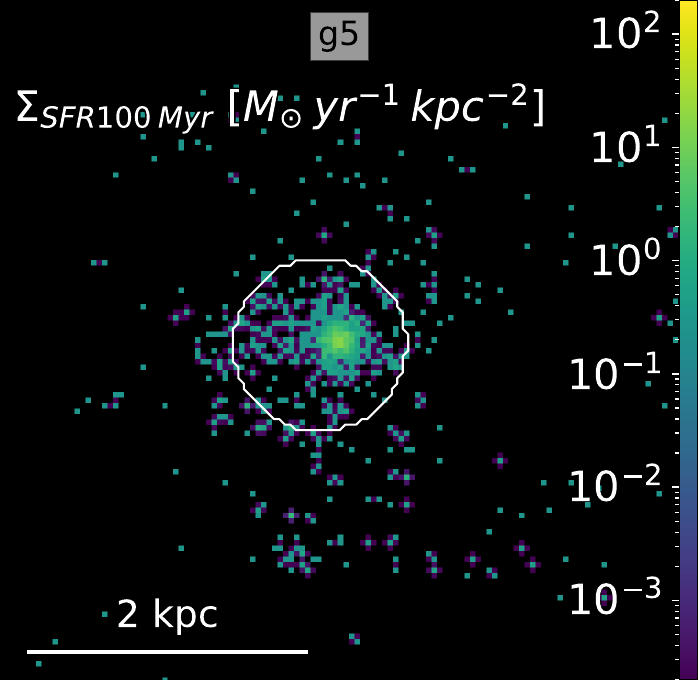}
\includegraphics[width=0.44\linewidth]{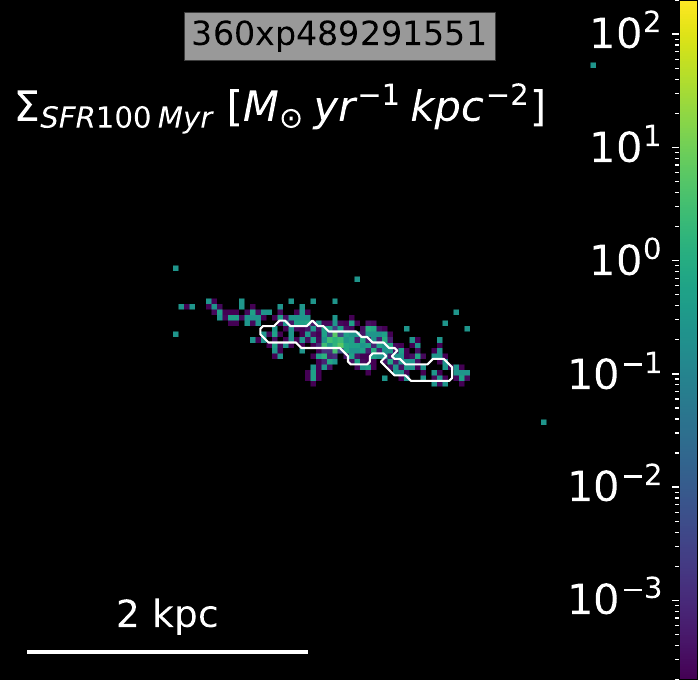}
\includegraphics[width=0.44\linewidth]{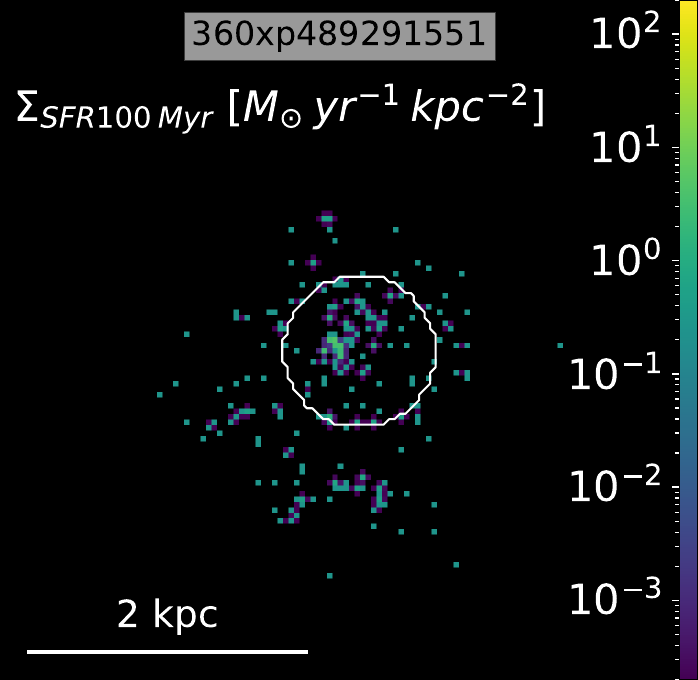}
\caption{Comparison of SFR surface density maps for galaxies g5 and 360xp489291551. Each row shows the same galaxy viewed in edge-on (left) and face-on (right) projections.}
\label{fig:inclination_maps}
\end{figure}

To illustrate this, Fig.~\ref{fig:inclination_maps} compares the $\Sigma_{\rm SFR}$ maps of two representative \textsc{Azahar} galaxies in edge-on and face-on projections, highlighting the increased clumpiness of star formation in the face-on view.

\begin{figure}[htbp]
\centering
\includegraphics[width=\linewidth]{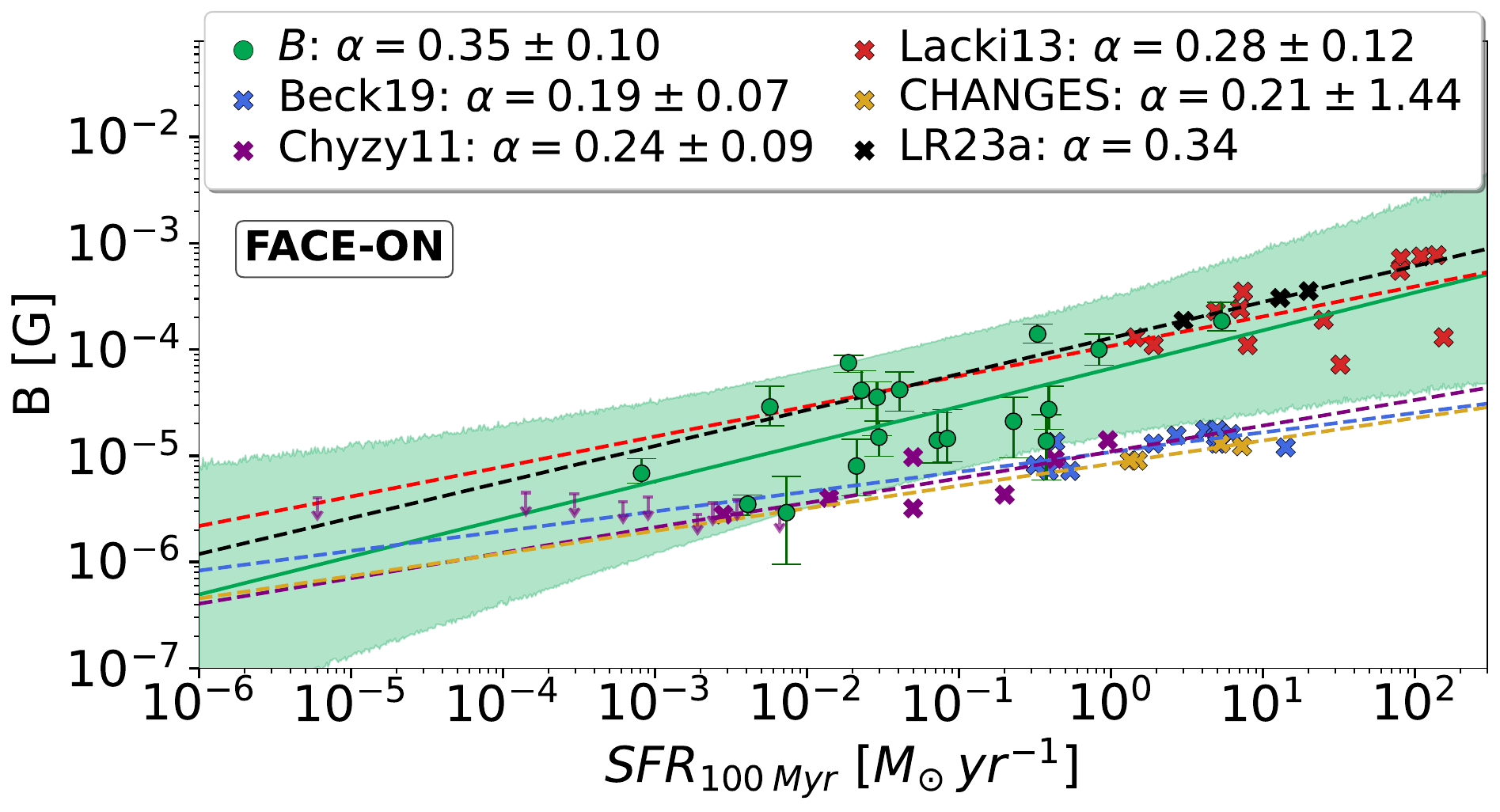}
\caption{Magnetic field strength as a function of integrated SFR over the last 100\,Myrs.
Each panel shows the median magnetic field strength $B$ measured in face-on projections of the simulated \textsc{Azahar} galaxy sample (green solid circles). The magnetic field is computed within a circular aperture whose radius is set equal to the semi-major axis of the disc region used for the corresponding edge-on projection (measured as described in Section~\ref{The simulated galaxy sample}). Error bars show IQR range. The solid green line shows the best fit, and the shaded region 
indicates the $95\%$ prediction interval, with their fitted slopes and the standard error reported in the legends.
Samples of observed galaxies taken from the literature and their best-fit relations (dashed lines), are overlaid with different color crosses: red \citep{Lacki2013}, black \citep{Lopez-Rodriguez2023}, purple \citep{Chyzy2011}, navy blue \citep{Beck2019}, and goldenrod (\citealp[CHANGES survey]{Heald2022, Stein2020, Stein2019, Mora-Partiarroyo2019}). Their respective best-fit relations are shown as dashed lines. Upper limits from \citet{Chyzy2011} are shown as downward arrows, and are excluded from the fits.}
\label{fig:Bfield_strength_SFR_faceon}
\end{figure}

\end{appendix}

\end{document}